\documentclass[12pt,final,fleqn]{article}

\usepackage[margin=1in] { geometry }
\usepackage{amssymb,amsmath, bm}
\usepackage{verbatim}
\usepackage[utf8]{inputenc}
\usepackage{setspace}
\usepackage{enumitem}
\usepackage[bottom]{footmisc}
\usepackage{url}
\usepackage[font={bf}]{caption}
\usepackage{float}
\usepackage{setspace}
\usepackage{latexsym}
\usepackage{graphicx}
\usepackage{marvosym}
\usepackage{amsmath} 
\usepackage{authblk}
\usepackage{xcolor}
\usepackage{blindtext}
\usepackage{changepage}

\usepackage[natbibapa]{apacite}
\bibpunct{(}{)}{;}{a}{}{,}

\usepackage{booktabs}
\usepackage{longtable}
\usepackage{array}
\usepackage{multirow}
\usepackage{wrapfig}
\usepackage{float}
\usepackage{colortbl}
\usepackage{pdflscape}
\usepackage{tabu}
\usepackage{threeparttable}
\usepackage{threeparttablex}
\usepackage[normalem]{ulem}
\usepackage{makecell}
\usepackage{xcolor}
\usepackage{siunitx}

\newcolumntype{d}{S[
    input-open-uncertainty=,
    input-close-uncertainty=,
    parse-numbers = false,
    table-align-text-pre=false,
    table-align-text-post=false
 ]}

\usepackage{dcolumn}
\newcolumntype{.}{D{.}{.}{-1}}
\newcolumntype{d}[1]{D{.}{.}{#1}}
\usepackage{multirow}
\usepackage[figuresright]{rotating}
\usepackage{pdflscape}
\usepackage{subcaption}

\usepackage{CJKutf8}

\usepackage{grffile}
\usepackage{afterpage}
\usepackage{float}
\usepackage[section]{placeins}

\usepackage{theorem}
\theoremstyle{plain}
\theoremheaderfont{\scshape}

\expandafter\def\expandafter\normalsize\expandafter{%
    \normalsize%
    \setlength\abovedisplayskip{2pt}%
    \setlength\belowdisplayskip{2pt}%
    \setlength\abovedisplayshortskip{-8pt}%
    \setlength\belowdisplayshortskip{2pt}%
}

\usepackage{pdflscape}

\usepackage[compact]{titlesec}

\usepackage{color}
\usepackage{hyperref}
\usepackage{xcolor}
\hypersetup{
    colorlinks,
    linkcolor={blue!50!black},
    citecolor={blue!50!black},
    urlcolor={blue!80!black}
}

\newcommand{\aref}[1]{\hyperref[#1]{Appendix~\ref{#1}}}

\usepackage{minitoc}
\usepackage[toc,page,header]{appendix}

\usepackage{chngcntr}
\usepackage{etoolbox}
\usepackage{lipsum}



\allowdisplaybreaks[4]
\titleformat{\subsection}
  {\itshape\large}{\thesubsection}{1em}{}

\begin{document}
\singlespace
\title{\textbf{How firms, bureaucrats, and ministries benefit from the revolving door: \\ Evidence from Japan}\vspace{-1ex}}

\author{Trevor Incerti\thanks{Assistant Professor, Department of Political Science, University of Amsterdam. t.n.incerti@uva.nl}\vspace{-1ex}}
\date{September 25, 2025}
\maketitle
\pagenumbering{gobble}

\vspace{-0.25cm}
\begin{center}
Forthcoming, \textit{American Political Science Review}
\end{center}
\vspace{0.25cm}

\begin{abstract}
\noindent A growing literature finds high returns to firms with legislative connections. Less attention has been paid to returns from bureaucratic connections and to organizations beyond for-profit firms. Using data recording the first post-bureaucracy position occupied by all former civil servants in Japan, I reveal a bifurcated job market for former bureaucrats. High-ranking officials from elite economic ministries are more likely to join for-profit firms, where they generate returns such as increased government loans and positive stock market reactions. Lower-ranking officials are more likely to join nonprofits linked to government ministries, which receive higher-value contracts when former bureaucrats are in leadership roles. These patterns suggest that while firms wish to hire bureaucrats who can deliver tangible benefits, ministries also shape revolving door pathways by directing benefits to ensure long-term career value for civil servants. These findings reframe revolving door dynamics as the result of both firm-driven demand and bureaucratic incentives.
\end{abstract}

\vspace{0.5cm}
\begin{center}
Keywords: Business and government; political connections; bureaucracy; procurement; money in politics; nonmarket strategy
\end{center}

\pagebreak
\doublespace
\pagenumbering{arabic}
\setcounter{page}{1}


A growing literature has established the high value of legislative connections to firms. However, elected office is not the only form of government connection that firms can leverage. Firms also regularly hire civil servants---a  practice commonly referred to as the bureaucratic revolving door. Despite the typically far larger number of civil servants in central government agencies compared to elected office holders,\footnote{For example, there are 537 federal elected officials and over 2 million civil servants in the United States, and 722 central government elected officials and over 250,000 civil servants in Japan.}  examination of the value of bureaucratic connections to firms is notably lacking from the political connections literature. Are bureaucrats similarly in demand by employers? If so, \textit{which} bureaucrats are in demand and what benefits do they bring to the firms that hire them? 

In theory, firms have incentives to hire former bureaucrats only when those individuals can deliver tangible benefits such as technical expertise, access to government loans or contracts, influence over regulatory decisions, or other political rents. Bureaucrats without such value should hold little appeal to private employers. By contrast, bureaucratic institutions have incentives to place as many former officials as possible into post-bureaucracy roles as the revolving door: (1) helps ministries maintain informal channels of influence, and (2) acts as a credible signal to prospective recruits that a relatively low-paid public sector career will be rewarded with lucrative post-retirement opportunities.

Using comprehensive data on the initial post-retirement placements of all former civil servants in Japan, I provide empirical evidence consistent with the theoretical predictions outlined above.  A bifurcated market exists for former bureaucrats, in which the highest ranking officials are naturally in demand by large, publicly traded corporations. These officials then deliver benefits such as low-interest government loans to their new employers. However, for lower ranking bureaucrats who cannot offer these kinds of benefits to firms, alternate sources of employment exist at nonprofit organizations bolstered by government contracts. Though these connections are under-examined in the existing literature, I show that roughly half of Japanese bureaucrats are re-hired by public corporations or nonprofits, and that these organizations leverage their bureaucratic connections to increase the size of government contracts they receive.  In the context of a weak welfare state with relatively low pay for civil servants, the promise of such revolving door positions---regardless of private sector demand---is essential for successful bureaucratic retainment and recruitment.

I evaluate the concrete benefits that former officials provide to their new employers, and show that flows of bureaucrats to different sectors of the economy lead to sector-specific benefits. First, using a matched differences-in-differences (DiD) approach \citep{imai2019matching}, I show that private firms hiring senior officials from elite economic ministries receive increased volumes of government loans in subsequent years. Second, using an interrupted time series design, I find that investors respond positively to appointments of high-ranking economy ministry bureaucrats. Third, I leverage novel data on nonprofit leadership and DiD approaches that account for the negative weighting issues highlighted in recent literature \citep{dechaisemartin2020twoway} to show that nonprofits with ex-bureaucrats in director roles receive more lucrative government contracts. These contracts also exhibit financial irregularities according to common forensic accounting techniques, patterns not observed in comparable nonprofits without bureaucratic ties.

Most research on the bureaucratic revolving door relies on theoretical models, focusing on how post-government hiring shapes regulatory leniency \citep{dal2006regulatory, che1995revolving, salant1995behind} or the conditions under which bureaucratic connections should be expected to bring value to firms \citep{bils2020working}. Empirical evidence is limited and often drawn from convenience samples or single-agency case studies. Little attention has been paid to employment destinations beyond for-profit firms or to the proactive role bureaucratic agencies may play in cultivating revolving door opportunities.

This study addresses these gaps using comprehensive administrative data and a broader view of the revolving door, spanning both for-profit and nonprofit destinations. I combine newly constructed datasets of all initial revolving door hires, all government loans to private firms, stock prices of all firms that make high-level bureaucratic hires, and all government contracts with nonprofits in Japan over a period of one decade to test for benefits that accrue to organizations that hire former bureaucrats. To identify the kinds of benefits these hires may generate, I supplement the quantitative analysis with interviews with current and former bureaucrats, business leaders, and nonprofit executives. These interviews point to two recurring themes: former bureaucrats offer valuable political connections that may help secure contracts or loans, but are often perceived as lacking technical or managerial expertise. 

Collectively, interviews, descriptive analysis, and causal estimates reveal a bifurcated job market for former bureaucrats in which only a select few are highly valued by for-profit firms. Top bureaucrats from ministries that control the levers of finance, industrial policy, and regulation are in high demand, and these individuals are in turn able to drive benefits to for-profit firms. However, for those less desired by the private sector, positions exist in nonprofits, where former bureaucrats appear able to leverage their connections to help these agencies secure continued government funding, as well as maintain employment opportunities for future generations of retiring bureaucrats. 

These findings shed light on how the public and private sectors interact in advanced democracies, particularly those with large public-private pay gaps. While existing research on the revolving door focuses on rent-seeking firms hiring former officials for their expertise or connections, this paper highlights an additional dynamic. Only a subset of bureaucrats---typically those with influence, status, or control over key sectors---are in demand by private firms. This generates a supply of would-be revolving door candidates for whom no natural market demand exists. However, to sustain elite recruitment amid declining public sector compensation and prestige, ministries must offer credible guarantees of future income and employment to all recruits.  Ministries may therefore respond to this imbalance by actively constructing revolving door pathways---e.g., outsourcing  contracts and establishing informal pipelines to nonprofit or quasi-public roles---to provide employees with a credible path to long-term career value. These patterns suggest that the state is more than a passive participant in the revolving door phenomenon, which is in fact a function of both firm-driven rent seeking and state-led institutional maintenance. 

\section{Theory and hypotheses} \label{sec: theory}

\subsection{The value of political connections to firms}

A large literature explores the economic value of legislative political connections to firms. For example, \citet{blanes2012revolving} show that lobbying revenues generated by ex-congressional staffers fall sharply when their former employers leave office. In China, firms with CEOs who serve in the National People’s Congress exhibit higher stock prices and operating profits \citep{truex2014returns}. Similarly, \citet{faccio2006politically} and \citet{faccio2006political} find that firms benefit from the political ascension of major shareholders or executives, through stock price increases and higher likelihood of bailouts, respectively. Campaign donations also appear to yield returns, as Brazilian firms that support winning candidates receive more government contracts \citep{boas2014spoils}.

Alongside this legislative literature, growing attention has been paid to the revolving door between the bureaucracy and industry. Legislators and bureaucrats have different incentives for entering the revolving door, but both connections may be of value to firms. Classic models of regulation emphasize the role of \textit{electoral} incentives, theorizing that legislators balance the preferences of voters and organized producers, who provide campaign contributions and lobbying pressure in addition to rents \citep{stigler1971theory, peltzman1976toward, grossman2001special}. By contrast, bureaucrats are not subject to re-election pressures, nor do they occupy dual roles at corporations that may influence their behavior while in office as elected officials at times do (see for example \citet{weschle2024politicians})---their primary incentive is to secure attractive post-retirement employment. This creates incentives to foster relationships with firms and to shape policy or resource allocation in ways that increase future employability. Such actions may also enable regulatory capture, where civil servants favor future employers, or signaling, where regulators act stringently to demonstrate competence to potential industry recruiters \citep{dal2006regulatory}. Empirical studies offer evidence for both mechanisms: the capture hypothesis is supported by \citet{cohen1986dynamics}, \citet{gormley1979test}, \citet{spiller1990politicians}, and \citet{tabakovic2018revolving}, while \citet{dehaan2015revolving} provide evidence for the signaling model.

Despite theoretical overlap, relatively few studies directly link bureaucratic connections to firm-level rents. Notable exceptions include \citet{lee2023bureaucratic}, who show that U.S. firms with connections to the Office of the U.S. Trade Representative reduce their lobbying activity; \citet{barbosa2020value}, who find that Brazilian medical firms hiring ex-bureaucrats offer lower prices to government; and \citet{hong2016capture}, who demonstrate that Korean universities employing former officials receive more grants. In Japan, \citet{asai2021regulatory} show that firms hiring ex-infrastructure ministry officials are more likely to win public contracts, while \citet{luechinger2014value} observe stock price boosts for U.S. defense firms hiring Department of Defense officials.

Taken together, these studies suggest that firms value former civil servants for their ability to shape regulatory outcomes, influence government decision-making, and facilitate access to public resources such as contracts, subsidies, and loans. However, only select civil servants---such as those with regulatory oversight or access to public resources---can deliver these benefits. These civil servants should therefore be in demand by firms, but others should be of little value. In this sense, select bureaucratic connections may yield similar benefits to legislative ones—yet empirical work remains fragmented, typically focusing on single agencies or economic sectors due to data limitations.

\subsection{The value of post-government positions to bureaucrats and ministries}

A related literature explores the ``returns to office'' enjoyed by politicians, demonstrating that legislators often gain significant wealth after being elected to office \citep[e.g., ][]{eggers2009mps, fisman2014private}. By contrast, civil servants typically experience suppressed earnings relative to their private-sector counterparts and rely on the revolving door for deferred compensation. For example, in Japan, \citet{ramseyer2009japan} estimated that civil servant salaries were 11 percent lower than the monthly mean national wage in 1989. Given these bureaucrats' elite educations, this represents a substantial gap in income when compared to both elected officials and the private sector. 

Prior research and journalistic accounts document the substantial rents associated with revolving door movements in advanced economies with large pay differentials between high-skilled public and private sector workers, such as Japan, the United Kingdom, and the United States \citep{usui1995government, blumenthal1985practice, kalmenovitz2022closing, ft2024top, mizoguchi2012amakudari, ramseyer2009japan}. Accordingly, the revolving door functions as a primary institutional mechanism through which material returns to office are (later) realized for civil servants.

Yet little is known about the structure of the job market for former bureaucrats. Existing work has focused on politicians, offering limited insight into the broader population of civil servants. While it is intuitive that senior bureaucrats with policy influence and extensive networks would be desirable hires, the employment prospects for mid- and lower-level bureaucratic officials remain largely unexamined, in part due to data constraints.

Theoretically, firms should only seek to hire former bureaucrats who offer clear value—whether through specialized expertise, access to regulatory influence, or facilitation of rents such as contracts, loans, or bailouts. This implies that private sector hiring should be concentrated among officials from ministries with substantial regulatory or fiscal authority. In support of this view, empirical work from the United Kingdom finds that the revolving door is most active in departments with control over major policy levers, while officials from lower-capacity ministries are less likely to transition into private-sector roles \citep{andrews2024revolving}. 

Yet while the private sector exercises selective demand for bureaucrats, ministries face broader institutional incentives. Because  earnings during public service are constrained, ministries rely on post-government career pathways as a form of deferred remuneration. Facilitating these placements serves two primary functions. First, it preserves ministerial influence by embedding former officials in positions across the public and private sectors. And second, it signals to prospective hires that a relatively underpaid career in the civil service will be compensated in the future with a lucrative post-bureaucracy position. Indeed, contemporary policymakers have recognized this logic explicitly. For example, the head of the UK government’s legal department recently stated that she is ``all in favor of the so-called revolving door'' because of its utility for recruitment \citep{ft2024top}.

However, ministries face a problem of timing and imperfect information---it is impossible to predict \textit{ex ante} who will reach senior ranks at the time of recruitment. To address this information asymmetry problem, ministries may pursue a strategy of broad-based post-retirement placement, treating deferred compensation as an institution-wide guarantee rather than a reward contingent on individual career trajectory. This logic is especially salient in Japan, where civil servants retire before becoming eligible for pensions, heightening the need to secure post-retirement employment. In this context, revolving door placements may function as a quasi-contractual obligation: an implicit guarantee of post-career income and security in exchange for long-term bureaucratic loyalty. Like other informal labor institutions—such as lifetime employment in Japan, \textit{guanxi} in China, or the traditional labor protections conferred on federal bureaucrats in the United States—there may be severe consequences for institutional reputation and recruitment in the event that this informal contract is violated. 

I generate two main predictions from these theoretical insights. First, private-sector demand for former bureaucrats should be concentrated among high-ranking officials from prestigious ministries, whose networks and influence are of most value to firms. Second, ministries should endeavor to place as many officials into post-retirement roles as possible---including those who might not be in high demand from firms---in order to sustain the revolving door as a recruitment and retention tool.

\subsection{Hypotheses}

The revolving door offers firms access to insider knowledge, regulatory influence, and potential rents, while providing bureaucrats with deferred compensation and ministries with a mechanism for recruitment and retention. 

These dynamics generate three primary hypotheses. First, private firms should be more likely to hire high-ranking officials from prestigious and economically connected ministries. Second, if ministries use the revolving door as a tool to attract and retain talent, they should endeavor to place lower-ranking officials into outside positions as well, supplementing the natural market for high-ranking hires and resulting in a bifurcated job market for former civil servants. Third, if firms derive value from political connections, hires should be associated with observable benefits to hiring organizations, such as increased government loans and contracts or favorable investor responses. 

In the following sections, I (1) provide an overview of the specific context of study---the institutionalized revolving door in Japan, (2) introduce the data used for empirical analyses, and (3) test the propositions above using the described data and a variety of differences-in-differences methods.

\section{Case selection: \textit{amakudari} in Japan} \label{sec: case_japan}

The revolving door is well-known in the Japanese context, where it is referred to as \textit{amakudari}---literally ``descent from heaven.'' \textit{Amakudari} is the institutionalized practice of civil servants retiring into the private or public sector at the end of their careers,\footnote{A subset of \textit{amakudari} involving moves into public sector firms, called \textit{yokosuberi} or ``sliding sideways,'' is also discussed in prior literature. For simplicity, I use \textit{amakudari} to refer to both types of moves.} typically near age 60.\footnote{Retirement typically follows an ``up or out'' process with a pyramidal promotion system \citep{aoki1988information}. There are fewer available positions at each step up the promotion ladder (e.g., section chief to bureau chief positions to director general to vice-minister) for each ministry and those who are not promoted resign.} While not truly ``revolving'' since bureaucrats rarely return and mid-career private sector hires into the bureaucracy are uncommon, I adopt the term to align with the broader political connections literature.\footnote{For multiple movers (as well as those who originally come from industry), socialization into industry is theorized to make regulators more amenable to industry concerns. For one-time movers (such as in Japan), the mechanism is different (enhancing post-retirement marketability), but whether the effect on outcomes differs is unclear. Existing empirical work also does not examine true revolvers, but single-direction moves.} These post-retirement positions serve two primary functions: they provide deferred compensation for relatively low-paid bureaucrats \citep{colignon2003amakudari, mizoguchi2012amakudari},\footnote{The majority of the literature (including the formal model by \citet{mizoguchi2012amakudari}) assumes that bureaucrats want to go to the organizations that will provide them with the ``maximum remuneration possible'' (p. 822), and that possible salary rises with civil service rank. Future research could therefore consider more explicitly modeling revolving door movements as a function of broader individual and institutional incentives in addition to remuneration.} and offer continued employment in a country with low old-age cash transfers and high late-life labor force participation \citep{estevez2008welfare}. The institutionalized practice of retirement around 60 thus creates a shock that sheds light on how agencies and individuals manage sudden job insecurity.

While the supply-side motivations for \textit{amakudari} are well established, less is known about the universe of opportunities available to former bureaucrats and the benefits conferred on hiring firms. Existing studies suggest that \textit{amakudari} may offer firms privileged regulatory access \citep{calder1989elites, schaede1995old, colignon2003amakudari}, reduce oversight \citep{grimes2005reassessing}, or facilitate access to government loans and contracts \citep{blumenthal1985practice, mizoguchi2012amakudari, jones2013amakudari, the_economist_2010, usui1995government, japantimes_2017, woodall1997japan}. \textit{Amakudari}-staffed nonprofit and public organizations have also been implicated in scandals linked to bid rigging and the receipt of public subsidies \citep{mizoguchi2012amakudari, carlson2018political}. I examine whether these suspected benefits represent systematic phenomena by testing whether \textit{amakudari} leads to increased granting of government loans to for-profit firms, and increased granting of contracts to nonprofits.

Among Japan’s ministries, the Ministry of Finance (MOF) and the Ministry of Economy, Trade and Industry (METI) are widely regarded as the most prestigious and powerful \citep{mizoguchi2012amakudari, vogel2021rise, noble2025japan, calder1989elites, aoki1988information, usui1995government}. Their central roles in fiscal policy, financial regulation, and industrial planning have historically positioned them as dominant institutions in Japan’s developmental state \citep{johnson1982miti, rosenbluth2010japan}. These ministries recruit heavily from Japan's most prestigious university, the University of Tokyo, reinforcing their elite status. As a result, bureaucrats from MOF and METI are among the most desirable hires due to their close ties to powerful institutions that control the levers of financial and industrial policy.\footnote{The Ministry of Land, Infrastructure, and Transport (MLIT) is also considered relatively prestigious and additionally controls a large number of infrastructure and construction related contracts.} 

Several features of \textit{amakudari} make Japan a useful case for studying bureaucratic political connections more broadly. First, relatively low public sector pay creates strong incentives for deferred compensation through post-retirement jobs. Second, institutionalized early retirement introduces a predictable shock to job security. Third, Japanese officials are typically generalists rather than technical experts, suggesting their value lies in information, networks, and influence rather than specialized skills. Finally, long tenures and stable careers enhance the credibility and durability of the connections they offer, as theorized by \citet{bils2020working}. Taken together, these conditions provide a compelling context for testing broader theories about the formation, value, and consequences of political connections between firms and the bureaucracy—particularly in settings where bureaucrats are career generalists and connections must be valuable beyond technical expertise.

\section{Data}

\subsection{Case-specific interviews}

In addition to reviewing existing literature, I conducted semi-structured interviews to explore how the bureaucratic revolving door is perceived to benefit both firms and bureaucrats in Japan. I draw on 19 interviews with current and former bureaucrats as well as executives involved in revolving door hiring in Tokyo between 2019 and 2022.\footnote{Subject recruitment and engagement adhered to the APSA Principles and Guidance for Human Subjects Research. Prior to interviews, participants were provided with a document describing the purpose of the research project, potential risks, and efforts taken to ensure anonymity. Voluntary and informed consent was then obtained through verbal consent to participate in the study. The research did not intervene in any political processes, involve any vulnerable participants, or engage in deception. The interview and informed consent protocols were reviewed and approved by an Institutional Review Board at Yale University (protocol 2000026886).} These interviews serve primarily as a source of evaluating hypothesized mechanisms in the Japanese context, offering insights into the Japanese case rather than presuming cross-national equivalence in theoretical mechanisms.

Interviews revealed that receipt of public contracts, regulatory benefits, and government financial assistance (especially in times of crisis) were viewed as potential benefits by directors of firms and bureaucrats in Japan. For example, an official in one of Japan's ministries stated that while the revolving door ``does not necessarily result explicitly in subsidies or contracts, it is definitely beneficial'' (Author Interview D1a). A director in a major consulting firm noted that the revolving door ``is most beneficial in industries where regulations are most strict'' (Author Interview D1b). A corporate finance expert stated that bureaucratic hires tend to increase in the banking sector when banks are in trouble (Author Interview N1a). Finally, a corporate governance expert noted that ``investors would definitely notice'' high level appointments (Author Interview N1b). At the same time, interviewees often viewed the revolving door as more prevalent among older or weaker firms, and emphasized that the value of hires derived from connections and access rather than technical or managerial expertise.

Interviewees also echoed the logic that widening pay gaps between the public and private sectors make civil service recruitment increasingly difficult, and that the promise of post-bureaucracy jobs is one of the few remaining tools to attract top talent. Current and former bureaucrats emphasized stark wage differentials between entry-level civil servants and peers entering sectors such as finance and technology, noting that graduates of Japan's most prestigious universities are increasingly likely to join these firms or retire from the bureaucracy early citing low pay (Author Interviews J1a, N1c, S1a).\footnote{A recent report from the Cabinet Bureau of Personnel Affairs corroborates this view, with low pay and the desire to move to job with more career growth potential the most cited reasons for wanting to leave the bureaucracy among those in their 20s and 30s \citep{copa2022survey}.}

In sum, case-specific context and interview data supports the notion that the previously introduced theoretical predictions are applicable to \textit{amakudari} in Japan. High-ranking officials from prestigious ministries with valuable connections should be most in-demand by private firms, but ministries are also incentivized to ensure as many employees as possible secure outside employment as they rely on the revolving door as a tool to attract and retain talent. Finally, there may be tangible benefits to organizations that hire former bureaucrats, such as increased receipt of government loans and contracts or favorable investor responses. The administrative data described below allows for explicit testing of these hypotheses. 

\subsection{Administrative data: records of civil servant re-employment} \label{sec: data}

Pressure to regulate \textit{amakudari}\footnote{\textit{Amakudari} has been blamed for multiple regulatory and policy failures in Japan and the inability to enact structural economic reforms. For example, scholars and government reports have blamed amakudari for crises such as the savings and loan bailout \citep{carlson2018political, mishima2013missing}, the HIV-contaminated blood scandal \citep{carlson2018political, mishima2013missing}, and the Fukushima Daiichi nuclear plant disaster \citep{independentfukushima, mishima2013missing}.} culminated in reform of the National Public Service Act (NPSA) in 2008 \citep{mishima2013missing, terada2019changing, sota2017root}. The reform mandated that civil servants report post-government employment to the Cabinet Office, with appointments disclosed publicly each year \citep[Articles 106-23-25]{publicservice1947}.\footnote{The reforms also prohibited ministries from directly brokering employment \citep[Article 106-2]{publicservice1947} and established a surveillance commission to monitor compliance \citep[Article 106-5]{publicservice1947}. It is possible that the reporting requirements and/or reforms may have influenced firm behavior, including the possibility that firms began hiring former bureaucrats in part to signal alignment with sectoral norms or shareholder expectations. This is an interesting theoretical possibility that merits further exploration. The empirical analysis in the paper is, however, limited to the post-2009 period, and consequently the patterns observed are representative of the equilibrium that emerged after the reform, even if different dynamics may have prevailed prior to 2009.}

These disclosures provide the basis of a new dataset covering all \textit{amakudari} placements from 2009 to 2019. Each disclosure includes the bureaucrat’s former agency and title, along with their new employer and job title  (see \autoref{tab: amakudari_example}). Disclosures include only the first post-retirement job and omit subsequent placements.\footnote{As such, it excludes serial reemployment or ``\textit{wataridori}'' (literally ``migratory birds'').} As such, the strongest ties that might lead to \textit{quid pro quo} exchanges may be underrepresented, since officials are barred from joining organizations they directly oversaw for two years post-retirement. In addition, we do not observe bureaucrats who may have sought but failed to secure post-retirement positions, but the set of bureaucrats who do not receive placements is expected to be small.\footnote{In the Japanese context, the overwhelming majority of eligible bureaucrats actively seek and receive post-retirement placements. A 2000 government report revealed that out of 538 high-ranking bureaucrats who retired between August 1999 and August 2000, 485 (approximately 90\%) secured new positions \textit{within three months}, indicating a high placement rate for retiring officials \citep{colignon2003amakudari}.} Unlike earlier studies that rely on convenience samples, however, this dataset captures the full universe of initial revolving door appointments. The dataset is available online as \textit{Amakudata} \citep{incerti2024amakudata}.

To evaluate the consequences of \textit{amakudari}, I merge these records with outcome data across three domains: government loans to for-profit firms, stock market reactions to hires at publicly traded firms, and public contracts to nonprofit organizations. Loan data come from the NEEDS database, Japan’s largest source of firm-level financial information, and are merged with firm attributes to compare \textit{amakudari} and \textit{non-amakudari} firms on observables. Stock price data are daily adjusted closing prices from Yahoo Finance. Contract data are drawn from 93 publicly released reports, comprising roughly 25,000 records of public works projects, subsidies, and contracts issued by the national government to nonprofit organizations. These records include agency and recipient names, contract details, award dates, values, and auction types. For non-competitive (negotiated) contracts, the number of former civil servants on staff at the recipient organization is also reported. See \autoref{tab: data} for a summary of all data sources. 

I now turn to tests of each of the proposed hypotheses, along with the corresponding estimation strategies and results. Each of the upcoming sections also discusses the relevant datasets in greater detail, in relation to the specific empirical strategies employed.

\begin{table}[]
\caption{\label{tab: data}Overview of data sources}
\resizebox{\linewidth}{!}{
\begin{tabular}{@{}llll@{}}
\toprule
\multicolumn{1}{c}{Data} & Source & Information & Use \\ \midrule
\rowcolor[HTML]{EFEFEF} 
Civil servant re-employment & Cabinet Office reports & Bureaucratic rehires & Independent variable \\
Nikkei NEEDS & Nikkei Inc. & Firm attributes and financials & Matching covariates \\
\rowcolor[HTML]{EFEFEF} 
Nikkei NEEDS & Nikkei Inc. & Government loans & Outcome variable \\
Stock prices & Yahoo Finance & Stock performance & Outcome variable \\ 
\rowcolor[HTML]{EFEFEF} 
Nonprofit contracts \& subsidies & Cabinet Office reports & Nonprofit contracts \& subsidies & Outcome variable \\ \bottomrule
\end{tabular}}
\end{table}

\begin{singlespace}
\section{Is there a bifurcated job market for former civil servants?} \label{sec: descriptive_amakudari}
\end{singlespace}

\noindent To examine the dual expectation that firms prefer high-ranking officials from prestigious ministries, while ministries also seek placements for less in-demand officials, I begin by analyzing descriptive patterns of post-bureaucratic employment. These descriptive analyses confirm many insights from decades of qualitative work on \textit{amakudari}, but also highlight new patterns. Notably, in line with these hypotheses, the data reveal a job market bifurcated by nonprofit vs. for-profit corporations, top vs. lower level officials, and ministry prestige. 

\subsection{Half of former bureaucrats join nonprofit organizations} \label{sec: descriptive_nonprofits}

\begin{figure}[H]
\includegraphics{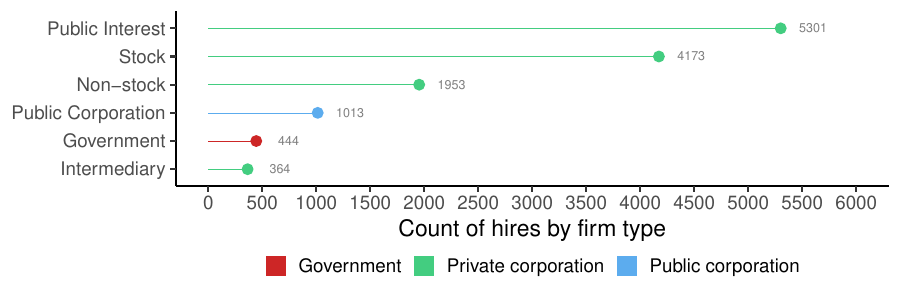}
\caption{\textit{Amakudari} destinations by firm type, all hires 2009-2019}
\small
\vspace{-0.3cm}
\label{fig: firm_type}
\end{figure}

First, I examine where bureaucrats seek reemployment following retirement from the bureaucracy, and demonstrate that the job market for former bureaucrats is bifurcated by destination type and position level. 6314---roughly one-half of---bureaucrats retired into nonprofit ``public interest'' (5301) or public (1013) corporations, compared with 6126 bureaucrats who retired into for-profit firms (i.e., stock and non-stock corporations) as expected (see \autoref{fig: firm_type}).\footnote{The potential for government waste stemming from high salaries paid to former bureaucrats at ``public interest corporations'' is a promising area for research, but one which is outside of the scope of this paper.} Previous scholars discussed cases of this phenomenon \citep{colignon2003amakudari, carlson2018political, jones2013amakudari}, but we can now confirm that public\footnote{Japanese: \textit{dokuritsugyōseihōjin}. This includes destinations such as patent offices, courts, notaries, etc.} and public interest corporation hiring of former bureaucrats is much more common as a percentage of total appointments than previously appreciated, and in fact even represents a slight majority.\footnote{Previous filings such as those analyzed by \citet{colignon2003amakudari} did not require reporting of most public interest hires.}  

Nonprofit corporations (NPOs) are a Japanese legal entity that is largely analogous with the term non-governmental organizations (NGOs) used elsewhere, or 501(c)(3) organizations in the United States. As of 2021, there were approximately 51,000 NPOs in Japan \citep{cao2021npo}. These can range from grassroots civic organizations, to religious organizations, to foundations and ``public interest corporations'' that conduct government-sanctioned public interest projects. Movements from the bureaucracy are primarily to foundations and public interest corporations, many of which are heavily or even entirely reliant on government funding. This has led some to argue that many Japanese NPOs are ``quasi-governmental organizations,'' or that the government outsources public work to these organizations \citep{ogawa2009failure}. Taking on former bureaucrats may therefore be viewed by NPOs as a method to ensure continued funding. 

\begin{figure}[H]
\begin{centering}
\hspace{-0.5cm}
\includegraphics[width = 0.85\textwidth]{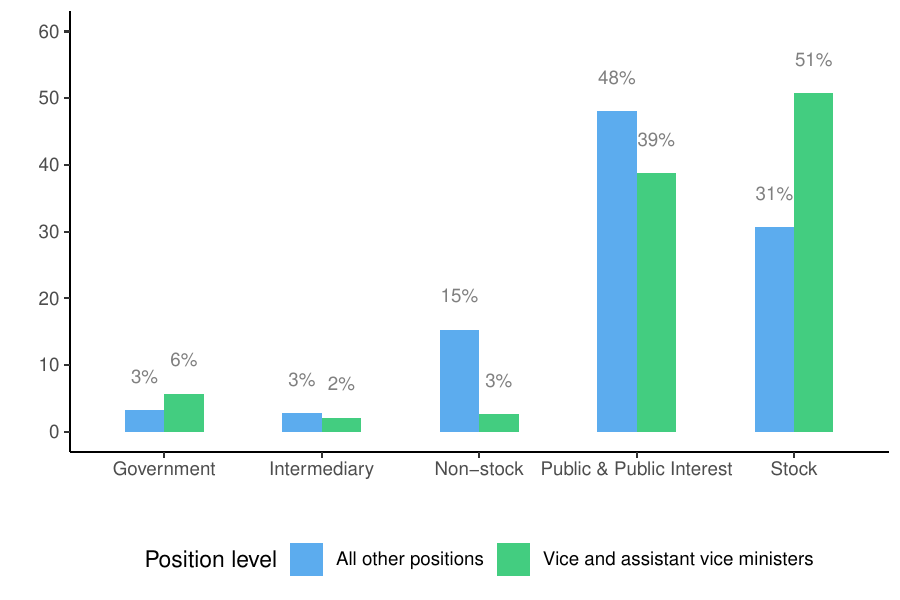}
\vspace{0.3cm}
\caption{Distribution of destinations (firm types) among all former bureaucrats who were re-hired, by position level}
\label{fig: vm_firm_dest}
\end{centering}
\end{figure}

Examination of hiring by position level reveals that only the highest ranking officials (i.e., vice ministers and assistant vice ministers) retire predominantly (51\%) into large, publicly traded firms, while roughly half (48\%) of bureaucrats below the rank of assistant vice-minister move to public interest corporations (see \autoref{fig: vm_firm_dest}). By contrast, only 39\% of vice and assistant vice ministers move to public interest corporations, and only 31\% of individuals below the level of assistant vice minister move to publicly traded firms. Further, top hirers in the public interest sector draw primarily from a single ministry (see \autoref{fig: ministry_public_interest}), suggesting that there are direct pipelines from individual ministries to public interest corporations. No vice or assistance vice minister placements exist within the top ten public interest hirers for the period observed, nor are these hirers drawing from MOF or METI.\footnote{With the exception of the METI Patent Office, an external office (\textit{gaikyoku}) comprised of lower ranking officials.} Subsequent analyses will demonstrate that these individuals drive contract receipt to their new places of employment, perpetuating a system in which the government drives funds to firms that are used as revolving door placements for certain employees.

\subsection{Higher ranking officials join for-profit and publicly traded firms}

Turning to for-profit firms, we again see a market bifurcated by position level. Higher ranking officials are more likely to be hired by large, publicly traded firms, while lower ranking officials are more likely to be hired by smaller, private firms (see \autoref{fig: vm_firm_dest}). Industries reliant on government contracts---such as transportation---and highly regulated industries---such as finance, banking and insurance---are overrepresented in hiring compared to the overall economy (see \autoref{tab: amakudari_firms_short} and \autoref{tab:amakudari_firms_full}),\footnote{This adds additional evidence to previous theories highlighting the importance of regulatory benefits from \textit{amakudari}, and represents a promising area for future research not addressed in this paper.} and the top for-profit hirers belong to highly regulated industries such as insurance, transportation, and finance.\footnote{This confirms \citet{schaede1995old}'s insight.} The most common posts bureaucrats take in for-profit companies are tax advisors, consultants, auditors, lawyers, board members (internal and external), and executives, with top officials more likely to take board member or executive roles in publicly traded firms. 

\FloatBarrier

\begin{figure}[!htb]
\begin{centering}
\hspace{-0.5cm}
\includegraphics[width = \textwidth]{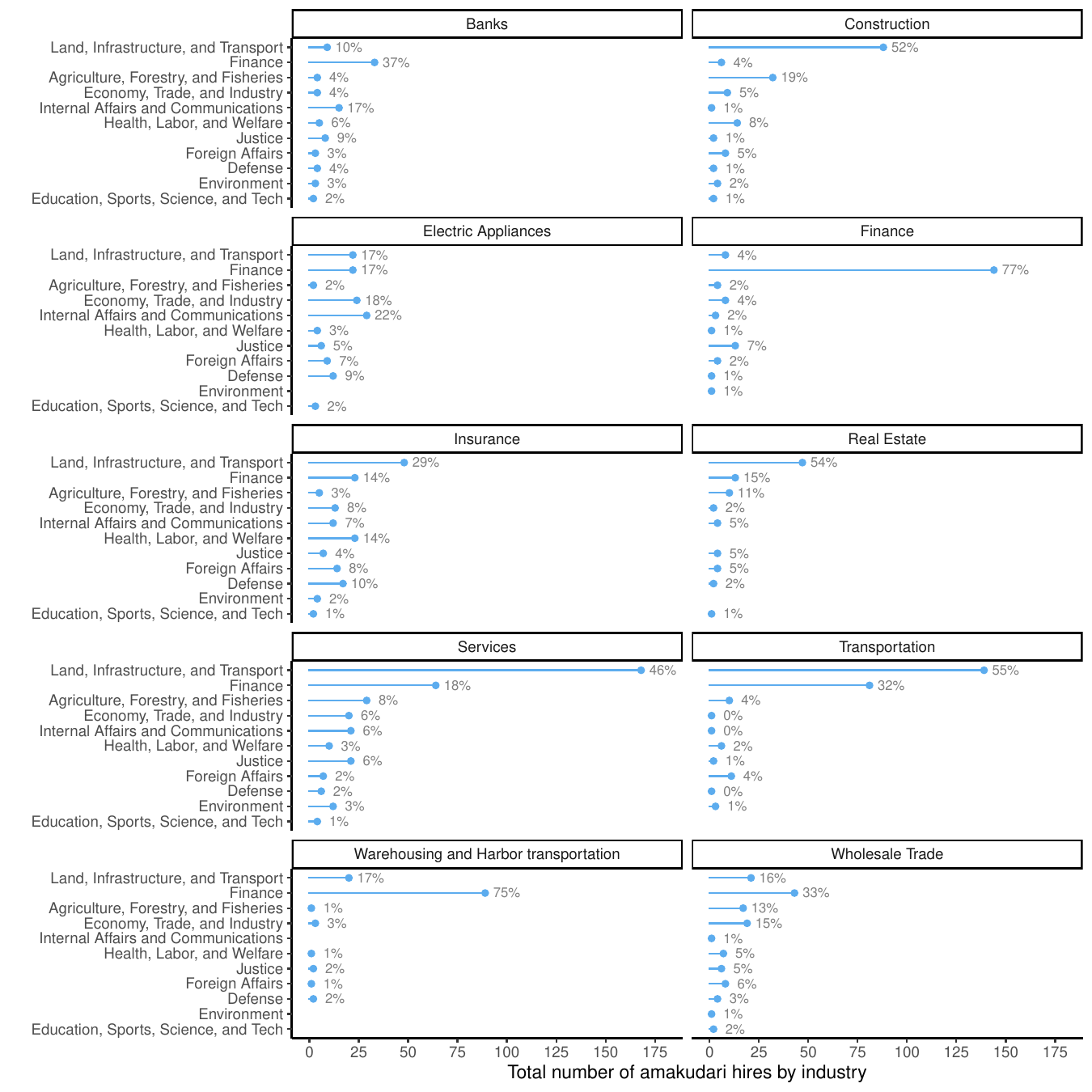}
\vspace{0.3cm}
\caption{Private sector hires by industry and ministry (2009 - 2019)}
\small
\label{fig: ministry_industry}
\end{centering}
\singlespace
\noindent
\footnotesize
\textit{Note}: Top 10 industries by number of hires. Includes appointments from ministries only. Independent agencies not included.
\end{figure}

In contrast with public interest corporations, for-profit hirers draw from multiple ministries (see \autoref{fig: ministry_private}). Some industries, however, draw overwhelmingly from ministries with direct connections. For example, the construction, electric power, transportation, and transport equipment sectors hire predominantly from the infrastructure and transport ministry (MLIT), the majority of banking and finance hires are from the Ministry of Finance (MOF), and the majority of information and communication hires are from the Ministry of Internal Affairs and Communications (MIAC) (see \autoref{fig: ministry_industry}).\footnote{These patterns exist despite a stipulation banning bureaucrats from taking positions in sectors they used to supervise. For example, 28 MOF officials retired into private sector banks since these regulations were passed. 115 retired into regional credit unions known as \textit{shinkin} banks, including 90 from regional finance bureaus. A further four officials retired into \textit{shinkin} banks from their direct regulator---the Financial Services Agency.\footnote{Note that the Financial Services Agency is not located with the Ministry of Finance.}} 

\subsection{Prestigious ministries place more bureaucrats in for-profit firms}

Next, I examine which ministries \textit{amakudari} come from and whether there is variation in placements by ministry. The largest number of hires come from the largest ministries in terms of number of employees.\footnote{Specifically, MLIT, followed by the Ministry of Health, Labour and Welfare (MHLW), the Ministry of Justice (MOJ), the MOF, and the Ministry of Economy, Trade, and Industry (METI) (see \autoref{fig: ministry_bar}).} Adjusted for ministry size, METI and the Ministry of Education, Culture, Sports, Science and Technology (MEXT) are the largest suppliers of former bureaucrats (see \autoref{fig: ministry_bar_adjusted}). 

The most prestigious ministries---the METI and MOF--- place the highest percentage of total bureaucrats into for-profit firms. The MOF places the largest percentage of its retirees in publicly traded firms (59\%), followed by the Ministry of Defense (MOD), Ministry of Foreign Affairs (MOFA), and MLIT (see \autoref{fig: ministry_firm_type}). By contrast, the MHLW, MEXT, and Ministry of Agriculture, Forestry and Fisheries (MAFF) place the largest percentages of their retirees into public interest corporations (72\%, 71\%, and 66\%, respectively). METI's share of employees in publicly traded firms appears low at first glance. However, this is due to the existence of the METI Patent Office, an external office (\textit{gaikyoku}) of METI with a large number of employees who move to public interest corporations.\footnote{Note that there is a distinction within ministries between ``career'' bureaucrats who passed the most rigorous civil service exam and ``noncareer'' bureaucrats who passed only lower level exams. As such, this represents another level of job market bifurcation within ministry.} In fact, the Industrial Property Cooperation Center---a patent advisory firm---is the largest public interest hirer (see \autoref{fig: ministry_public_interest}). Excluding the Patent Office, the same percentage of METI bureaucrats retire in to publicly traded firms as MLIT. 

Bureaucrats from more prestigious ministries also tend to retire at a younger age. As the mandatory retirement age is 60 for most civil service positions, the mean age at which an individual leaves the civil service is 59 and there is little variation by firm type (see \autoref{tab: age}). However, again there is variation by ministry, with younger bureaucrats more likely to leave more prestigious (e.g., METI and MOF) ministries (see \autoref{fig: age}).

These patterns are consistent with higher demand for officials from prestigious ministries---particularly given the concentration of these officials in more desirable post-retirement roles (e.g., high-paying corporate board appointments) and their relative scarcity.

\begin{singlespace}
\section{Do firms that hire bureaucrats secure more government loans?}
\end{singlespace}

\noindent To investigate the hypothesis that private sector firms derive tangible value from hires of high-ranking former bureaucrats, I first examine whether private sector hires are associated with increases in government loans received by firms. I therefore examine the value of government loans granted to firms before and after their first \textit{amakudari} hire observed in the data. The analysis shows that the value of government loans received by firms that make \textit{amakudari} hires increase relative to their matched controls in the years following the hire, and that these effects are driven by hires from prestigious economic ministries. 

\subsection{Data overview: government loans and corporate financials}

To compare how for-profit firms that hire \textit{amakudari} officials compare with those that do not, I use the universe of time-series-cross-sectional (TSCS) data on corporate attributes, financials, and loans from 2009-2019 from the NEEDS financial database. This includes data on assets, liabilities, revenue, earnings,\footnote{EBITDA, or earnings before interest, taxes, depreciation, and amortization} and number of employees, allowing for comparison of firms of similar size and performance. This dataset includes 5809 unique firms across all years, and was merged with the data on bureaucratic rehires (i.e., \textit{Amakudata}). 

\begin{figure}[H]
\begin{centering}
\hspace{-0.5cm}
\includegraphics[width = \textwidth]{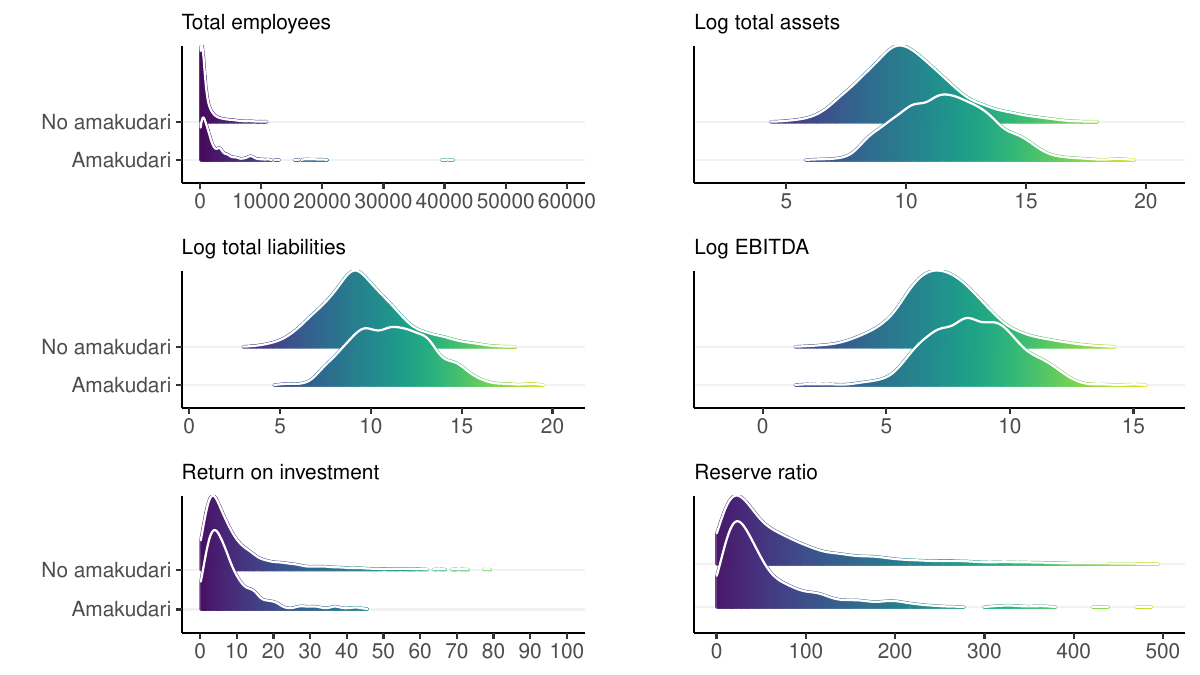}
\vspace{0.3cm}
\caption{Distributions of financial indicators by \textit{amakudari} status (2009-2019)}
\small
\label{fig: amakudari_financials}
\footnotesize
\textit{Note}: Includes all firms for which financial data exists in NEEDS.
\normalsize
\end{centering}
\end{figure}

Firms that hire \textit{amakudari} are different from firms that do not across a number of metrics (see \autoref{fig: amakudari_financials}). First, \textit{amakudari} hirers tend to be older and larger.\footnote{In terms of assets, liabilities, earnings, and number of employees.} Second, firms that make \textit{amakudari} hires have roughly 12 times the amount of debt from public sources as firms that do not make hires (see \autoref{tab: loans_balance}). While not surprising given that firms that hire former officials are on average larger, it is notable as \textit{amakudari} hirers have only 6 times the amount of private sector debt as firms that do not make such hires. Third, there is empirical support for the characterization that \textit{amakudari} is more-often practiced by lower-performing firms---as suggested in previous research by \citep{horiuchi2001did, van2002informality} and echoed by interviewees in business and finance---with \textit{amakudari} firms exhibiting less than half of the return on investment and possessing roughly half of the capital reserves when compared to firms that do not engage in bureaucratic rehiring (see \autoref{tab: amakudari_balance} for this data in tabular form). Finally, while only 4\% of \textit{amakudari} firms matched with the NEEDS financial database are missing financial information, 19\% of non-\textit{amakudari} firms are missing the same data,\footnote{See \autoref{tab: amakudari_balance}. Firms in the NEEDs database also hire more former officials over the observed period than firms that do not appear in the database (see \autoref{fig: hires_ecdf}).} once again implying that \textit{amakudari} firms tend to be larger and more well-known. However, while financial data missingness is highly correlated with bureaucratic hiring, it does not vary highly across industries or years (see \autoref{fig: loan_missing}). 

\noindent

\subsection{Empirical strategy: time-series cross sectional matching}

As noted above, the baseline financials of firms that make \textit{amakudari} hires differ from those that do not. To investigate the impact of hiring former bureaucrats on government loans received, I therefore combine a DiD approach with matching methods in order to compare firms ``treated'' with former bureaucrats with similar ``control'' firms that do not make a hire \citep{imai2019matching}. Given that \textit{amakudari} firms differ from non-\textit{amakudari} firms across their firm fundamentals, I use mahalanobis distance matching to create matches that are close on these covariate values. 

I code all years prior to the first \textit{amakudari} hire observed for each firm as 0 or ``control,'' and the year of hire and all subsequent years as 1 or ``treated.''\footnote{\autoref{fig: loan_treat_control} depicts the treatment or control status of each firm by year. Note that majority of firms remain in ``control'' for all time periods as the majority of firms do not make \textit{amakudari} hires.} As firms are considered always treated following their first \textit{amakudari} hire, the percentage of firms treated is a strictly increasing function of time. As we can only observe the year in which an \textit{amakudari} hire was made, not how many former bureaucrats are currently on staff at a firm at a given time, this likely underestimates the actual effect of \textit{amakudari} on size of government loans overall. If a firm already possesses former bureaucrats on staff, the estimates will capture the effect of an additional hire, rather than any hire.  

After matching control and treatment firms on covariates from other units with the same treatment status in the year prior to treatment ($t_{-1}$), I apply a DiD estimator to account for a time trend. This allows me to estimate short and long-term average treatment effects (ATT) of a bureaucratic hire for  treated firms. I therefore estimate the change in loan volume among firms that switch from no hires in the year prior to one or more hires ($t_{-1}$) vs. the year of the hire ($t_{+0}$) and the subsequent five years ($t_{+1}$...$t_{+5}$), controlling for firm fundamentals via matching. I also conduct the analysis requiring matching on covariates from other units with the same treatment status for additional years prior to treatment (e.g., $t_{-1}$ \textit{and} $t_{-2}$). Note, however, that increasing this required number of ``lags'' will necessarily increase uncertainty by reducing the number of matches and effectively decreasing sample size. 

\subsection{Results}

\begin{figure}[H]
\centering
\begin{subfigure}[b]{\textwidth}
    \centering
    \includegraphics[width=0.85\textwidth]{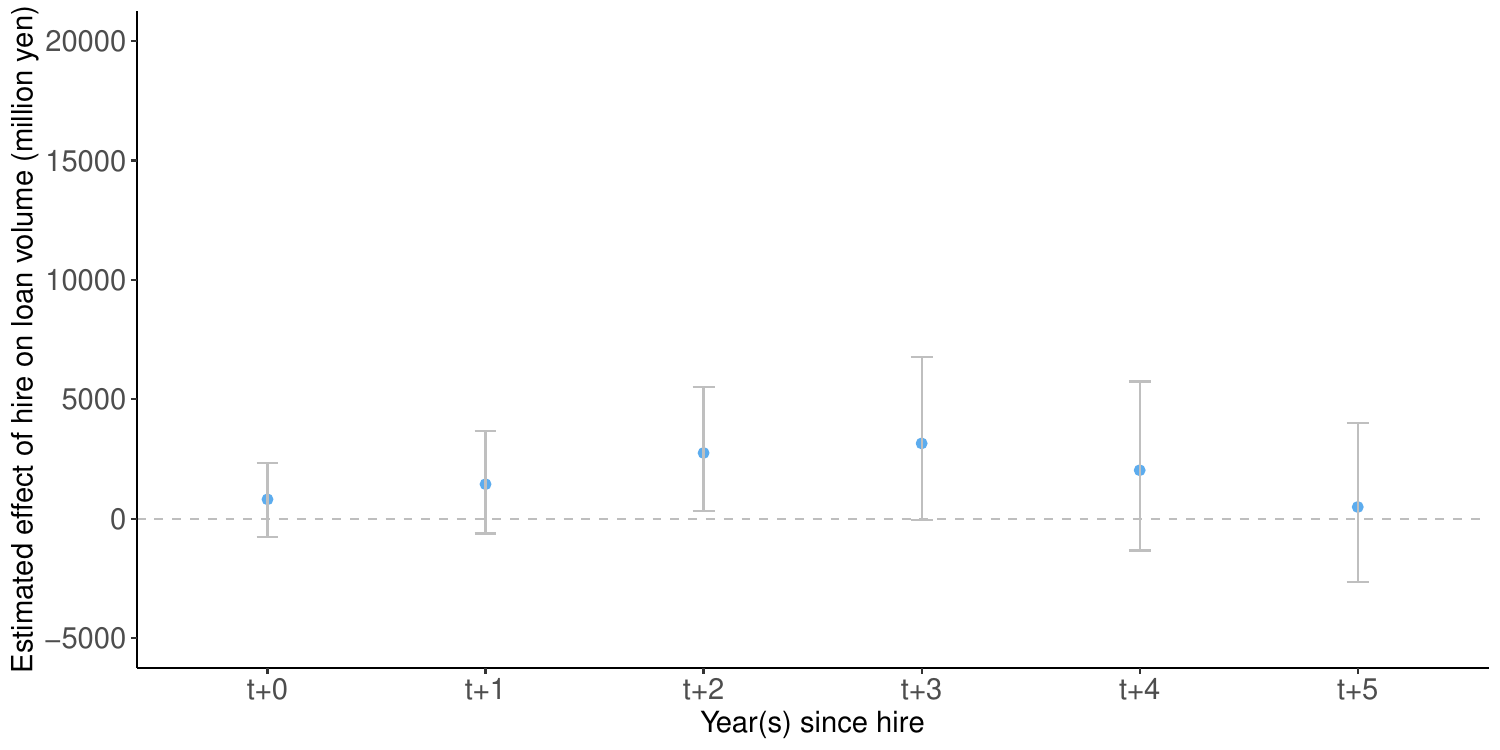}
    \caption{All bureaucratic hires}
    \label{fig: tscs_loan}
\end{subfigure}
\vspace{0.3cm} 
\begin{subfigure}[b]{\textwidth}
    \centering
    \includegraphics[width=0.85\textwidth]{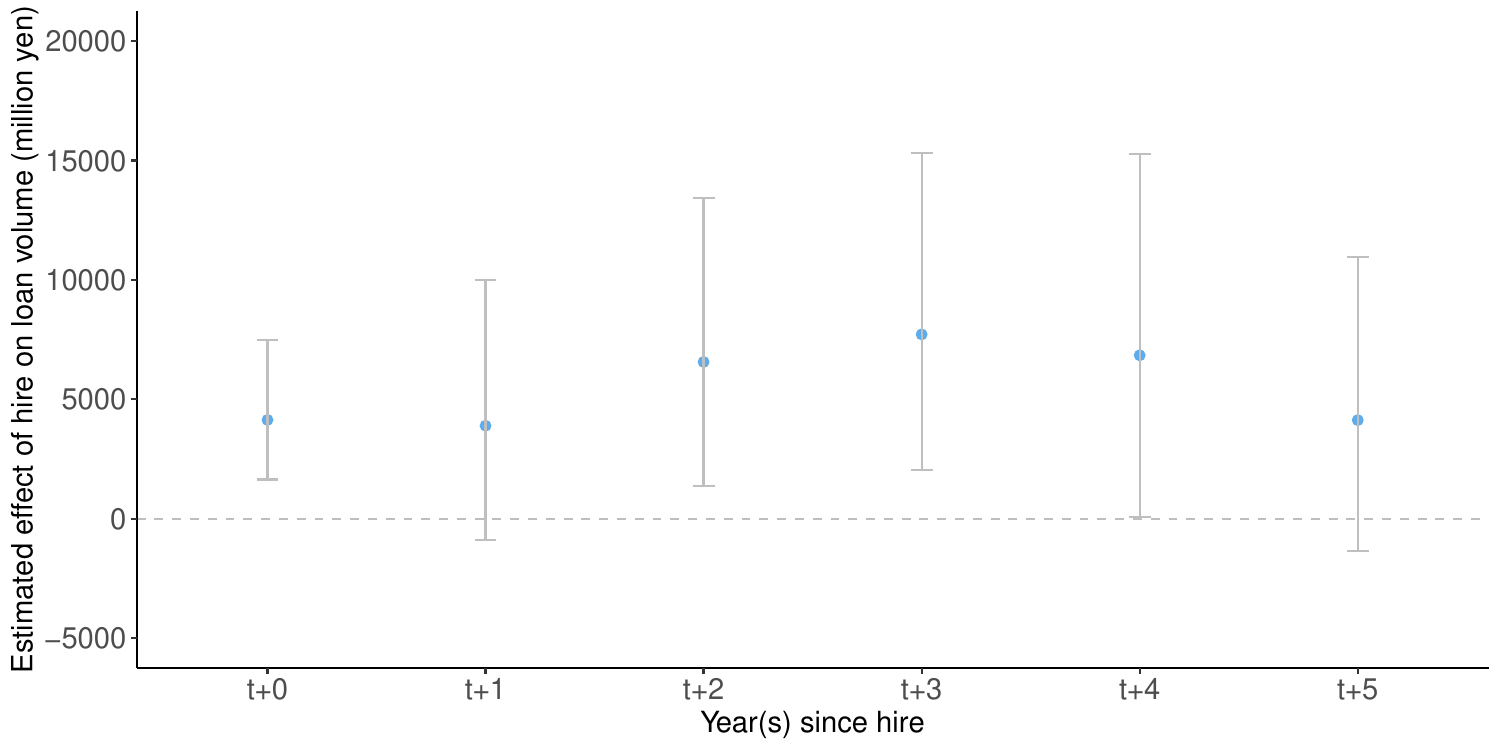}
    \caption{METI and MOF hires only}
    \label{fig: tscs_loan_METI_MOF}
\end{subfigure}
\caption{Estimated effect of bureaucratic hires on size of government loan received, by year after hire.}
\label{fig: tscs_loan_combined}
\vspace{-0.25cm}
{\footnotesize
\raggedright Note: Tabular results can be found in \autoref{tab: tscs_loan} and \autoref{tab: tscs_loan_meti_mof}. \par}
\end{figure}

\autoref{fig: tscs_loan_combined} shows that the value of government loans received by firms that make \textit{amakudari} hires begins to increase relative to their matched control pairs until the third year following the hire, then government loan receipts begins to decrease and returns to baseline levels by year five.\footnote{\autoref{fig: tscs_balance} shows the degree to which covariate balance is improved by matching.} This increase is sizable, with the point estimate for years two and three following an \textit{amakudari} hire representing an increase of roughly 3 billion yen in total loans held.\footnote{Note, however, that the increase is not so large as to be qualitatively unreasonable, translating to 0.16 of a standard deviation of total liabilities among treated firms.} 

In keeping with their roles as the purveyors of industrial policy and domestic financing, breaking apart the results by ministry suggests that hires from the METI and MOF are the most valuable in terms of securing government loans (see \autoref{fig: tscs_loan_METI_MOF}, \autoref{fig: tscs_loan_meti}, \autoref{fig: tscs_loan_mof}, \autoref{tab: tscs_loan_meti_mof}, \autoref{tab: tscs_loan_meti}, and \autoref{tab: tscs_loan_mof}). By contrast, hires from other ministries do not appear to have any effect on the amount of government loans received in the years following an \textit{amakudari} hire (see \autoref{fig: tscs_loan_other} and \autoref{tab: tscs_loan_other}). 

Applying the same analysis to private sector loans reveals a negative---albeit not significant at conventional levels---relationship between hiring former bureaucrats and future receipt of private loans (see \autoref{fig: tscs_loan_private} and \autoref{tab: tscs_loan_private}). This suggests that firms may hire former bureaucrats in order to substitute away from private sector loans and towards public financing. However, it does not appear that government loans are necessarily going to Japan's famous ``zombie" firms, as these firms are primarily supported by their private sector main banking partners \citep{nakamura202350year}.\footnote{I thank Jun-ichi Nakamura for providing me with the data to investigate the correlation between zombie firms and \textit{amakudari} hiring.}\textsuperscript{,}\footnote{Note, however, that this is somewhat tautological as most existing definitions of zombie firms use low-interest main bank loans as an indicator for zombie status.} 

\subsection{Robustness}

To assess the robustness of the main results, I examine potential threats from violations of identification assumptions and model misspecification. Specifically, I evaluate how results vary with the choice of covariate set and matching algorithm, conduct sensitivity checks using alternative model specifications, and examine pre-trends.

To examine whether results are sensitive to the choice of matching covariates, I extend the set of matching covariates to additional variables that are plausibly post treatment (leverage, reserve ratio, roe, and roi) (see \autoref{fig: tscs_loan_posttreatment}). Results remain significant at the 5\% level when matching on these covariates. Patterns of loan receipt remain constant regardless of choice of matching/refinement method.\footnote{Mahalanobis, covariate balancing propensity score, marginal structural model covariate balancing propensity score weighting, or traditional propensity score.} Increased loan receipt from METI and MOF in time period $t_{+0}$ remains significant at the 5\% level across all specifications, and at either the 5\% or 10\% level in time period $t_{+3}$ for all specifications that improve covariate balance (see \autoref{fig:meti_mof_matching_methods} and \autoref{fig: tscs_balance_meti_mof}). 

To address the possibility that estimated effects are driven by short-term fluctuations or longer-run dynamics not captured within the originally specified window, I assess the sensitivity of the estimates to the lead time---i.e., the number of years post-treatment included in the analysis. Results also remain significant at the 1\% level when expanding the lead window and at the 6\% level when reducing the lead window (see \autoref{fig: tscs_loan_leads}). The results for METI and MOF at $t_{+0}$ also remain significant at the 5\% level after transforming the loan outcome variable into either a binary outcome or taking the inverse hyperbolic sine (IHS).\footnote{This test is added due to potential concerns about the skewed nature of the distribution of government loans across firms. However, as the non-parametric matching-based estimator does not rely on distributional assumptions about the outcome variable, a transformation of the outcome variable is not required for unbiased estimation. Additionally, untransformed the ATT remains interpretable in the original units of the outcome — in this case, loan volume, and the level-based functional form is chosen based on the expectation of additive common trends \citep{mcconnell2024s}.}

An additional concern is that firms that hire bureaucrats are on a different pretreatment trajectory than those that do not. For example, a firm with declining value might hire bureaucrats to stave off further losses, or a firm might hire bureaucrats when it is already on an increasingly successful growth path. In other words, it is important to ensure that hiring firms and non-hiring
firms follow parallel trends before the treatment of hiring a bureaucrat. To address this concern, I (1) require matches on additional pre-treatment years, (2) conduct placebo tests on pre-treatment periods, and (3) descriptively examine whether treatment and control firms' covariates diverge in the (non-matched) years prior to hiring ex-bureaucrats. 

Requiring matches for two years prior to treatment yields a similar pattern of effects, although estimates are no longer significant at conventional levels due to increased uncertainty stemming from fewer matched pairs (see \autoref{fig: tscs_loan2} and \autoref{tab: tscs_loan_lag}). For METI and MOF hires only, the results remain significant at the 5\% level in time period $t_{+0}$ when requiring matches for two years prior to treatment, and at the 10\% level  when requiring matches for three years prior to treatment, but are not significant at conventional levels for subsequent periods (see \autoref{fig:meti_mof_effects_all} and \autoref{tab: tscs_loan_lag_meti_mof}). Once again, uncertainty increases due to the smaller number of possible matches with increased lags and fewer possible outcome years, but this analysis ensures by design that (matching) covariates do not exhibit diverging pre-trends up to and including three years prior to treatment. 

Additionally, placebo tests examine the effect of treatment at time $t$ on the difference in the outcome between the treated and control units for the pre-treatment periods (i.e., $t_{-2}$ vs. $t_{-1}$, $t_{-3}$ vs. $t_{-1}$, and $t_{-4}$ vs. $t_{-1}$) for years $t_{-2}$ for all bureaucratic hires (\autoref{fig: all_ministries_2lags_placebo}), as well as $t_{-2}$, $t_{-3}$, $t_{-4}$ for METI and MOF hires (\autoref{fig:meti_mof_placebo_faceted}). In no instances is the effect of treatment on government loans at time $t$ significantly different from zero for the treated and control units in any pre-treatment period. Finally, I examine whether treatment and control firms' covariates diverge in the years prior to hiring ex-bureaucrats for analyses with only one lag period (\autoref{fig: pre_trends} and \autoref{fig: pre_trends_meti_mof}),\footnote{As this is not possible to do naively for control firms (as they don't hire ex-bureaucrats), I conduct this analysis \textit{within} the matched sets of firms, using the date of hire for treatment firms as the pre-trend cutoff (i.e., as different firms are treated at different times I shift the pre-period dates to always be $t-1$, $t-2$, etc. from treatment) and weight the pre-treatment means among control firms by their weights used in the calculation of the ATT.} and demonstrate that loans and pre-treatment covariates are not highly divergent in terms of pre-trends.\footnote{To the degree that there is divergence, the divergence is in the direction of growth, rather than firms on a downward trend seeking loans for rescue. This makes it unlikely that the loans occurred simply because firms on a downwards trajectory are more likely to seek them out. However, this raises the possibility that higher-performing firms may be more attractive to bureaucrats or have more capital on hand to hire bureaucrats.}

\section{Do investors reward firms for bureaucratic hires?} \label{sec: reputation}

Next, I investigate the hypothesis that publicly traded firms may additionally derive tangible value from hires of high-ranking former bureaucrats through favorable investor responses. If top former bureaucrats are perceived as beneficial to firms, we may observe boosts in stock prices in response to their hiring as investors reward firms for their recruitment. I therefore conduct an interrupted time series/event study analysis in which I test for abnormal changes in stock prices on the day a high-profile hire is made. However, more prestigious ministries such as METI and MOF may also be perceived as more valuable to firms due to their abilities to secure financing and contracts, and to influence economic and financial regulations. 

There may also be variation in returns in terms of the type of position a bureaucrat occupies at their new firm. Interviews revealed differential expectations for the value of \textit{amakudari} hires by the type of role they occupy at a firm. A director at an executive consultancy suggested that outside directors were ``probably negatively correlated with the profitability of a company'' (Author Interview N1c), an executive at a major consulting firm suggested that ``government outside directors have no meaning'' as they lack business experience (Author Interview D1c), and analysts from a boutique investment firm claimed that government outside directors lowered return on equity (Author Interview J1a).  Top bureaucrats are hired in four primary capacities according to our data: as advisors, executives, managers, and outside directors.\footnote{A 2019 law requires Japanese corporations to have at least one outside director on their executive board. The justification for this law is that outside directors provide more independent management oversight, improving corporate governance and providing more objective feedback on strategic decisions. An increasingly large number of outside directors have been drawn from the public sector.} I therefore conduct the analysis separately for internal (advisor, manager, and executive) and corporate governance (i.e., directors) related appointments due to these different expectations of the usefulness of these positions. 

\subsection{Data overview: hiring announcements and daily stock prices} \label{sec: stock_data}

To estimate the financial value of political connections to firms that make \textit{amakudari} hires, I first examine the full sample of high-profile hires (i.e., vice minister or assistant vice ministers) into publicly traded firms. I restrict the sample to vice-ministerial and assistant vice-ministerial appointments for two reasons: (1) these individuals are likely to have the largest impact due to their high level of influence, and (2) top appointments are reported in newspapers, and such announcements are necessary to identify an event day for an interrupted time-series estimation strategy. I therefore examine changes in stock returns on the day these hires are announced in Japan's largest business newspaper, the \textit{Nihon Keizai Shimbun}. In total, I identified 47 events made public in newspaper reports. Stock price data are daily adjusted closing prices from \textit{Yahoo Finance} for publicly traded firms.

\subsection{Empirical strategy: market model event study} \label{sec: design_stock}

I estimate cumulative abnormal returns using a market-model event study approach, which measures the stock valuation effects of a corporate event at the time of the event (i.e. a local average treatment effect). This is an interrupted time series model 
\[ R_{it}= \alpha_{i} + \beta_{i}R_{Mt} + \epsilon_{it} \] 
where $R_{it}$ captures the returns to firm $i$ at time $t$, $R_{Mt}$ is the return on the market portfolio (here the Nikkei 225 index) at time $t$, and $\epsilon_{it}$ captures returns to firm $i$ at time $t$ that can be considered ``abnormal'' (above and beyond changes in the market porfolio $R_{Mt}$). The key quantities of interest are therefore the cumulative $\epsilon_{it}$ time series, conventionally referred to as cumulative abnormal returns (CARs), and specifically the CAR on the day of the hiring announcement. I calculate 95\% confidence intervals using the bootstrap as it is free from distributional assumptions. The short time window of one day mitigates endogeneity concerns as confounding events would need to occur on the same day, and do so for a large portion of the independently tested events to influence the estimates.\footnote{For example, for the estimates to be driven by the effect of competitor bankruptcies, this would imply that independent competitors would need to go bankrupt \textit{on the day of the hire} for a large enough sample of hires---e.g. 20 times---to influence the aggregate estimate.} 

\subsection{Results} \label{sec: results_stock}

\begin{figure}[H]
\includegraphics{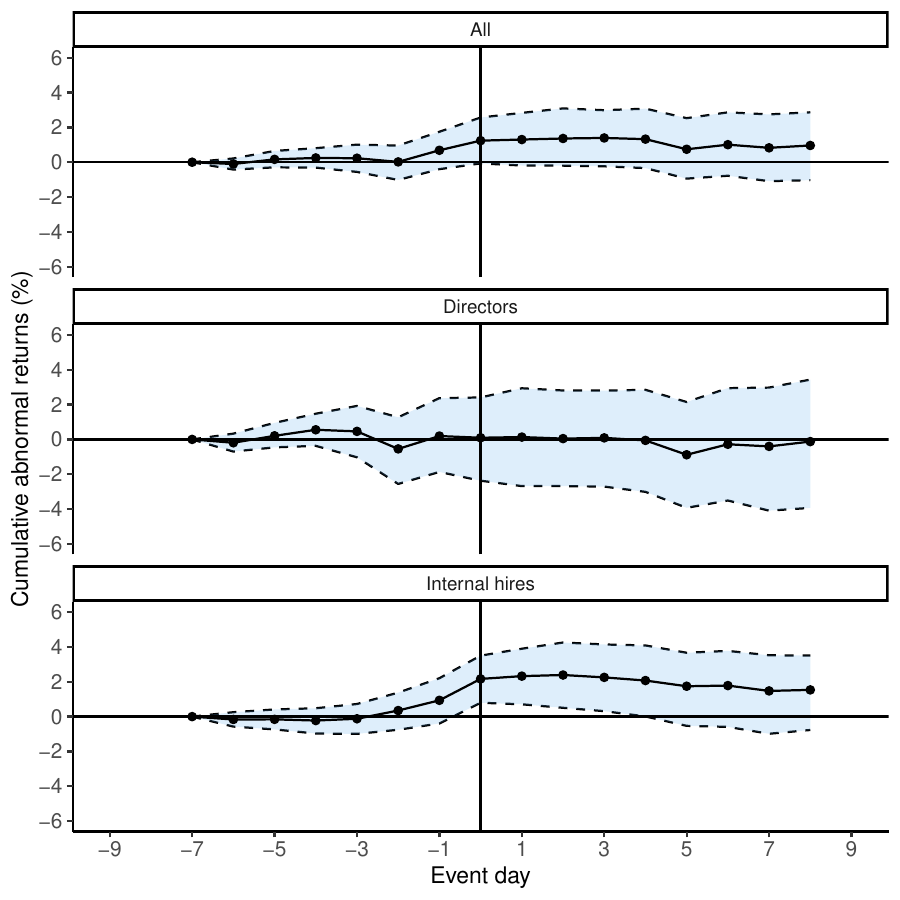}
\vspace{0.2cm}
\caption{Cumulative abnormal returns from assistant vice-minister and vice-minister appointments}
\small
\vspace{-0.3cm}
\label{fig: event_study}
\vspace{-0.25cm}
{\footnotesize
\raggedright Note: Tabular results can be found in \autoref{tab: event_study_all}, \autoref{tab: event_study_director}, and \autoref{tab: event_study_internal}. \par}
\end{figure}

Overall, \autoref{fig: event_study} shows a slightly positive impact of top hiring on firm returns (event day CAR = +1.24, 95\% CI = [-0.1, 2.6]). However, this aggregate analysis masks important variation by both the ministry that was the source of the hire, as well as the type of position the official was recruited for. In terms of ministries, investors also appear to react positively to recruitment from METI relative to other ministries (CAR = +2.34 , 95\% CI = [0.71, 4.01]) see \autoref{fig: event_study_meti} and \autoref{tab: event_study_meti}).  

The event study results corroborate claims that different types of hires may be valued differently. \autoref{fig: event_study}---which depicts cumulative abnormal returns on the day a hire appears in Japan's largest financial newspaper---provides suggestive evidence that direct hires are perceived favorably by investors.  However, an aggregate analysis masks a near zero and null effect of director appointments, and a larger positive effect of roughly 2.2\% for direct internal roles (event day CAR = +2.16, 95\% CI = [0.71, 3.5]).\footnote{The sample size of successful events is 19 for directors (17 outside and 2 internal), and 25 for internal hires (11 consultant, 11 executive, and 3 managerial positions).} 

These findings are notable as previous research has found that markets react favorably to the appointment of outside directors, especially those perceived as independent \citep{rosenstein1990outside, nguyen2010value}. Directors from the bureaucracy may therefore not be perceived as offering the same kind of effective independent oversight and industry expertise, but rather as ``yes-men'' for the corporation. By contrast, internal hires may bring tangible benefits such as regulatory expertise, connections to contract granting agencies, or expertise regarding loan receipt stemming from their ministerial connections. 

In short, I find evidence that investors may view high level bureaucratic hires as indicative of positive future financial performance, but through the mechanism of internal connections rather than corporate governance and oversight. In addition, the most prestigious ministries such as METI again appear to be the drivers of value. 

\subsection{Robustness} \label{sec: event_robustness}

Three potential inferential threats to the event study estimates are: (1) the CARs are driven by factors unrelated to \textit{amakudari} hires, (2) the effects are underestimated as investors knew about the appointments prior to the news releases, and (3) model misspecification. 

To address endogeneity concerns, I re-estimate CARs while substituting the real event dates with time-shifted placebo dates. I shift the actual event days forward and backward by the following daily increments: -200, -100, -50, -25, -10, -5, 5, 10, 25, 50, 100, 200. We should not observe significant abnormal returns when performing an identical test on dates where no hire occurred, as this would raise concerns that the abnormal returns were caused by factors other than the hires. There are no significant CARs on any shifted dates except when shifted backwards by 50 days and forward by five days. The results at -50 days are negative and sensitive to changes in the event window, and results at +5 days are at a time in which abnormal returns are still positive and volatile (see \autoref{fig: event_study_placebo} and \autoref{fig: event_study}). 

There is some evidence that investors may have information about hires prior to the dates identified from newspaper reports given positive trends in the pre-event period. However, while such \textit{a priori} information may call our exact point estimates into question, it would cause an underestimation of the magnitude of the effect on the event day.  

To gauge the sensitivity of the estimates to changes in model specification, I re-calculate all estimates using a constant mean return model (i.e., with no market index control), calculate confidence intervals using the classic t-test and the Wilcoxon rank-test,\footnote{The Wilcoxon rank test is a non-parametric statistical technique that can be used to compare differences between matched samples.} and estimate effects using additional event windows.\footnote{This analysis ensures that results are not only valid for the specific time window chosen in the primary estimates.} Estimates remain virtually unchanged using the t-test, Wilcoxon rank test, and using different event windows (\autoref{fig: event_study_classic_wilcox}, \autoref{fig: event_study_event_windows}), and significance levels increase using a constant mean return model (\autoref{fig: event_study_cmr}). 

\begin{singlespace}
\section{Do nonprofits with bureaucratic connections receive more lucrative contracts?} \label{sec: nonprofits}
\end{singlespace}

\noindent Roughly half of civil servants initially take up posts at nonprofits after leaving the bureaucracy (see \autoref{fig: firm_type}). To investigate whether politically connected nonprofit organizations also derive tangible value from these connections, I examine whether the presence of former bureaucrats is associated with increases in the value of government contracts received by these organizations. 

\subsection{Data overview: government contracts with nonprofits} \label{sec: npo_data}

The Japanese Cabinet Office (CAO) collects and reports data on all subsidies and contracts granted to NPOs in a given year. These reports were scraped and compiled into a dataset of approximately 25,000 contracts and subsidies from ministries to NPOs over a 10 year period. The data includes contract values and names of organizations in months in which a contract was granted. For non-competitive-bid (i.e., negotiated) contracts, the data also includes the total number of government re-employees from the ministry that granted the contract at the NPO at the time of contract receipt.\footnote{This data was collected following the 2009 Accounting Inspectorate Act. See \textit{Kakufushō shokan no kōeki hōjin ni kansuru kaikei kensa no kekka ni tsuite} [results of accounting inspections for public interest corporations under the jurisdiction of each prefecture and ministry] for the government's own audit, and section \begin{CJK}{UTF8}{min}(ア)\end{CJK} \textit{Shokan fushō taishokusha no saishūshokusha no gaikyō} [Overview of re-employed people who have retired from ministries]. Data on number of re-employees does not exist for subsidies or competitive bid contracts.} This data therefore has additional benefits over the dataset of initial \textit{amakudari} appointments used in the previous loan and stock price analyses, as it allows us to view the total number of former bureaucrats at the nonprofit at the time the contract was granted.\footnote{The data of yearly \textit{amakudari} appointments used in the loan and stock price analyses only allows us to observe how many individuals took their first post-bureaucracy appointment in a firm in a given year, not the total number of \textit{amakudari} on staff in a given year. Recall that officials cannot join organizations with which they had a direct working relationship for two-years following their initial retirement from the bureaucracy. NPOs are subject to this same two-year cooling off period, during which time they cannot hire former bureaucrats who were directly involved in the disbursement of government contracts or subsidies.  Knowing the total number of bureaucrats on staff at any given time therefore allows us to view the value of contracts granted both when nonprofits have no government re-employees on staff, as well as when they do. This includes bureaucrats whose initial appointment was not with the NPO, but instead joined the NPO after their two year cooling off period ended.} 

As noted earlier and seen in \autoref{fig: ministry_public_interest}, there are direct pipelines of civil servants that flow from specific ministries to specific NPOs each year. 
The contract data shows that the same organizations that receive regular flows of bureaucrats from specific ministries often receive large volumes of contracts from those ministries. For example, the Japan Forest Foundation\footnote{Japanese name: \begin{CJK}{UTF8}{min}一般財団法人日本森林林業振興会\end{CJK} (\textit{ippan shadan hōjin nippon shinrinringyō shinkōkai})} hired 41 officials from the MAFF from 2009-2019, and in the same period received 305 contracts from MAFF totaling over 2.5 billion yen.  The Japan Construction Information Center\footnote{Japanese name: \begin{CJK}{UTF8}{min}一般財団法人日本建設情報総合センター\end{CJK} (\textit{ippan shadan hōjin nippon kensetsu jyōhō sōgō sentā})} hired 21 officials from MLIT and received 67 contracts totaling over 1.15 billion yen, 48 contracts and 1.07 billion yen of which came from MLIT. I next use a differences-in-differences estimation strategy to systematically investigate if nonprofits receive higher value contracts when former officials are in director positions at their organizations.  

\subsection{Empirical strategy: $DID_M$ estimator and robust TWFE approaches}

The contract data takes the form of an unbalanced panel in which NPO-contract dates are observed at unevenly spaced intervals. In other words, we only observe months in which a contract was granted, and contracts are not granted to all NPOs in all months. In addition, units are ``treated'' with former bureaucrats on staff at different points in time, and units can switch from control to treatment \textit{and} from treatment to control. 

As I have no covariates for NPOs\footnote{NPOs are not listed in NEEDS, and no central database of NPO attributes exists according to nonprofit experts (Author Interview J2b).} I cannot employ the same matching-adjusted DiD design that was used for government loans, placing us in the realm of traditional TWFE estimators. However, recent findings have shown that coefficients from traditional TWFE models may not represent an average of unit-level treatment effects when effects are heterogeneous across time or units (as in this case). \citet{dechaisemartin2020twoway} show that TWFE models can even lead to coefficients having the opposite sign of each of the unit-level treatment effects, as TWFE estimates are a weighted average of unit-level treatment effects and these weights can sometimes be negative due to differences in the timing of treatment or heterogeneity amongst units. 

I therefore adjust my estimation strategy and use the $DID_M$\footnote{$M$ here stands for ``multiple.''} estimator proposed by \citet{dechaisemartin2020twoway} to estimate the ATT. The $DID_M$ estimator compares outcomes among groups whose treatment status switches between time $t-1$ and $t$, and control groups whose treatment status remains constant in time $t-1$ and $t$. The $DID_M$ estimator therefore accounts for the data structure as it relies on first differences only. Formally, the $DID_M$ estimator can be described as 
\[ ATT = \mathbb{E}[Y_{it}(1) - Y_{it}(0)|(D_{i, t-1} = 0, D_{i,t} = 1$ \textit{or} $D_{i, t} = 1, D_{i, t+1} = 0)] \] where $Y$ are potential outcomes and $D$ is the treatment status of unit $i$ in time $t$. The $DID_M$ estimator is then equivalent to the average of the $DID$s across all pairs of consecutive time periods and across all values of the treatment. The estimator also accommodates both binary and continuous treatments, allowing us to estimate the effect of any \textit{amakudari} appointments on contract value, as well as the marginal effect of an additional appointment on contract value. The treatment effect can therefore be interpreted as the average effect of the treatment on the units that experienced a change in treatment status—specifically, ``switchers in'' to treatment or ``switchers out'' of treatment. Note that this does not necessarily imply an immediate effect within a given month, but rather the effect on contract value compared to previous periods when no bureaucrats were on staff.\footnote{The timing of the bureaucratic hire compared to the timing of contract negotiation is unknown---the bureaucrat could have been hired at any point between the negotiation of a previous contract and the current---and the treatment effect therefore represents the average of effects across units with different treatment timings.}

Next, I apply Benford's Law to the value of contracts with \textit{amakudari} bureaucrats as well as those without.\footnote{I thank Yusaku Horiuchi for this suggestion.} Benford's Law is used in forensic accounting to examine discrepancies in the natural probability of leading digits appearing in data---i.e. numbers beginning with 1, 2, 3, etc. If contracts negotiated without former bureaucrats on staff conform with Benford's Law while contracts negotiated with former bureaucrats on staff do not, this would suggest a more competitive negotiation process for non-connected NPOs on the one hand, and evidence of contract price fixing for connected NPOs on the other. To investigate this possibility, I examine the mean absolute deviation (MAD) of leading digits in contract values compared to the predicted frequency according to Benford's Law, for which \citet{nigrini2012benford} has proposed critical scores for conformity and nonconformity with Benford's Law.

\subsection{Results}

\begin{figure}[!htb]
\begin{centering}
\includegraphics[width = \textwidth]{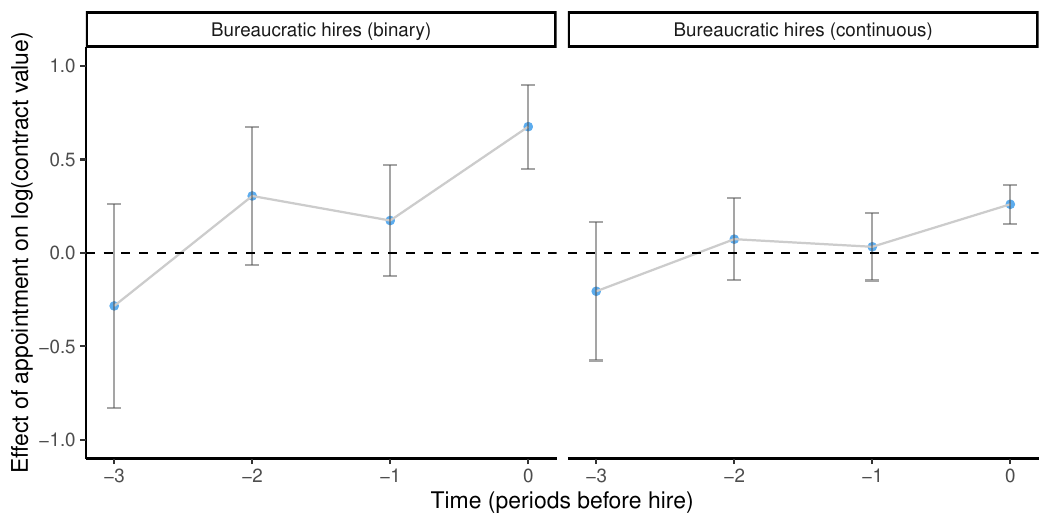}
\caption{Effect of \textit{amakudari} appointments on NPO negotiated contract value}
\small
\vspace{-0.3cm}
\label{fig: npo_govt_reemploy}
\end{centering}
{\footnotesize
\raggedright Note: Tabular results can be found in \autoref{tab: npo_binary} and \autoref{tab: npo_cont}. \par}
\end{figure}

When examining initial appointments immediately following retirement from the bureaucracy only, I find a near zero and null effect of bureaucratic rehires on government contract value for both all subsidies and contracts as well as negotiated contracts only. However, when using the total number of bureaucratic employees at the time of the contract as the dependent variable, I find a positive and statistically significant increase in the value of negotiated contracts granted to nonprofits with \textit{amakudari} appointees (see \autoref{fig: npo_govt_reemploy}). 

These results suggest that bureaucrats may be waiting until after the end of their two-year cooling off period to join nonprofits, at which time they can use their connections to exhibit influence on contract negotiation. As the CAO data only includes director level appointments, it is also possible that only  high-level appointees have the connections needed to negotiate higher contract values. 

Additional evidence of the potential influence of former bureaucrats on contract value can be found through the application of Benford's Law. Based on \citet{nigrini2012benford}'s critical MAD scores, competitive bid contracts exhibit ``close conformity'' with Benford's Law, while negotiated contracts exhibit ``marginally acceptable conformity,'' and negotiated contracts with former bureaucrats on staff exhibit ``nonconformity.'' This suggests that negotiated contracts between former officials in particular may not be subject to hard bargaining.  Visual depictions of the leading digit of contract values as predicted by Benford's Law and as observed in the NPO contract data can be found in \autoref{fig: benford}. 

\subsection{Robustness}

To address concerns about pre-trends in contract value, I include ``placebo'' estimates of the ATT for -3, -2, and -1 periods before treatment as a check of whether outcomes for the treated and comparison groups move in parallel prior to the staggered treatment periods. Placebo estimates are never significantly different from zero at conventional levels. 

Given recent concerns with TWFE estimators---particularly reliance on strict exogeneity, functional form assumptions, and negative weighting in the presence of treatment effect heterogeneity---I implement a number of alternative estimators which also allow for treatment status to vary over time and address the potential for negative weighting. Specifically, I use the estimators referred to by \citet{liu2024practical} as: the fixed effects counterfactual estimator (FEct), the interactive fixed effects counterfactual estimator (IFEct), and the matrix completion (MC) estimator,\footnote{I do not go into detail regarding these estimators here. However, FEct was proposed by \citet{liu2024practical}, \citet{borusyak2024revisiting}, and \citet{butts2021did2s}; IFEct was proposed by \citet{gobillon2016regional} and \citet{xu2017generalized}; and MC methods were proposed by \citet{athey2021matrix} and \citet{kidzinski2018longitudinal}.} in addition to a traditional TWFE estimator. Finally, to assess robustness to functional form assumptions and allow for interpretation of effects in both relative (percent) and absolute terms, I run all models including both a log-transformed outcome variable and in levels.

Estimates using yearly aggregated data, alternate functional forms, and using the FEct, IFEct, MC, or traditional TWFE estimators corroborate the estimates from the $DID_M$  estimator in terms of sign, magnitude, and pre-trends, and can be found in Figures \ref{fig: npo_robust_binary}-\ref{fig: npo_twfe} and Tables \ref{tab: twfe_month}-\ref{tab: twfe_year}.\footnote{Analysis using non-transformed data benefit of being interpretable in terms of levels, and indicate that the presence of a bureaucrat in a director position is worth approximately 50 million yen in additional contract value.} 

\section{Discussion and conclusion} \label{sec: conclusion}

This paper provides the first systematic analysis of all initial revolving door placements from the bureaucracy to post-bureaucratic employment in any country, leveraging a novel, comprehensive dataset of post-bureaucracy employment in Japan. By combining this data with government loan and contract records, financial outcomes, and qualitative interviews, I analyze both the structure and economic consequences of Japan's revolving door.

Theoretically, this paper puts forward a new framework for understanding the role of the revolving door in bureaucratic systems with relatively limited compensation when compared to the private sector---institutional features of bureaucratic systems in many advanced democracies. Rather than viewing post-bureaucratic employment solely as a means for firms to gain influence, I argue that the revolving door is an important recruitment tool for ministries. Because information about a bureaucrat’s rank at retirement is unavailable at the time of hiring, the system virtually guarantees all recruits some form of lucrative or secure post-retirement employment. This helps the civil service attract top talent who might otherwise choose more lucrative private sector careers from the outset. In support of this theory, I show that higher-ranking bureaucrats are more likely to be hired into for-profit firms that may value their prestige or influence, while lower-ranking bureaucrats are more likely to move into quasi-governmental and nonprofit organizations, many of which receive significant government contracts or funding. This makes the revolving door not only a possible post-bureaucratic career benefit, but also an integral part of the bureaucratic compensation structure and recruitment pipeline.

Empirically, I find that nearly half of all revolving door hires occur in the nonprofit or public sector, and that these hires are disproportionately lower-ranking bureaucrats and those from less prestigious ministries. Ministries appear to create revolving door opportunities for lower-ranking staff by channeling funds into connected nonprofits, creating demand where private sector demand is absent. Difference-in-differences methods show that nonprofits with ex-bureaucrats on staff receive more generous government contracts. In the private sector, firms that hire high-ranking former bureaucrats—particularly from powerful economic and financial ministries such as METI and MOF—receive larger volumes of government loans, consistent with utilizing the revolving door for rent extraction. Event studies also show that these appointments generate positive stock market reactions, indicating that investors recognize their value. These findings reveal the consequences of an institutionalized bureaucratic revolving door system where individuals do not cycle back into public office.

The one-way nature of Japan's revolving door distinguishes it from other well-studied systems. For example, in  France, the United Kingdom, and the United States, bureaucrats and politicians may move repeatedly between the public and private sector, or hold dual roles (if they are politicians), raising concerns about regulatory capture due to socialization, future political ambition, or conflict of interest. In Japan, however, bureaucrats exit public service upon retirement, and their incentive structure while in office is driven by the need to secure stable post-retirement employment—especially given early mandatory retirement and comparatively low public salaries. As a result, Japanese bureaucrats may tailor policy, resource allocations, or regulatory decisions to appeal to potential post-retirement employers, not because of reelection incentives or campaign contributions, but for employment security.

Although Japan’s system is shaped by unique institutional and cultural factors, the underlying mechanisms are not likely to be unique. Countries with shrinking bureaucratic prestige, rising public-private pay gaps, and tightening retirement benefits may increasingly rely on revolving doors to sustain bureaucratic recruitment. Indeed, calls to institutionalize revolving door systems have recently emerged in countries like the United Kingdom, where officials note that without such mechanisms, government service will fail to attract top talent. Research shows that more prestigious UK departments are also more likely to feed into revolving door positions in the private sector \citep{andrews2024revolving}. As protections for civil servants are eroded in the United States, Departments may increasingly need to market the possibility of post-bureaucracy positions to retain recruitment. These trends suggest that Japan’s model may foreshadow broader patterns in advanced democracies where the state’s capacity to compete for expertise increasingly depends on informal compensation structures.

In the Japanese context, these dynamics intersect with broader issues such as the welfare state and economic stagnation. Government-connected nonprofits may serve as a public-sector analog to the ``zombie lending'' practices of providing low-interest government loans to uncompetitive firms in Japan's private sector. Informal ``contracts'' such as quasi-guaranteed post-retirement positions may function as a backdoor safety net and alternative to expanded programmatic welfare transfers—providing employment to former bureaucrats while channeling funds into politically connected organizations. This echoes previous arguments that Japan's response to economic stagnation has involved sustaining employment and institutional stability through informal channels, even at the cost of efficiency \citep{caballero2008zombie}. Like the private sector case of providing zombie lending to ensure stable employment, it is unclear whether costs in terms of efficiency or resource misallocation outweigh the benefits of reduced unemployment, more successful bureaucratic recruitment, and increased worker motivation. Nonetheless, as the bureaucracy has lost prestige compared to the heyday of Japan's rapid economic development and the pay gap with the private sector has grown for elite bureaucrats, theoretical expectations are that the maintenance of this system will become even more critical in order to guarantee new hires lucrative post-retirement positions.

From a policy perspective, this paper also highlights how transparency is undermined not only by secrecy but by fragmented and inconsistent reporting. While Japan formally discloses bureaucratic retirement data, it is dispersed across thousands of files and databases with irregular formatting. This makes it difficult for the public to trace patterns of influence or accountability. Reforming such data systems is an essential step toward meaningful transparency and oversight—not just in Japan, but in any system looking to trace the influence of institutions.

Finally, this research opens several avenues for future investigation. First, while I examine domestic outcomes such as loans and contracts, I do not study how revolving door hires affect international business strategies, regulatory alignment or language, or trade outcomes. The effects of the revolving door in international arenas and on regulatory policy decisions remain an important area for further theoretical development and empirical analysis. Second, cross-national comparisons of revolving door systems can help clarify the conditions under which revolving doors serve as tools of influence, mechanisms of state capacity, or both. I hope the public data and theoretical contributions offered here serve as a foundation for future research.

\vspace{1cm}
\subsubsection{Data availability statement}

\begin{singlespace}
Research documentation and data that support the findings of this study are openly available at the American Political Science Review Dataverse: \url{https://doi.org/10.7910/DVN/JGHDYL}. Limitations on data availability are discussed in the appendix.
\end{singlespace}

\subsubsection{Acknowledgements}

\begin{singlespace}
\noindent I extend a special thank you to Sayumi Miyano, Diana Stanescu, and Hikaru Yamagishi, my partners in constructing a dataset that made this project possible, and to my mentor in all things Japanese politics, Frances Rosenbluth---this paper is dedicated to her memory. I also thank: Ayumi Sudo, Ken Tanaka, and Junyao Zhang for excellent research assistance; Jason Anastasopoulos, P.M. Aronow, In Song Kim, Miriam Golden, Mina Pollmann, Ulrike Schaede, Kenneth Scheve, Tara Slough, Seiki Tanaka, Yuhua Wang, and Hye Young You for invaluable comments; participants at Columbia's Horizons in Japanese Politics conference, Copenhagen Business School's Money in Politics conference, the 2024 APSA Annual Meeting, and seminars at EUI, Georgetown, Harvard Weatherhead Center, Stanford APARC, the University of Amsterdam, the University of Toronto Munk School of Global Affairs, and Yale; the UTokyo Institute for Social Science and Waseda Institute of Political Economy for database access; and the Japan Foundation for generous financial support. Any and all errors are my own.
\end{singlespace}

\subsubsection{Funding statement}

\begin{singlespace}
This research was funded by a Japan Foundation Center for Global Partnership Grant. 
\end{singlespace}

\subsubsection{Conflict of interest}

\begin{singlespace}
The author declares no ethical issues or conflicts of interest in this research. 
\end{singlespace}

\subsubsection{Ethical standards}

\begin{singlespace}
The author declares the human subjects research in this article was reviewed and approved by an Institutional Review Board at Yale University and certificate numbers are provided in the text/appendix.   
The author affirms that this article adheres to the principles concerning research with human participants laid out in APSA's Principles and Guidance on Human Subject Research (2020).
\end{singlespace}

\pdfbookmark[1]{References}{References}
\bibliography{bibliography}

\singlespace
\newpage
\appendix
\addcontentsline{toc}{section}{Appendix} 
\part{\large{Supporting Information} }

\vspace{0.5cm}
\begin{center}
\large{\textbf{How firms, bureaucrats, and ministries benefit from the revolving door: Evidence from Japan}} 
\end{center}
\begin{center}
\large{Trevor Incerti} \\ 
\end{center}
\vspace{0.25cm}
\parttoc 

\setcounter{table}{0}
\renewcommand{\thetable}{A\arabic{table}}
\setcounter{figure}{0}
\renewcommand{\thefigure}{A\arabic{figure}}
\pagenumbering{arabic}
\renewcommand*{\thepage}{A\arabic{page}}

\normalsize
\newpage

\section{Limitations on data availability}

Full replication data is available for all descriptive statistics and estimates in the ``Data'', ``Is there a bifurcated job market for former civil servants?'', ``Do investors reward firms for bureaucratic hires?'', and ``Do nonprofits with bureaucratic connections receive more lucrative contracts?'' sections. Replication data is not available for descriptive statistics and estimates in the ``Do firms that hire bureaucrats secure more government loans?'' section, as these analyses rely on proprietary data from Nikkei Inc. Analyses in this section can be replicated using the provided replication code and the universe of firm financial information and government loan data from Nikkei NEEDS for years 2009 - 2019. Please contact the author at \url{t.n.incerti@uva.nl} for more detailed information. 

Replication files for the creation and cleaning of the \textit{Amakudata} dataset can be found at: \url{https://github.com/tincerti/amakudata}. The cleaned \textit{Amakudata} dataset is provided at Harvard Dataverse. A dashboard that allows users to explore and download the data can also be found at \url{https://trevorincerti.shinyapps.io/amakudashboard/}. Replication files for the creation and cleaning of the jNPO nonprofit contract and subsidy dataset can be found at: \url{https://github.com/tincerti/jNPO}. The cleaned jNPO dataset is provided at Harvard Dataverse.

Interview protocols are provided but full interview transcripts are not made publicly available for the purpose of confidentiality. 

\pagebreak

\section{Additional literature on \textit{amakudari}} \label{sec: additional_case}
\subsection{Previous literature --- empirics}

Few empirical studies examine correlations between \textit{amakudari} and specific outcomes, and those that do rely on convenience samples. These studies find that: between 2001-2004 firms with former bureaucrats from the Ministry of Land, Infrastructure, Transport and Tourism (MLIT) on staff were more likely to win bids for government contracts from MLIT \citep{asai2021regulatory}, that 125 regional banks that hired 200 officials from the Ministry of Finance (MOF) between 1977 and 1991 tend have reduced capital adequacy levels and more non-performing loans \citep{horiuchi2001did}, and that 266 banks with 204 MOF and Bank of Japan \textit{amakudari} officials on their boards of directors between 1977-1993 had lower profits and engaged in more risky lending \citep{van2002informality}. 

Past empirical analyses therefore suggest that \textit{amakudari} is not a practice regularly undertaken by the highest performing or most dynamic firms---a view shared by many interviewees. A review of the literature on \textit{amakudari} concludes that  ``despite the longstanding interest and sometimes heated debate of scholars, one of the most striking things about this literature is the lack of serious data analysis'' \citep{grimes2005reassessing}. Theoretical benefits of \textit{amakudari} to firms therefore remains a subject of debate, and empirical adjudication is limited. 

\subsection{Previous literature --- institutional details}

Additional literature addresses which bureaucrats and/or ministries want to send officials to which organizations, which organizations desire
which bureaucrats, and how the process in which bureaucrats get matched to organizations happens. \citet{mizoguchi2012amakudari} model \textit{amakudari} as an auction in which the ministry asks each of the firms that are interested in obtaining the service of the retiring bureaucrat for its valuation of the bureaucrat, and then chooses the firm with the highest virtual valuation. They conclude that firms that make risky investments and/or need bailouts or desire government contracts should be willing to pay top dollar for high-ranking Ministry of Finance and Ministry of Economy, Trade, and Industry bureaucrats. \citet{blumenthal1985practice} similarly concludes that ``when companies get into trouble, outside managers are sought to solve the problems,'' and that top economy ministry bureaucrats are desired as managers in these cases. \citet{usui1995government} argue that \textit{amakudari} is valuable for firms that wish to ``absorb the uncertainties of markets and government contract allocations,'' and that ``the high quality of MoF and MITI [now METI] officials and their networks of information make retiring bureaucrats from these two ministries attractive for private companies to hire.'' 

In terms of which bureaucrats want to go to which organizations and the process in which bureaucrats get matched to organizations, the majority of the literature assumes that bureaucrats want to go to the organizations that will pay them the most. \citet{usui1995government} suggest that future analysis could ``examine the supply and demand or push and pull sides of the process by identifying the relationships between each ministry and each private firm, for each \textit{amakudari} placement,'' but to my knowledge such a study has not been conducted. 

\pagebreak

\section{Examples of data sources} \label{sec: data_examples}

\begin{table}[H]

\caption{\label{tab: amakudari_example}Amakudari dataset example}
\centering
\resizebox{\linewidth}{!}{
\begin{tabular}[t]{lllllr}
\toprule
date\_ret & agency & ministry\_short & firm\_dest\_en & firm\_type1\_en & tse\_code\\
\midrule
\cellcolor{gray!6}{2012-11-01} & \cellcolor{gray!6}{Aeronautical Safety College} & \cellcolor{gray!6}{MLIT} & \cellcolor{gray!6}{AIRCRAFT SAFE OPERATIONS SUPPORT CENTER} & \cellcolor{gray!6}{Foundation} & \cellcolor{gray!6}{-99}\\
2013-10-01 & Securities and Exchange Surveillance Commission & MOF & JAPAN SECURITIES DEALERS ASSOCIATION & Other association & -99\\
\cellcolor{gray!6}{2015-06-23} & \cellcolor{gray!6}{Fisheries Agency} & \cellcolor{gray!6}{MAFF} & \cellcolor{gray!6}{KENKO MAYONNAISE} & \cellcolor{gray!6}{Stock company} & \cellcolor{gray!6}{2915}\\
2016-06-23 & National Tax Agency & -99 & JAPAN BANK FOR INTERNATIONAL C & Stock company & -99\\
\cellcolor{gray!6}{2015-11-01} & \cellcolor{gray!6}{Minister's Secretariat} & \cellcolor{gray!6}{MLIT} & \cellcolor{gray!6}{ADVANCED CONSTRUCTION TECHNOLOGY CENTER} & \cellcolor{gray!6}{Foundation} & \cellcolor{gray!6}{-99}\\
\addlinespace
2018-07-01 & Japan Coast Guard & MLIT & NARITA INT.AIRPORT & Stock company & -99\\
\cellcolor{gray!6}{2016-10-01} & \cellcolor{gray!6}{Minister's Secretariat} & \cellcolor{gray!6}{METI} & \cellcolor{gray!6}{DAIDO STEEL} & \cellcolor{gray!6}{Stock company} & \cellcolor{gray!6}{5471}\\
2013-07-01 & Minister's Secretariat & MLIT & NARITA INT.AIRPORT & Stock company & -99\\
\cellcolor{gray!6}{2013-05-01} & \cellcolor{gray!6}{Public Employment Security Office} & \cellcolor{gray!6}{MHLW} & \cellcolor{gray!6}{OKAZAKI SHINKIN BANK} & \cellcolor{gray!6}{Shinkin bank} & \cellcolor{gray!6}{-99}\\
2018-07-01 & Japan Coast Guard & MLIT & JAPAN MARINE RECREATION ASSOCIATION & Foundation & -99\\
\addlinespace
\cellcolor{gray!6}{2015-10-01} & \cellcolor{gray!6}{Minister's Secretariat} & \cellcolor{gray!6}{MOF} & \cellcolor{gray!6}{SMBC CONSULTING} & \cellcolor{gray!6}{Stock company} & \cellcolor{gray!6}{-99}\\
2017-10-01 & Japan Customs & MOF & CANON & Stock company & 7751\\
\cellcolor{gray!6}{2015-04-01} & \cellcolor{gray!6}{Nature Conservation Bureau} & \cellcolor{gray!6}{MOE} & \cellcolor{gray!6}{REGIONAL COEXISTENCE AND SOCIETAL COOPERATION ASSOCIATION} & \cellcolor{gray!6}{Incorporated association} & \cellcolor{gray!6}{-99}\\
2017-08-01 & General & MAFF & MSandAD INSURANCE GROUP HOLDINGS & Stock company & 8725\\
\cellcolor{gray!6}{2016-06-22} & \cellcolor{gray!6}{Statistics Bureau} & \cellcolor{gray!6}{MIAC} & \cellcolor{gray!6}{INFOCOM RESEARCH} & \cellcolor{gray!6}{Stock company} & \cellcolor{gray!6}{-99}\\
\addlinespace
2015-12-01 & Administrative Evaluation Bureau & MIAC & NEC & Stock company & 6701\\
\cellcolor{gray!6}{2016-07-01} & \cellcolor{gray!6}{Regional Development Bureau} & \cellcolor{gray!6}{MLIT} & \cellcolor{gray!6}{COASTAL DEVELOPMENT INSTITUTE OF TECHNOLOGY} & \cellcolor{gray!6}{Foundation} & \cellcolor{gray!6}{-99}\\
2012-01-01 & Civil Aviation Bureau & MLIT & RELIABILITY ENGINEERING FOUNDATION FOR AIR NAVIGATION FACILITIES & Foundation & -99\\
\cellcolor{gray!6}{2011-06-01} & \cellcolor{gray!6}{Japan Coast Guard} & \cellcolor{gray!6}{MLIT} & \cellcolor{gray!6}{SANKYU} & \cellcolor{gray!6}{Stock company} & \cellcolor{gray!6}{9065}\\
2011-09-01 & National Tax Agency & -99 & TEIKYO UNIVERSITY & Educational institution & -99\\
\addlinespace
\cellcolor{gray!6}{2015-06-26} & \cellcolor{gray!6}{Minister's Secretariat} & \cellcolor{gray!6}{METI} & \cellcolor{gray!6}{SHIMADZU} & \cellcolor{gray!6}{Stock company} & \cellcolor{gray!6}{7701}\\
2012-01-01 & Japan Coast Guard & MLIT & COMPUTER INSTITUTE OF JAPAN & Stock company & 4826\\
\cellcolor{gray!6}{2018-01-01} & \cellcolor{gray!6}{National Tax Agency} & \cellcolor{gray!6}{-99} & \cellcolor{gray!6}{MITSUI FUDOSAN} & \cellcolor{gray!6}{Stock company} & \cellcolor{gray!6}{8801}\\
2015-01-01 & Vice-Minister for Policy Coordination & MIAC & AKTIO & Stock company & -99\\
\cellcolor{gray!6}{2012-10-01} & \cellcolor{gray!6}{Japan Customs} & \cellcolor{gray!6}{MOF} & \cellcolor{gray!6}{ALL NIPPON AIRWAYS} & \cellcolor{gray!6}{Stock company} & \cellcolor{gray!6}{-99}\\
\addlinespace
2011-07-01 & Maritime Affairs Bureau & MLIT & MARITIME HUMAN RESOURCE INSTITUTE & Foundation & -99\\
\cellcolor{gray!6}{2011-06-01} & \cellcolor{gray!6}{Regional Legal Affairs Bureau} & \cellcolor{gray!6}{MOJ} & \cellcolor{gray!6}{JAPAN ASSOCIATION FOR PUBLIC HUMAN RESOURCES DEVELOPMENT} & \cellcolor{gray!6}{Foundation} & \cellcolor{gray!6}{-99}\\
2018-07-01 & Japan Coast Guard & MLIT & WAKACHIKU CONSTRUCTION & Stock company & 1888\\
\cellcolor{gray!6}{2013-07-01} & \cellcolor{gray!6}{Regional Development Bureau} & \cellcolor{gray!6}{MLIT} & \cellcolor{gray!6}{JAPAN FEDERATION OF CONSTRUCTION CONTRACTORS} & \cellcolor{gray!6}{Incorporated association} & \cellcolor{gray!6}{-99}\\
2013-11-01 & Rural Development Bureau & MAFF & MAEDA & Stock company & 1824\\
\addlinespace
\cellcolor{gray!6}{2014-06-01} & \cellcolor{gray!6}{Industrial Science and Technology Policy and Environment Bureau} & \cellcolor{gray!6}{METI} & \cellcolor{gray!6}{TOKYO UNIVERSITY OF SCIENCE} & \cellcolor{gray!6}{Educational institution} & \cellcolor{gray!6}{-99}\\
2016-04-01 & Industrial Safety Supervisory Bureau & METI & MITSUBISHI MATERIALS & Stock company & 5711\\
\cellcolor{gray!6}{2014-09-01} & \cellcolor{gray!6}{Small and Medium Enterprise Agency} & \cellcolor{gray!6}{METI} & \cellcolor{gray!6}{OSAKA UNIVERSITY} & \cellcolor{gray!6}{Educational institution} & \cellcolor{gray!6}{-99}\\
2013-11-01 & Science and Technology Policy Research Institute & MEXT & NATIONAL GRADUATE INSTITUTE FOR POLICY STUDIES & Educational institution & -99\\
\cellcolor{gray!6}{2016-06-22} & \cellcolor{gray!6}{National Tax Agency} & \cellcolor{gray!6}{-99} & \cellcolor{gray!6}{JMS} & \cellcolor{gray!6}{Stock company} & \cellcolor{gray!6}{7702}\\
\addlinespace
2018-07-01 & National Research Institute of Fire and Disaster & MIAC & ENEOS HOLDINGS & Stock company & 5020\\
\cellcolor{gray!6}{2010-10-01} & \cellcolor{gray!6}{Manufacturing Industries Bureau} & \cellcolor{gray!6}{METI} & \cellcolor{gray!6}{INTERNATIONAL BUSINESS MACHINE} & \cellcolor{gray!6}{Stock company} & \cellcolor{gray!6}{-99}\\
2014-07-01 & Japan Coast Guard & MLIT & TOKYO GAS & Stock company & 9531\\
\cellcolor{gray!6}{2013-07-15} & \cellcolor{gray!6}{Kyushu Regional Agricultural Administration Office} & \cellcolor{gray!6}{MAFF} & \cellcolor{gray!6}{MIRAI GROUP} & \cellcolor{gray!6}{Stock company} & \cellcolor{gray!6}{-99}\\
2017-06-01 & Minister's Secretariat & -99 & J-OIL MILLS & Stock company & 2613\\
\addlinespace
\cellcolor{gray!6}{2018-05-24} & \cellcolor{gray!6}{Public Prosecutors Office} & \cellcolor{gray!6}{MOJ} & \cellcolor{gray!6}{FAMILYMART} & \cellcolor{gray!6}{Stock company} & \cellcolor{gray!6}{-99}\\
2017-11-01 & Japan Customs & MOF & SOJITZ & Stock company & 2768\\
\cellcolor{gray!6}{2015-10-01} & \cellcolor{gray!6}{Japan Fair Trade Comission} & \cellcolor{gray!6}{-99} & \cellcolor{gray!6}{NIPPON TELEGRAPH AND TELEPHONE} & \cellcolor{gray!6}{Stock company} & \cellcolor{gray!6}{-99}\\
2017-05-01 & Japan Meteorological Agency & MLIT & JAPAN METEOROLOGICAL BUSINESS SUPPORT CENTER & Foundation & -99\\
\cellcolor{gray!6}{2015-11-01} & \cellcolor{gray!6}{Labor Standard Bureau} & \cellcolor{gray!6}{MHLW} & \cellcolor{gray!6}{HEALTH AND SAFETY TECHNOLOGY EXAMINATION ASSOCIATION} & \cellcolor{gray!6}{Foundation} & \cellcolor{gray!6}{-99}\\
\addlinespace
2017-12-01 & General & MHLW & TORAY INDUSTRIES & Stock company & 3402\\
\cellcolor{gray!6}{2018-09-01} & \cellcolor{gray!6}{Japan Customs} & \cellcolor{gray!6}{MOF} & \cellcolor{gray!6}{SANKYU} & \cellcolor{gray!6}{Stock company} & \cellcolor{gray!6}{9065}\\
2011-04-01 & Japan Meteorological Agency & MLIT & JAPAN METEOROLOGICAL BUSINESS SUPPORT CENTER & Foundation & -99\\
\cellcolor{gray!6}{2018-06-01} & \cellcolor{gray!6}{Minister's Secretariat} & \cellcolor{gray!6}{MLIT} & \cellcolor{gray!6}{NATIONAL GRADUATE INSTITUTE FOR POLICY STUDIES} & \cellcolor{gray!6}{Educational institution} & \cellcolor{gray!6}{-99}\\
2017-10-30 & National Tax Agency & -99 & TKC & Stock company & 9746\\
\bottomrule
\end{tabular}}
\end{table}

\pagebreak
\section{Additional descriptive statistics} \label{sec: additional_stats}

\subsection{Retirements by firm and firm type} 

\begin{figure}[!htb]
\includegraphics{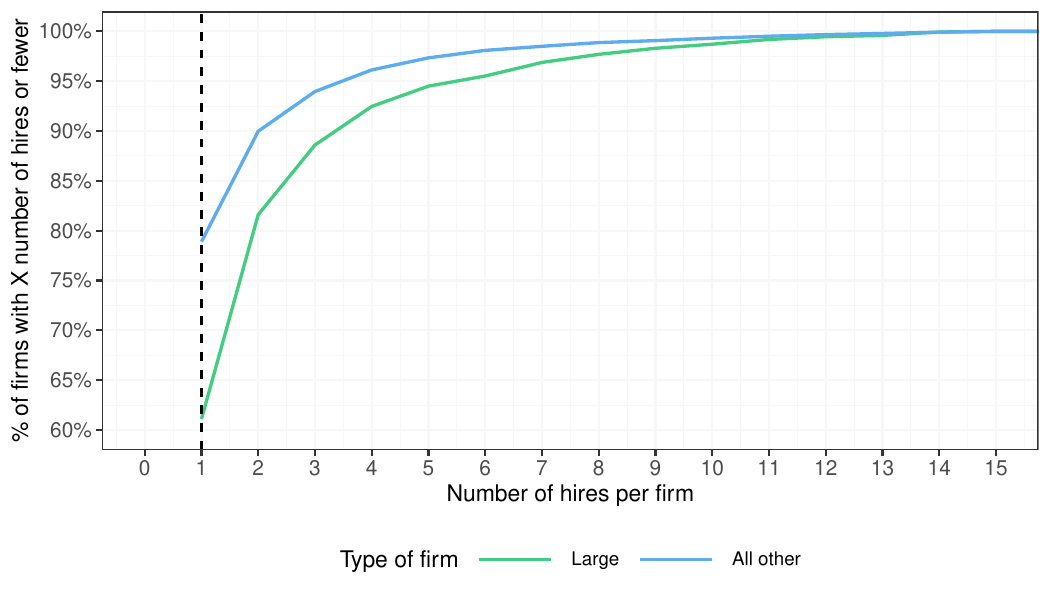}
\vspace{0.2cm}
\caption{Empirical cumulative distribution function of number of hires per firm}
\small
\vspace{-0.3cm}
\label{fig: hires_ecdf}
\end{figure}

\noindent
\small
Note: Large firms are those listed in the Nikkei NEEDS financial database.
\normalsize

\pagebreak 
\subsection{Retirements by industry} 

\begin{figure}[!htb]
\includegraphics[width=\textwidth]{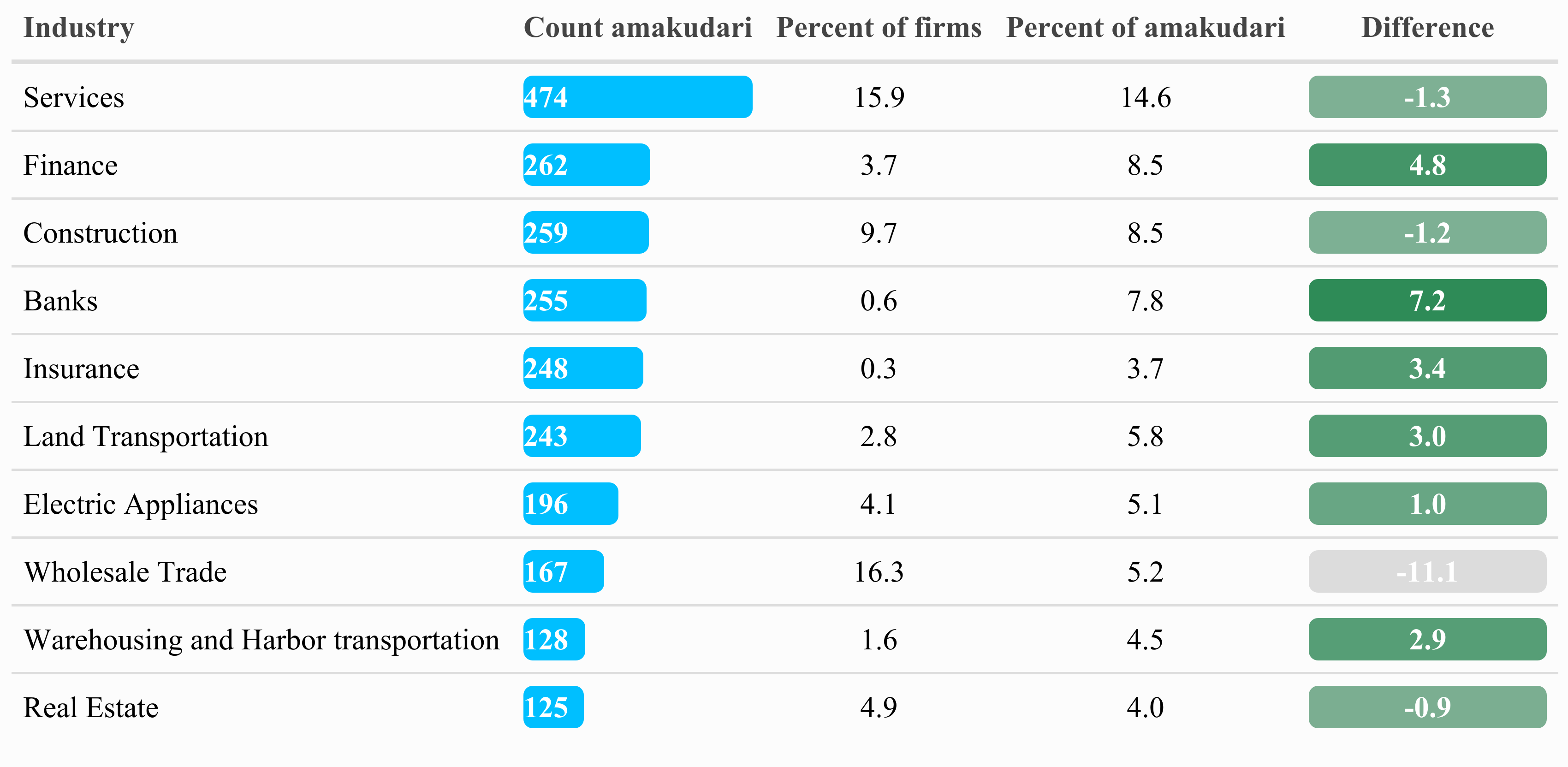}
\caption{Top 10 \textit{amakudari} destinations vs. overall economy}
\small
\vspace{-0.3cm}
\label{tab: amakudari_firms_short}
\textit{Note}: ``Percent of firms'' only includes firms with financial information in the NEEDS database. 
\end{figure}
\vspace{-1cm}
\noindent
\footnotesize
\normalsize
\vspace{0.5cm}

\begin{table}[H]

\caption{\label{tab:amakudari_firms_full}Amakudari industry destinations vs. overall economy}
\centering
\resizebox{\linewidth}{!}{
\begin{tabular}[t]{lcccc}
\toprule
Industry & Count amakudari & Percent of firms & Percent of amakudari & Difference\\
\midrule
\cellcolor{gray!6}{Services} & \cellcolor{gray!6}{474} & \cellcolor{gray!6}{15.9} & \cellcolor{gray!6}{14.6} & \cellcolor{gray!6}{-1.3}\\
Finance & 262 & 3.7 & 8.5 & 4.8\\
\cellcolor{gray!6}{Construction} & \cellcolor{gray!6}{259} & \cellcolor{gray!6}{9.7} & \cellcolor{gray!6}{8.5} & \cellcolor{gray!6}{-1.2}\\
Banks & 255 & 0.6 & 7.8 & 7.2\\
\cellcolor{gray!6}{Insurance} & \cellcolor{gray!6}{248} & \cellcolor{gray!6}{0.3} & \cellcolor{gray!6}{3.7} & \cellcolor{gray!6}{3.4}\\
\addlinespace
Land Transportation & 243 & 2.8 & 5.8 & 3.0\\
\cellcolor{gray!6}{Electric Appliances} & \cellcolor{gray!6}{196} & \cellcolor{gray!6}{4.1} & \cellcolor{gray!6}{5.1} & \cellcolor{gray!6}{1.0}\\
Wholesale Trade & 167 & 16.3 & 5.2 & -11.1\\
\cellcolor{gray!6}{Warehousing and Harbor transportation} & \cellcolor{gray!6}{128} & \cellcolor{gray!6}{1.6} & \cellcolor{gray!6}{4.5} & \cellcolor{gray!6}{2.9}\\
Real Estate & 125 & 4.9 & 4.0 & -0.9\\
\addlinespace
\cellcolor{gray!6}{Electric Power \& Gas} & \cellcolor{gray!6}{118} & \cellcolor{gray!6}{0.6} & \cellcolor{gray!6}{3.2} & \cellcolor{gray!6}{2.6}\\
Information \& Communication & 109 & 2.5 & 3.5 & 1.0\\
\cellcolor{gray!6}{Machinery} & \cellcolor{gray!6}{101} & \cellcolor{gray!6}{4.1} & \cellcolor{gray!6}{3.1} & \cellcolor{gray!6}{-1.0}\\
Transport Equipment & 88 & 1.9 & 3.2 & 1.3\\
\cellcolor{gray!6}{Retail Trade} & \cellcolor{gray!6}{83} & \cellcolor{gray!6}{5.7} & \cellcolor{gray!6}{3.1} & \cellcolor{gray!6}{-2.6}\\
\addlinespace
Chemicals & 72 & 3.0 & 2.5 & -0.5\\
\cellcolor{gray!6}{Air Transportation} & \cellcolor{gray!6}{70} & \cellcolor{gray!6}{0.2} & \cellcolor{gray!6}{2.0} & \cellcolor{gray!6}{1.8}\\
Foods & 62 & 3.3 & 2.2 & -1.1\\
\cellcolor{gray!6}{Marine Transportation} & \cellcolor{gray!6}{49} & \cellcolor{gray!6}{0.8} & \cellcolor{gray!6}{1.7} & \cellcolor{gray!6}{0.9}\\
Iron \& Steel & 38 & 1.1 & 0.9 & -0.2\\
\addlinespace
\cellcolor{gray!6}{Other Products} & \cellcolor{gray!6}{34} & \cellcolor{gray!6}{3.9} & \cellcolor{gray!6}{1.2} & \cellcolor{gray!6}{-2.7}\\
Nonferrous Metals & 25 & 0.8 & 0.9 & 0.1\\
\cellcolor{gray!6}{Pharmaceutical} & \cellcolor{gray!6}{23} & \cellcolor{gray!6}{0.7} & \cellcolor{gray!6}{0.8} & \cellcolor{gray!6}{0.1}\\
Glass \& Ceramics Products & 22 & 1.9 & 0.7 & -1.2\\
\cellcolor{gray!6}{Textile \& Apparels} & \cellcolor{gray!6}{22} & \cellcolor{gray!6}{4.2} & \cellcolor{gray!6}{0.8} & \cellcolor{gray!6}{-3.4}\\
\addlinespace
Metal Products & 21 & 2.4 & 0.8 & -1.6\\
\cellcolor{gray!6}{Precision Instruments} & \cellcolor{gray!6}{18} & \cellcolor{gray!6}{1.0} & \cellcolor{gray!6}{0.7} & \cellcolor{gray!6}{-0.3}\\
Oil \& Coal Products & 17 & 0.2 & 0.6 & 0.4\\
\cellcolor{gray!6}{Mining} & \cellcolor{gray!6}{5} & \cellcolor{gray!6}{0.3} & \cellcolor{gray!6}{0.2} & \cellcolor{gray!6}{-0.1}\\
Rubber Products & 4 & 0.4 & 0.2 & -0.2\\
\addlinespace
\cellcolor{gray!6}{Pulp \& Paper} & \cellcolor{gray!6}{2} & \cellcolor{gray!6}{0.9} & \cellcolor{gray!6}{0.1} & \cellcolor{gray!6}{-0.8}\\
Fishery, Agriculture \& Forestry & 1 & 0.3 & 0.0 & -0.3\\
\bottomrule
\multicolumn{5}{l}{\rule{0pt}{1em}\textit{Note: } ``Percent economy'' calculation is the total number of firms in each industry divided by all firms in the Nikkei NEEDS database.}\\
\end{tabular}}
\end{table}

\pagebreak

\subsection{Retirements by ministry} 

\begin{figure}[!htb]
\begin{centering}
\begin{subfigure}{\textwidth}
\centering
\includegraphics{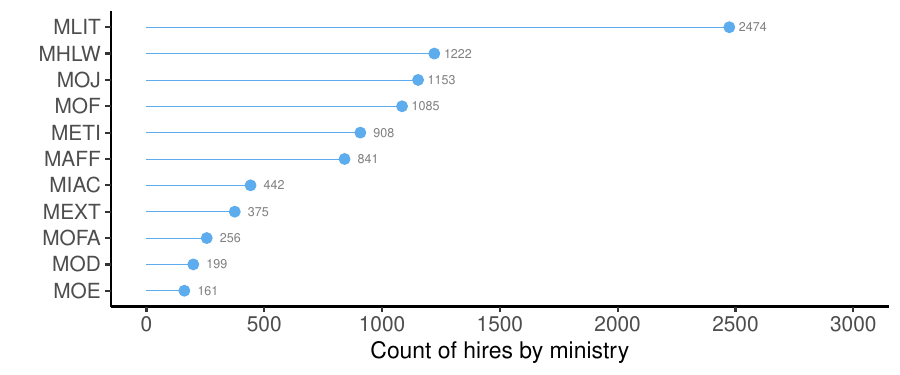}
\caption{Total \textit{amakudari} appointments by former ministry (all years)} 
\label{fig: ministry_bar}
\end{subfigure}
\begin{subfigure}{\textwidth}
\centering
\includegraphics{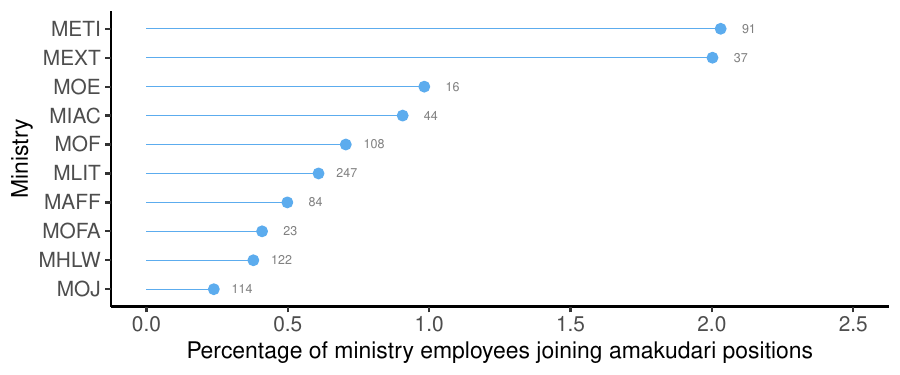}
\caption{Mean \textit{amakudari} appointments as a percentage of ministry employees (all years)}
\label{fig: ministry_bar_adjusted}
\end{subfigure}
\caption{Amakudari appointments by ministry}
\end{centering}
\small
Note: Ministry of Defense (MOD) excluded from adjusted figures. Data on total MOD employees is unavailable as Japan Statistical Yearbook data only includes only regular (8 hours per day) employees, and excludes members of the Japan Self Defense Forces (JSDF). Official numbers therefore exclude JSDF members and civilian MOD employees without 8-hour workdays. I thank Samuel Leiter and an anonymous MOD official for this insight. 
\vspace{-0.3cm}
\end{figure}

\pagebreak
\clearpage

\begin{figure}[!htb]
\includegraphics[scale=0.75, trim={2cm 2cm 2cm 2cm}, clip]{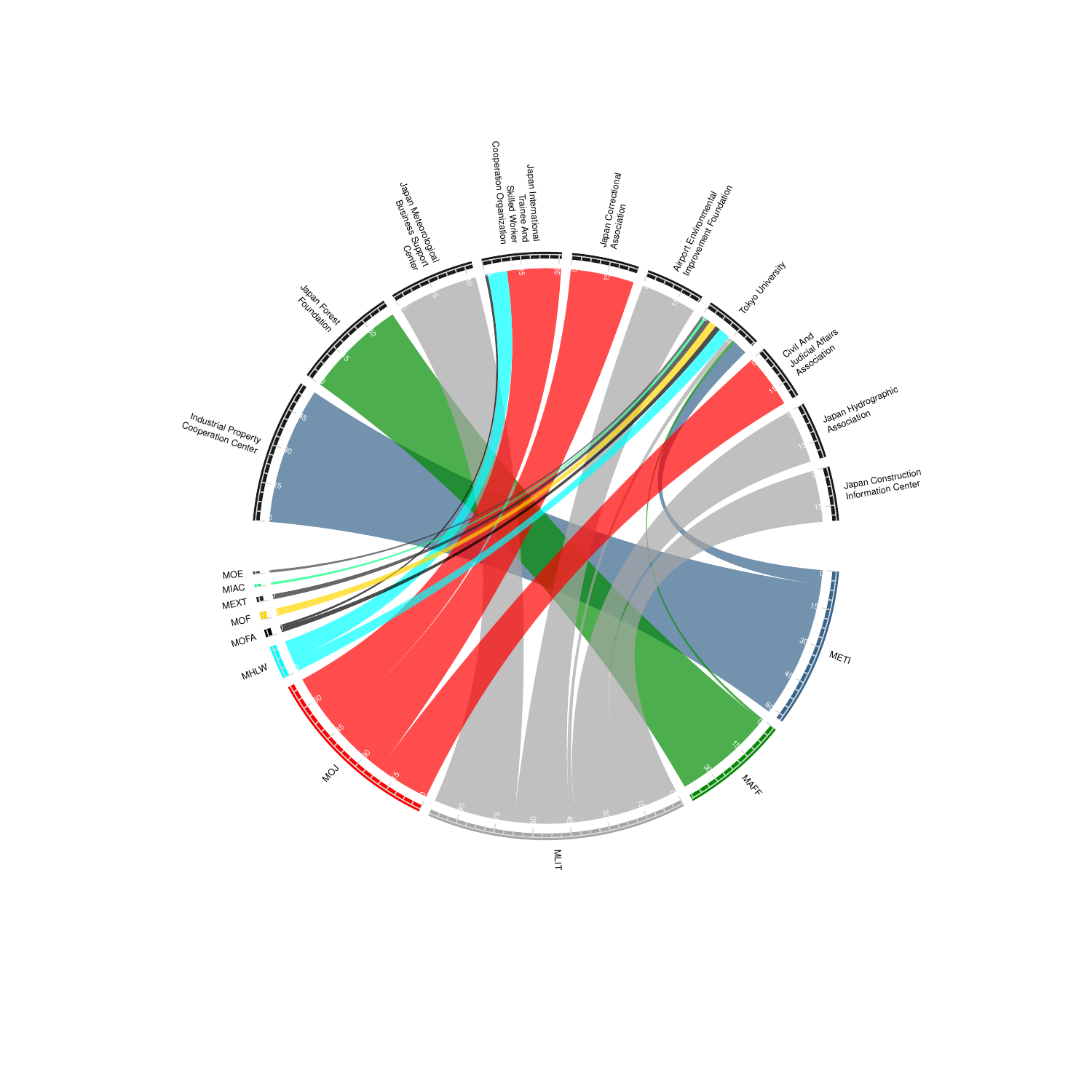}
\vspace{0.2cm}
\caption{Flows from ministries to top ten public interest corporations (all years)}
\small
\vspace{-0.3cm}
\label{fig: ministry_public_interest}
\end{figure}


\pagebreak

\begin{figure}[!htb]
\includegraphics[scale=1, trim={4cm 5cm 4cm 4cm}, clip]{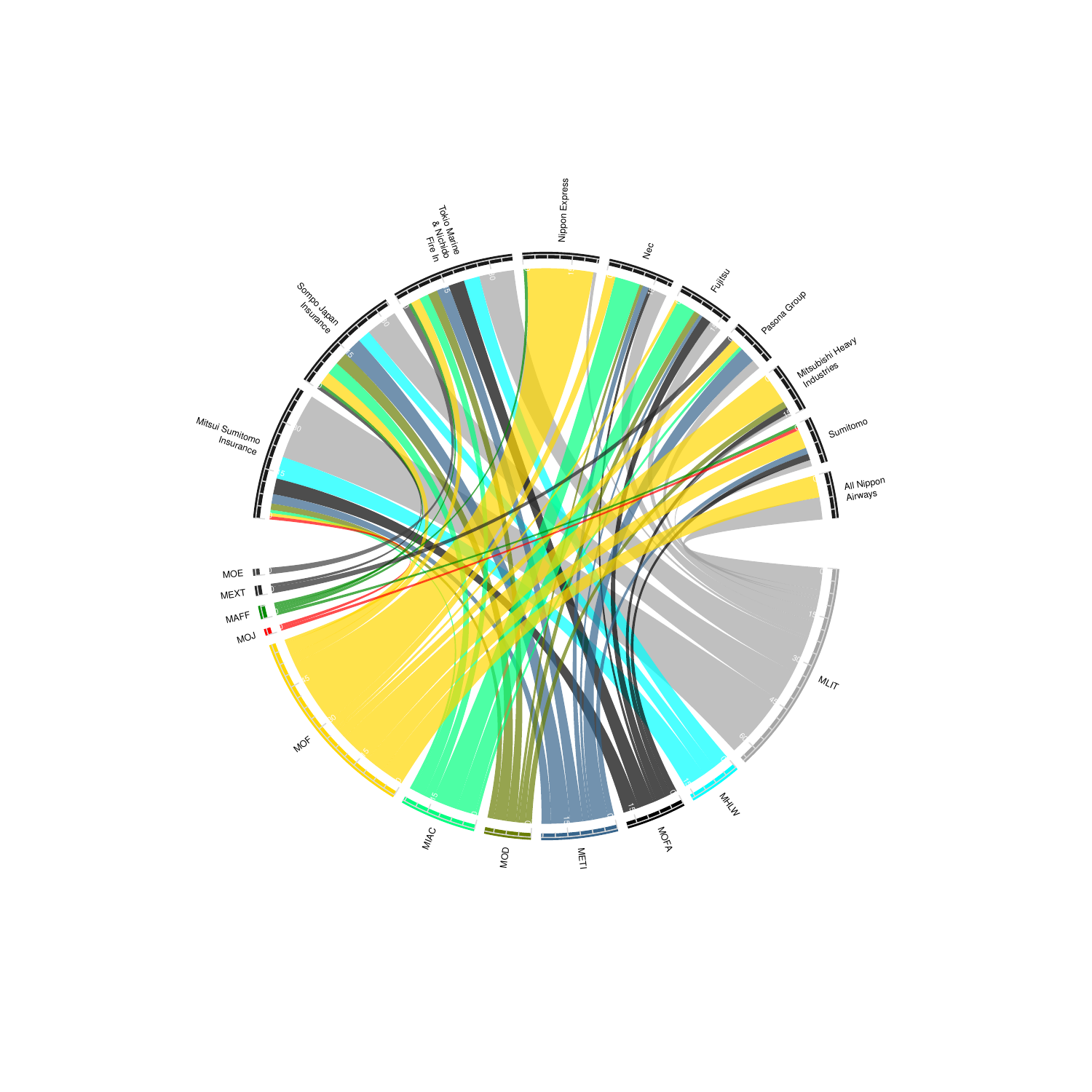}
\vspace{0.2cm}
\caption{Flows from ministries to top ten private corporations (all years)}
\small
\vspace{-0.3cm}
\label{fig: ministry_private}
\end{figure}
\begin{singlespace}
\noindent
\small
The top corporations by number of \textit{amakudari} hires tend to draw from a diverse array of ministries. 
\end{singlespace}
\normalsize

\clearpage

\begin{figure}[!htb]
\centering
\begin{subfigure}{\textwidth}
\centering
\includegraphics[width = 0.9\textwidth]{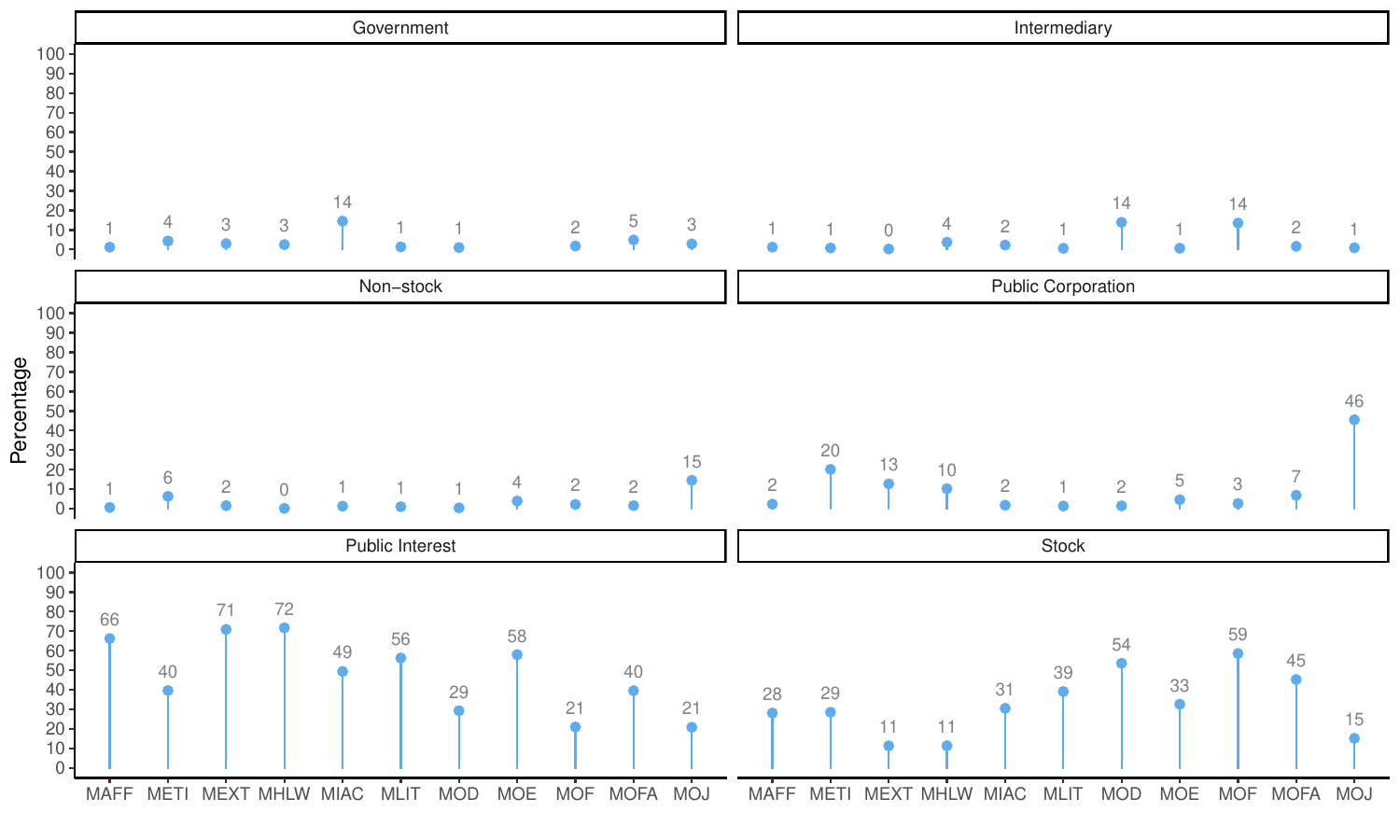}
\caption{Percentage of retirees in each firm type by ministry} 
\label{fig: ministry_firm_type_withpatent}
\end{subfigure}
\begin{subfigure}{\textwidth}
\centering
\includegraphics[width = 0.9\textwidth]{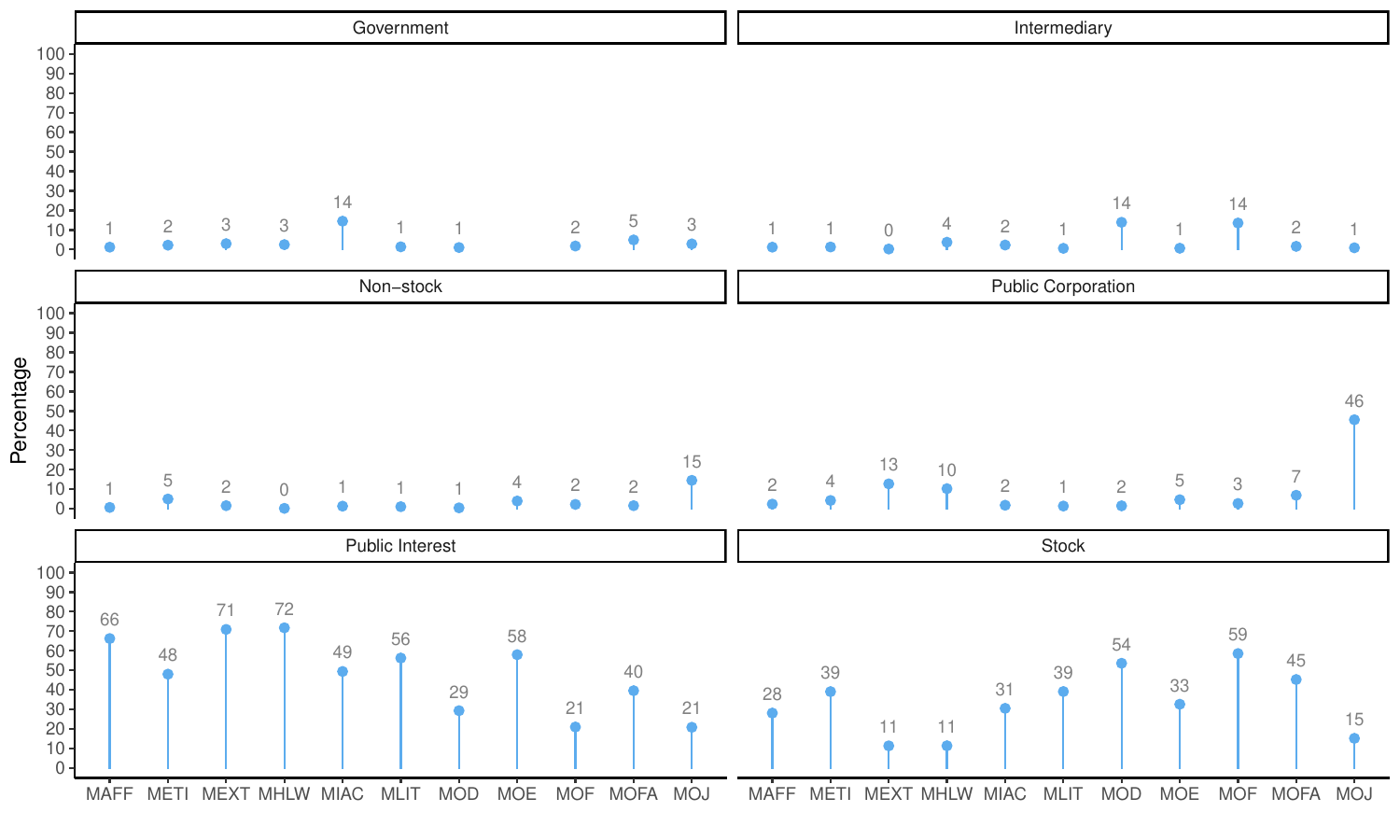}
\caption{Percentage of retirees in each firm type by ministry, excluding METI Patent Office}
\label{fig: ministry_firm_type_wo_patent}
\end{subfigure}
\caption{Percentage of retirees in each firm type by ministry}
\label{fig: ministry_firm_type}
\end{figure}

\subsection{Age of retirement} \label{sec: firm_financials}

\begin{table}[!htbp] \centering 
  \caption{Age of retirement: mean and quantiles (all years)} 
  \label{tab: age} 
\begin{tabular}{@{\extracolsep{0.5cm}} cccccccc} 
\\[-1.8ex]\hline 
\hline \\[-1.8ex] 
Firm type & Firm sub-type & Mean & 5 & 25 & Median & 75 & 95 \\ 
\hline \\[-1.8ex] 
Unclassified &  & 59 & 52 & 59 & 60 & 61 & 64 \\ 
Government &  & 58 & 45 & 58 & 60 & 60 & 63 \\ 
Private corporation & Intermediary & 59 & 56 & 57 & 59 & 60 & 61 \\ 
Private corporation & Non-stock & 59 & 46 & 60 & 60 & 60 & 61 \\ 
Private corporation & Public Interest & 59 & 55 & 58 & 60 & 60 & 63 \\ 
Private corporation & Stock & 59 & 55 & 58 & 60 & 60 & 62 \\ 
Public corporation &  & 59 & 54 & 58 & 59 & 60 & 63 \\ 
\hline \\[-1.8ex] 
\end{tabular} 
\end{table}

\begin{figure}[H]
\begin{centering}
\hspace{-0.5cm}
\includegraphics[width = \textwidth]{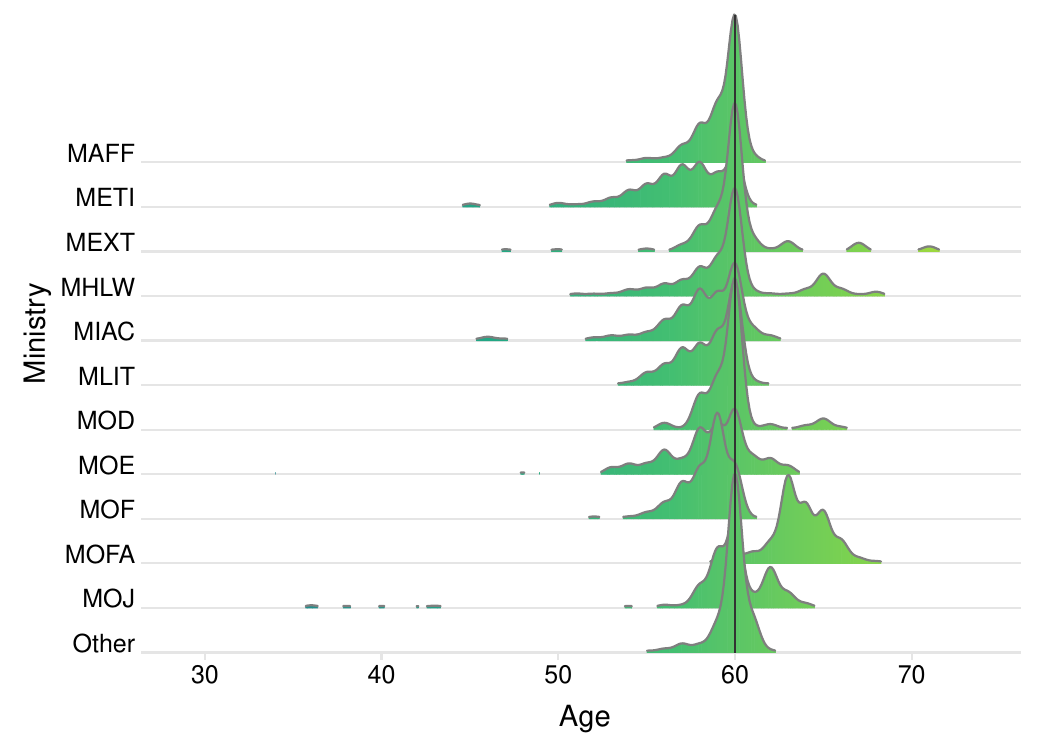}
\vspace{0.3cm}
\caption{Age of exit from ministry, by ministry}
\label{fig: age}
\footnotesize
\textit{Note}: Vertical line at ``mandatory'' retirement age of 60.
\end{centering}
\end{figure}

\pagebreak
\subsection{Firm financials} \label{sec: firm_financials}

\begin{table}[H]
\centering
\fontsize{10}{12}\selectfont
\begin{tabular}[t]{lrrrrrr}
\toprule
\multicolumn{1}{c}{ } & \multicolumn{2}{c}{Amakudari (N=711)} & \multicolumn{2}{c}{No amakudari (N=5100)} & \multicolumn{2}{c}{ } \\
\cmidrule(l{3pt}r{3pt}){2-3} \cmidrule(l{3pt}r{3pt}){4-5}
\textbf{ } & \textbf{Mean} & \textbf{Std. Dev.} & \textbf{Mean} & \textbf{Std. Dev.} & \textbf{Diff. in Means} & \textbf{p}\\
\midrule
Number of amakudari & 3.02 & 4.73 & 0.00 & 0.00 & -3.02 & $<$0.01\\
Total government loans & 4.29 & 18.88 & 0.39 & 6.20 & -3.89 & $<$0.01\\
Total private loans & 28.84 & 92.20 & 5.61 & 28.31 & -23.23 & $<$0.01\\
Total assets & 4024.22 & 21618.16 & 152.45 & 910.75 & -3871.77 & $<$0.01\\
Total liabilities & 3542.80 & 20529.55 & 111.66 & 828.90 & -3431.14 & $<$0.01\\
Operating revenue & 763.15 & 1903.71 & 78.04 & 269.44 & -685.11 & $<$0.01\\
Gross profit & 169.36 & 469.07 & 19.40 & 78.94 & -149.96 & $<$0.01\\
EBITDA & 85.05 & 246.94 & 7.02 & 39.40 & -78.03 & $<$0.01\\
Leverage & 3.24 & 4.84 & 3.03 & 4.44 & -0.22 & 0.30\\
Employees & 14729.80 & 35799.86 & 1698.24 & 4498.09 & -13031.55 & $<$0.01\\
Temporary employees & 4748.39 & 14918.97 & 893.74 & 2931.38 & -3854.66 & $<$0.01\\
Return on investment & 4.94 & 60.37 & 11.45 & 113.39 & 6.51 & 0.03\\
Return on equity & 4.05 & 20.44 & -0.05 & 89.58 & -4.10 & 0.01\\
Reserve ratio & 65.02 & 92.79 & 126.69 & 261.07 & 61.67 & $<$0.01\\
Missing & 0.04 & 0.18 & 0.19 & 0.39 & 0.15 & $<$0.01\\
\bottomrule
\end{tabular}
\caption{For-profit firm financial data by \textit{amakudari} status}
\label{tab: amakudari_balance}
{\raggedright Notes: Firm-level means across all years 2009-2019. Includes all firms for which government loan data exists in the NEEDS financial database. Loans, assets, liabilities, revenue, profit, and EBITDA in billion yen.}
\end{table}

\begin{table}[H]
\centering
\fontsize{10}{12}\selectfont
\begin{tabular}[t]{lrrrrrr}
\toprule
\multicolumn{1}{c}{ } & \multicolumn{2}{c}{No public loans (N=4820)} & \multicolumn{2}{c}{Public loans (N=991)} & \multicolumn{2}{c}{ } \\
\cmidrule(l{3pt}r{3pt}){2-3} \cmidrule(l{3pt}r{3pt}){4-5}
\textbf{ } & \textbf{Mean} & \textbf{Std. Dev.} & \textbf{Mean} & \textbf{Std. Dev.} & \textbf{Diff. in Means} & \textbf{p}\\
\midrule
Number of amakudari & 0.29 & 1.77 & 0.77 & 2.52 & 0.48 & $<$0.01\\
Total assets & 735.96 & 8449.58 & 391.20 & 1219.81 & -344.76 & $<$0.01\\
Total liabilities & 655.80 & 8036.91 & 272.00 & 950.07 & -383.80 & $<$0.01\\
Operating revenue & 124.20 & 645.87 & 306.06 & 1037.25 & 181.86 & $<$0.01\\
Gross profit & 28.42 & 168.80 & 61.34 & 190.15 & 32.92 & $<$0.01\\
Return on investment & 14.84 & 322.74 & 18.38 & 220.85 & 3.53 & 0.68\\
EBITDA & 11.83 & 88.03 & 29.63 & 93.25 & 17.80 & $<$0.01\\
Return on equity & -12.79 & 1424.56 & -2.33 & 155.09 & 10.46 & 0.64\\
Leverage & 4.32 & 37.71 & 4.28 & 6.42 & -0.04 & 0.95\\
Reserve ratio & 126.27 & 611.64 & 59.78 & 89.62 & -66.49 & $<$0.01\\
Employees & 2542.09 & 11590.18 & 5749.02 & 18528.88 & 3206.93 & $<$0.01\\
Temporary employees & 1204.17 & 5115.74 & 2447.99 & 10078.88 & 1243.81 & $<$0.01\\
Total government loans & 0.00 & 0.00 & 5.10 & 21.02 & 5.10 & $<$0.01\\
Total private loans & 3.23 & 23.58 & 33.84 & 84.16 & 30.61 & $<$0.01\\
\bottomrule
\end{tabular}
\caption{For-profit firm financial data by government loan status}
\label{tab: loans_balance}
{\raggedright Notes: Firm-level means across all years 2009-2019. Includes all firms for which government loan data exists in the NEEDS financial database. Loans, assets, liabilities, revenue, profit, and EBITDA in billion yen. \par}
\end{table}


\begin{figure}
\centering
\begin{subfigure}[b]{0.85\textwidth}
  \includegraphics[width=\textwidth]{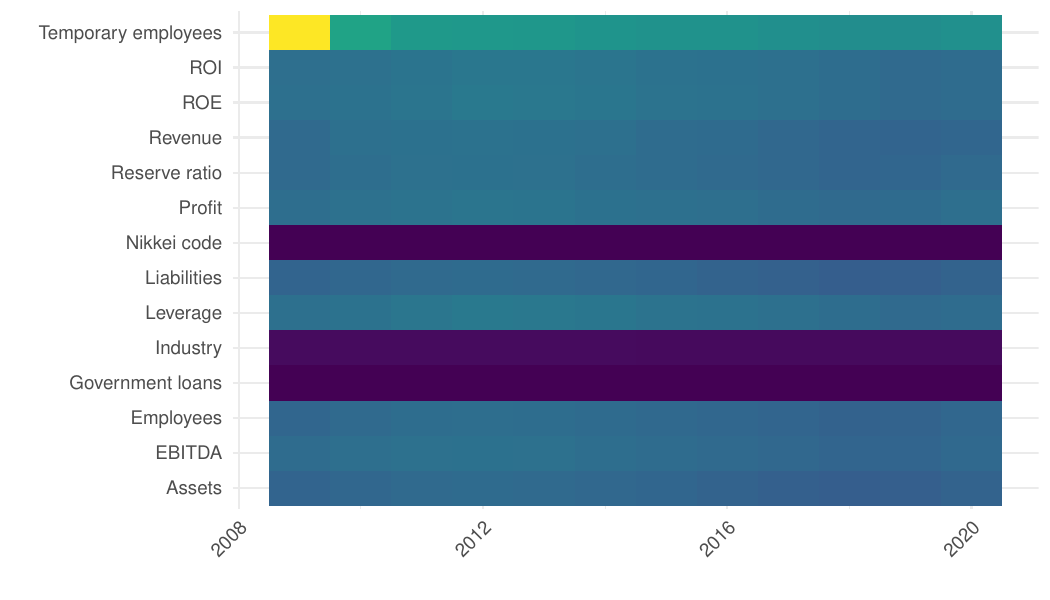}
  \caption{Financial data missingness by year}
  \label{fig: loan_missing_time} 
\end{subfigure}
\begin{subfigure}[b]{0.85\textwidth}
  \includegraphics[width=\textwidth]{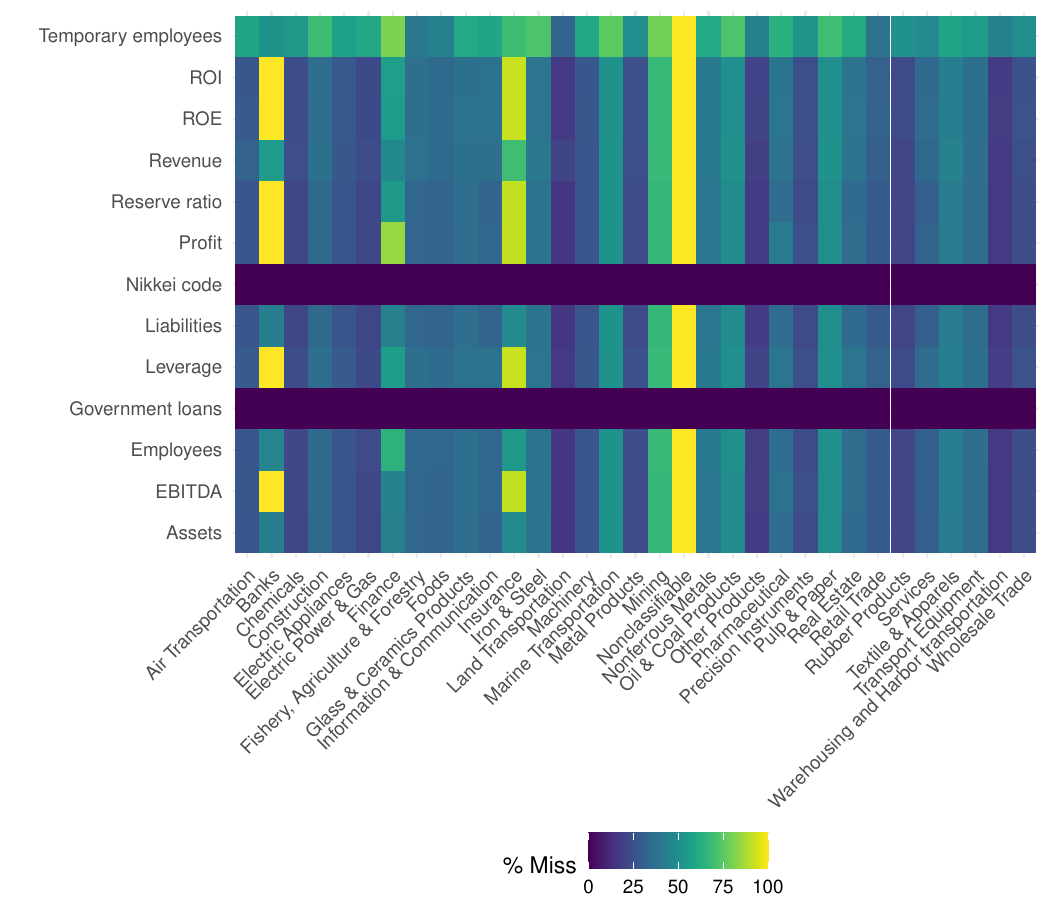}
  \caption{Financial data missingness by industry}
  \label{fig: loan_missing_industry}
\end{subfigure}
\caption{Financial data missingness}
\label{fig: loan_missing}
\end{figure}

\newpage
\clearpage
\section{Loan analysis} \label{sec: loan_analysis}

\subsection{Descriptive statistics} 

\begin{figure}[!htb]
\begin{centering}
\includegraphics[width = 0.8\textwidth]{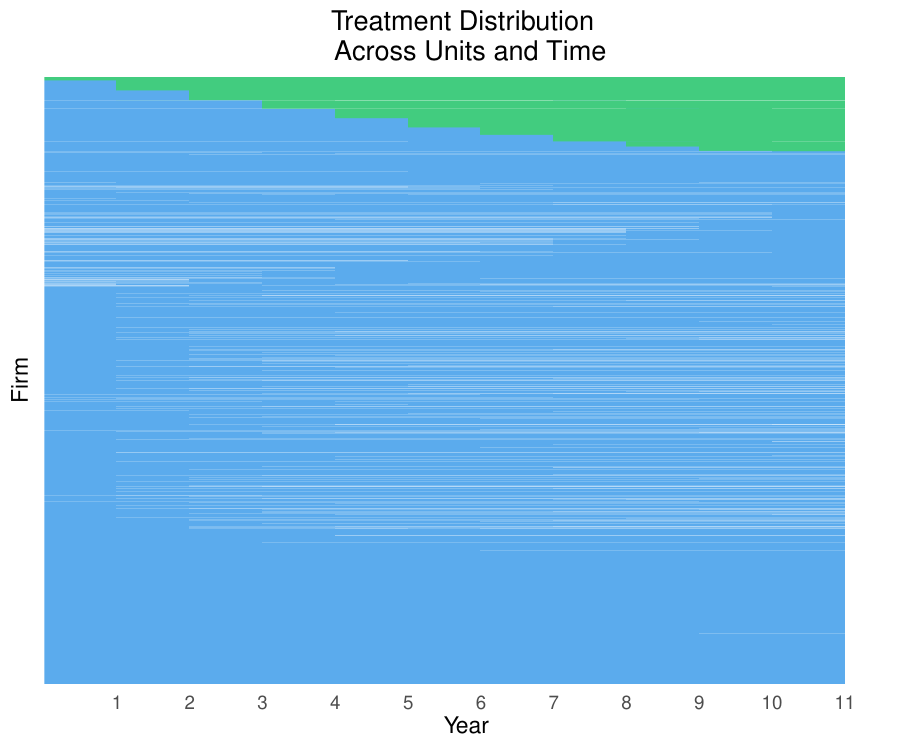}
\vspace{0.2cm}
\caption{Distribution of treatment and control status of all firms in loan analysis}
\small
Note: Treated firms in green and control firms in blue. White areas depict missing data. 
\vspace{-0.3cm}
\label{fig: loan_treat_control}
\end{centering}
\end{figure}

\pagebreak
\clearpage
\subsection{Effects by ministry} 

\begin{figure}[!htb]
\begin{centering}
\includegraphics[width = 0.8\textwidth]{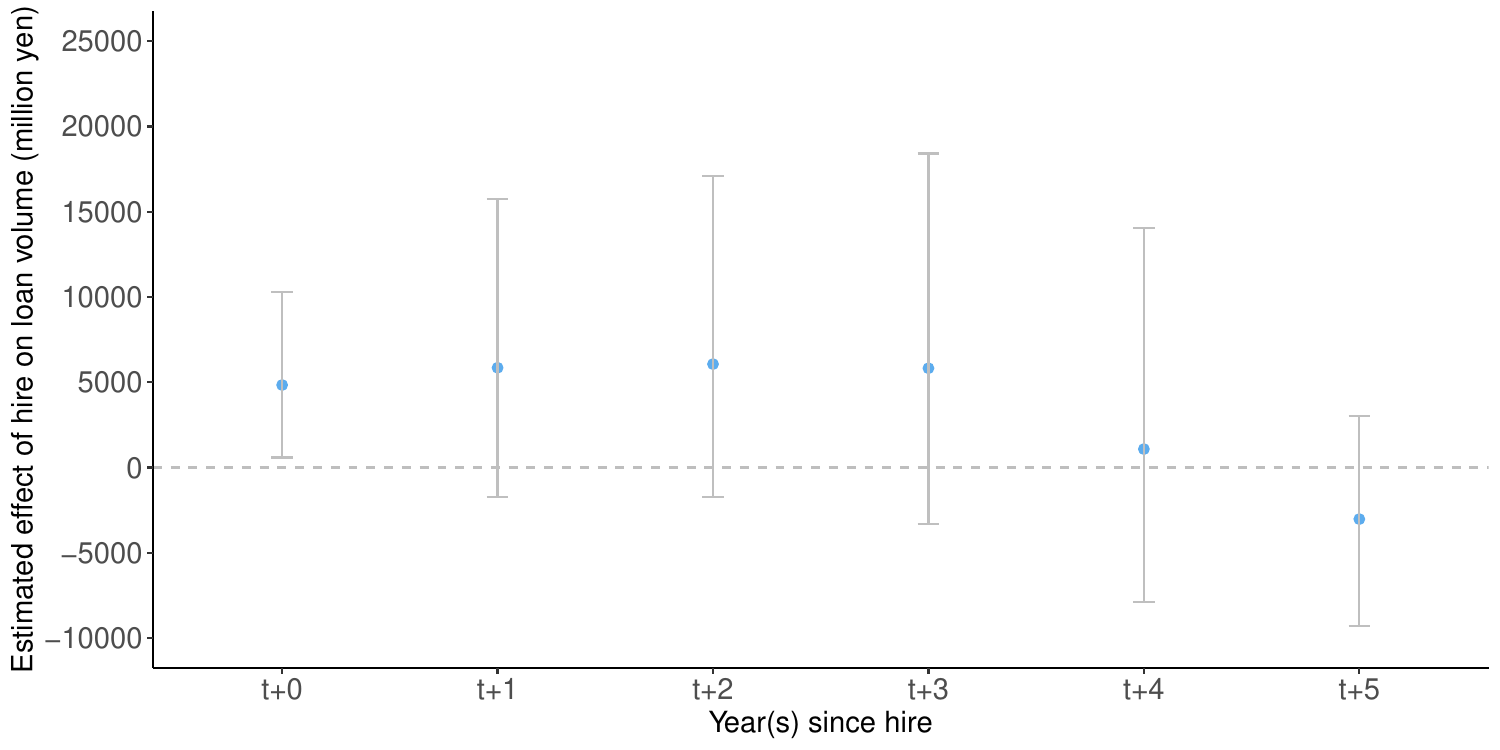}
\vspace{0.2cm}
\caption{Estimated effect of bureaucratic hires on size of government loan received, METI re-hires only}
\small
\vspace{-0.3cm}
\label{fig: tscs_loan_meti}
\end{centering}
{\footnotesize
\raggedright Note: Tabular results can be found in \autoref{tab: tscs_loan_meti}. \par}
\end{figure}

\begin{figure}[!htb]
\begin{centering}
\includegraphics[width = 0.8\textwidth]{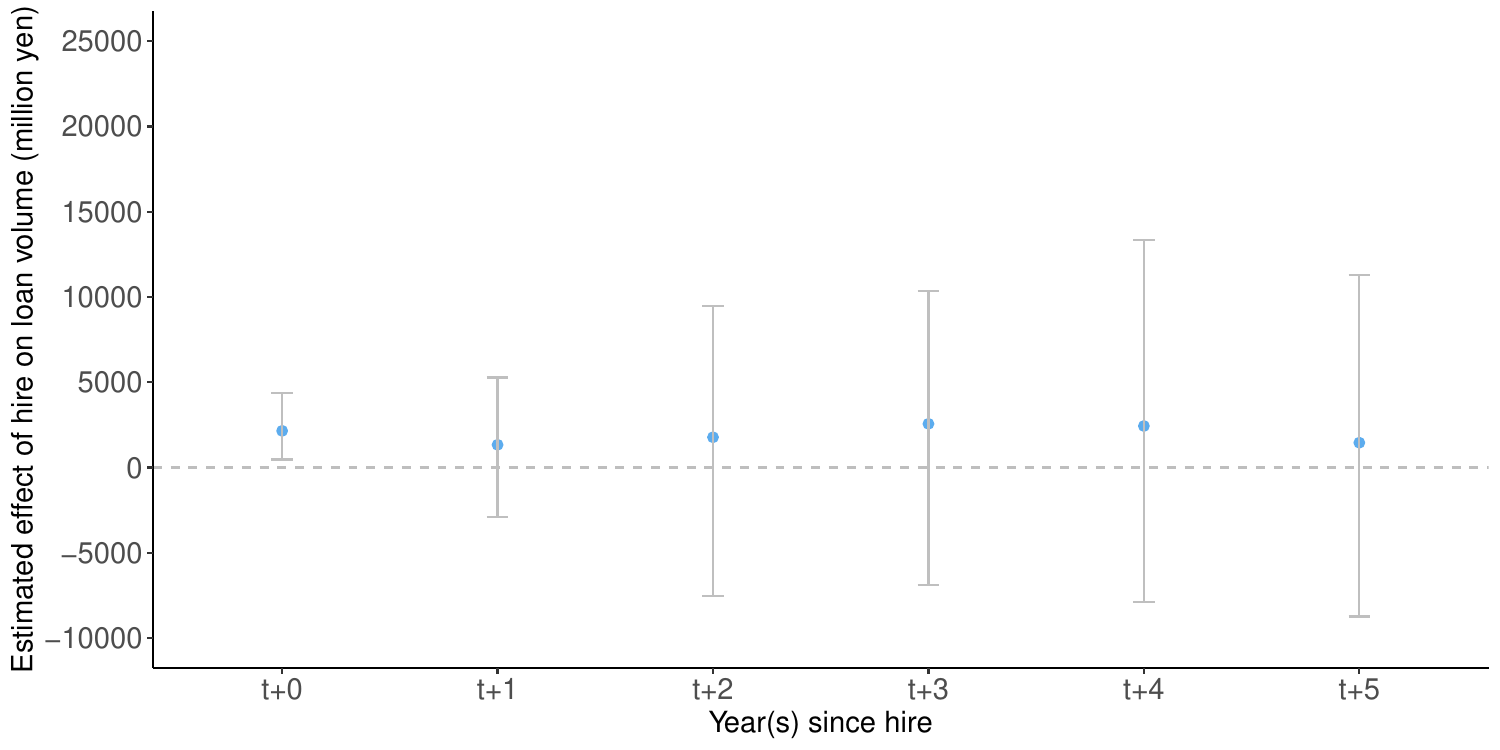}
\vspace{0.2cm}
\caption{Estimated effect of bureaucratic hires on size of government loan received, MOF re-hires only}
\small
\vspace{-0.3cm}
\label{fig: tscs_loan_mof}
\end{centering}
{\footnotesize
\raggedright Note: Tabular results can be found in \autoref{tab: tscs_loan_mof}. \par}
\end{figure}

\begin{figure}[!htb]
\begin{centering}
\includegraphics[width = 0.8\textwidth]{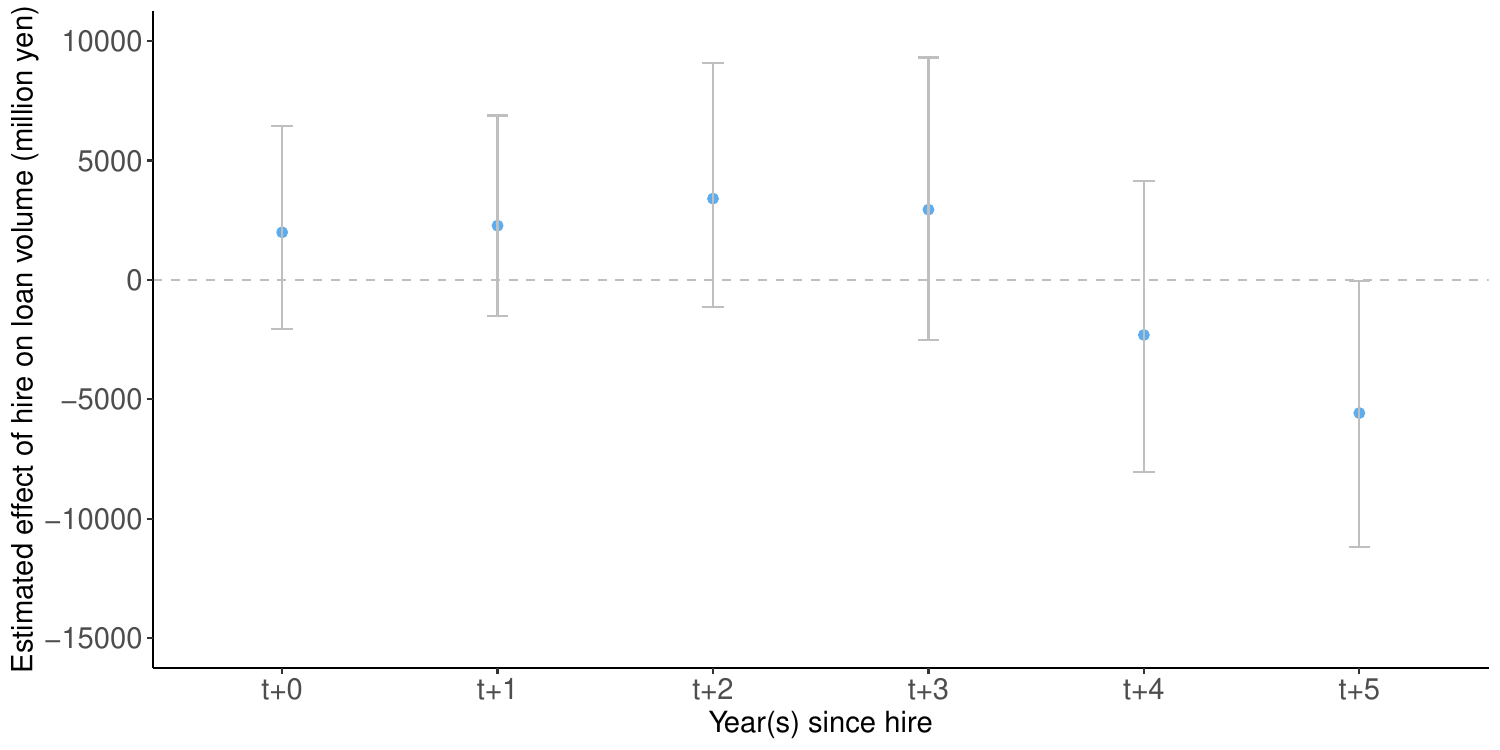}
\vspace{0.2cm}
\caption{Estimated effect of bureaucratic hires on size of government loan received, all ministries other than METI and MOF}
\small
\vspace{-0.3cm}
\label{fig: tscs_loan_other}
\end{centering}
{\footnotesize
\raggedright Note: Tabular results can be found in \autoref{tab: tscs_loan_other}. \par}
\end{figure}

\pagebreak
\clearpage
\subsection{Tabular results}

\begin{table}[!htbp] \centering 
  \caption{Estimated effect of bureaucratic hires on size of government loans received, by year afer hire} 
  \label{tab: tscs_loan} 
\begin{tabular}{@{\extracolsep{5pt}} ccccc} 
\\[-1.8ex]\hline 
\hline \\[-1.8ex] 
Time window & Estimate & SE & 95\% CI lower & 95\% CI upper \\ 
\hline \\[-1.8ex] 
t+0 & $810.22$ & $813.98$ & $$-$758.69$ & $2,340.21$ \\ 
t+1 & $1,444.96$ & $1,088.45$ & $$-$617.01$ & $3,686.78$ \\ 
t+2 & $2,753.18$ & $1,326.34$ & $312.63$ & $5,506.05$ \\ 
t+3 & $3,153.69$ & $1,710.44$ & $$-$46.38$ & $6,757.87$ \\ 
t+4 & $2,028.75$ & $1,805.32$ & $$-$1,335.91$ & $5,748.67$ \\ 
t+5 & $489.80$ & $1,641.46$ & $$-$2,639.04$ & $3,995.84$ \\ 
\hline \\[-1.8ex] 
\multicolumn{5}{l}{Note: Matched sets = 444} \\ 
\end{tabular} 
\end{table}

\begin{table}[!htbp] \centering 
  \caption{Estimated effect of bureaucratic hires on size of government loans received, METI re-hires only} 
  \label{tab: tscs_loan_meti} 
\begin{tabular}{@{\extracolsep{5pt}} ccccc} 
\\[-1.8ex]\hline 
\hline \\[-1.8ex] 
Time window & Estimate & SE & 95\% CI lower & 95\% CI upper \\ 
\hline \\[-1.8ex] 
t+0 & $4,839.34$ & $2,526.42$ & $582.92$ & $10,315.10$ \\ 
t+1 & $5,850.87$ & $4,584.92$ & $$-$1,732.86$ & $15,761.23$ \\ 
t+2 & $6,065.97$ & $4,848.95$ & $$-$1,705.76$ & $17,107.01$ \\ 
t+3 & $5,824$ & $5,677.01$ & $$-$3,333.77$ & $18,423$ \\ 
t+4 & $1,082.36$ & $5,669.50$ & $$-$7,906.57$ & $14,050.62$ \\ 
t+5 & $$-$3,024.54$ & $3,198.90$ & $$-$9,303.84$ & $3,010.77$ \\ 
\hline \\[-1.8ex] 
\multicolumn{5}{l}{Note: Matched sets = 67} \\ 
\end{tabular} 
\end{table}

\begin{table}[!htbp] \centering 
  \caption{Estimated effect of bureaucratic hires on size of government loans received, MOF re-hires only} 
  \label{tab: tscs_loan_mof} 
\begin{tabular}{@{\extracolsep{5pt}} ccccc} 
\\[-1.8ex]\hline 
\hline \\[-1.8ex] 
Time window & Estimate & SE & 95\% CI lower & 95\% CI upper \\ 
\hline \\[-1.8ex] 
t+0 & $2,151.10$ & $988.31$ & $477.10$ & $4,361.63$ \\ 
t+1 & $1,333.66$ & $2,062.18$ & $$-$2,888.21$ & $5,282.42$ \\ 
t+2 & $1,775.08$ & $4,101.66$ & $$-$7,542.83$ & $9,493.07$ \\ 
t+3 & $2,563.39$ & $4,281.40$ & $$-$6,891.22$ & $10,366.60$ \\ 
t+4 & $2,435.80$ & $5,093.48$ & $$-$7,893.75$ & $13,341.59$ \\ 
t+5 & $1,456.66$ & $4,976.09$ & $$-$8,734.25$ & $11,296.23$ \\ 
\hline \\[-1.8ex] 
\multicolumn{5}{l}{Note: Matched sets = 91} \\ 
\end{tabular} 
\end{table}

\begin{table}[!htbp] \centering 
  \caption{Estimated effect of bureaucratic hires on size of government loans received, METI and MOF re-hires only} 
  \label{tab: tscs_loan_meti_mof} 
\begin{tabular}{@{\extracolsep{5pt}} ccccc} 
\\[-1.8ex]\hline 
\hline \\[-1.8ex] 
Time window & Estimate & SE & 95\% CI lower & 95\% CI upper \\ 
\hline \\[-1.8ex] 
t+0 & $4,137.73$ & $1,457.14$ & $1,644.35$ & $7,484.99$ \\ 
t+1 & $3,894.44$ & $2,743.28$ & $$-$879.56$ & $9,998.44$ \\ 
t+2 & $6,565.70$ & $3,094.20$ & $1,383.07$ & $13,432.94$ \\ 
t+3 & $7,719.78$ & $3,396.88$ & $2,034.02$ & $15,331.04$ \\ 
t+4 & $6,845.47$ & $3,823.15$ & $73.34$ & $15,286.93$ \\ 
t+5 & $4,128.65$ & $3,102.98$ & $$-$1,348.69$ & $10,956.64$ \\ 
\hline \\[-1.8ex] 
\multicolumn{5}{l}{Note: Matched sets = 142} \\ 
\end{tabular} 
\end{table}

\begin{table}[!htbp] \centering 
  \caption{Estimated effect of bureaucratic hires on size of government loans received, all ministries other than METI and MOF} 
  \label{tab: tscs_loan_other} 
\begin{tabular}{@{\extracolsep{5pt}} ccccc} 
\\[-1.8ex]\hline 
\hline \\[-1.8ex] 
Time window & Estimate & SE & 95\% CI lower & 95\% CI upper \\ 
\hline \\[-1.8ex] 
t+0 & $1,996.94$ & $2,180.07$ & $$-$2,050.40$ & $6,435.91$ \\ 
t+1 & $2,274.63$ & $2,207.79$ & $$-$1,511.89$ & $6,888.23$ \\ 
t+2 & $3,407.35$ & $2,694.68$ & $$-$1,135.04$ & $9,087.35$ \\ 
t+3 & $2,947.01$ & $3,104.04$ & $$-$2,501.55$ & $9,316.04$ \\ 
t+4 & $$-$2,303.94$ & $3,086.54$ & $$-$8,058.49$ & $4,130.70$ \\ 
t+5 & $$-$5,575.68$ & $2,802.31$ & $$-$11,181.27$ & $$-$53.91$ \\ 
\hline \\[-1.8ex] 
\multicolumn{5}{l}{Note: Matched sets = 225} \\ 
\end{tabular} 
\end{table}

\begin{table}[!htbp] \centering 
  \caption{Estimated effect of bureaucratic hires on size of government loans received, requiring 2 lag periods} 
  \label{tab: tscs_loan_lag} 
\begin{tabular}{@{\extracolsep{5pt}} ccccc} 
\\[-1.8ex]\hline 
\hline \\[-1.8ex] 
Time window & Estimate & SE & 95\% CI lower & 95\% CI upper \\ 
\hline \\[-1.8ex] 
t+0 & $$-$639.83$ & $827.68$ & $$-$2,343.11$ & $837.51$ \\ 
t+1 & $$-$533.29$ & $930.44$ & $$-$2,442.88$ & $1,160.96$ \\ 
t+2 & $1,327.78$ & $1,304.65$ & $$-$1,199.34$ & $3,933.31$ \\ 
t+3 & $1,409.77$ & $1,832.70$ & $$-$1,879.36$ & $5,325.40$ \\ 
t+4 & $535.33$ & $1,982.26$ & $$-$3,001.40$ & $4,727.60$ \\ 
t+5 & $$-$116.82$ & $1,882.88$ & $$-$3,486.48$ & $3,683.78$ \\ 
\hline \\[-1.8ex] 
\multicolumn{5}{l}{Note: Matched sets = 349} \\ 
\end{tabular} 
\end{table}

\begin{table}[!htbp] \centering 
  \caption{Estimated effect of bureaucratic hires on size of government loans received, METI and MOF re-hires only, requiring 2 lag periods} 
  \label{tab: tscs_loan_lag_meti_mof} 
\begin{tabular}{@{\extracolsep{5pt}} ccccc} 
\\[-1.8ex]\hline 
\hline \\[-1.8ex] 
Time window & Estimate & SE & 95\% CI lower & 95\% CI upper \\ 
\hline \\[-1.8ex] 
t+0 & $2,433.57$ & $1,032.06$ & $585$ & $4,644.45$ \\ 
t+1 & $331.53$ & $2,072.43$ & $$-$3,374.64$ & $4,556.62$ \\ 
t+2 & $3,489.13$ & $2,627.73$ & $$-$1,076.92$ & $9,368.39$ \\ 
t+3 & $4,206.09$ & $2,819.80$ & $$-$672.19$ & $10,334.71$ \\ 
t+4 & $5,054.28$ & $3,606.21$ & $$-$785.86$ & $13,376.25$ \\ 
t+5 & $3,093.22$ & $3,329.61$ & $$-$2,785.24$ & $10,434.72$ \\ 
\hline \\[-1.8ex] 
\multicolumn{5}{l}{Note: Matched sets = 349} \\ 
\end{tabular} 
\end{table}

\begin{table}[!htbp] \centering 
  \caption{Estimated effect of bureaucratic hires on size of private loans received} 
  \label{tab: tscs_loan_private} 
\begin{tabular}{@{\extracolsep{5pt}} ccccc} 
\\[-1.8ex]\hline 
\hline \\[-1.8ex] 
Time window & Estimate & SE & 95\% CI lower & 95\% CI upper \\ 
\hline \\[-1.8ex] 
t+0 & $$-$3,328.28$ & $4,328.68$ & $$-$12,998.45$ & $4,325.57$ \\ 
t+1 & $$-$5,470.80$ & $5,206.24$ & $$-$16,754.42$ & $3,932.10$ \\ 
t+2 & $$-$7,131.97$ & $5,585.83$ & $$-$18,519.22$ & $2,768.17$ \\ 
t+3 & $1,790.87$ & $6,612.90$ & $$-$10,056.12$ & $15,683.24$ \\ 
t+4 & $$-$12,494.41$ & $8,889.79$ & $$-$29,776.99$ & $4,813.61$ \\ 
t+5 & $$-$13,167.75$ & $8,269.36$ & $$-$28,668.40$ & $3,532.50$ \\ 
\hline \\[-1.8ex] 
\multicolumn{5}{l}{Note: Matched sets = 444} \\ 
\end{tabular} 
\end{table}

\pagebreak
\clearpage
\subsection{Loan robustness} \label{loan_robustness}

\begin{figure}[H]
\begin{centering}
\includegraphics[width = 0.8\textwidth]{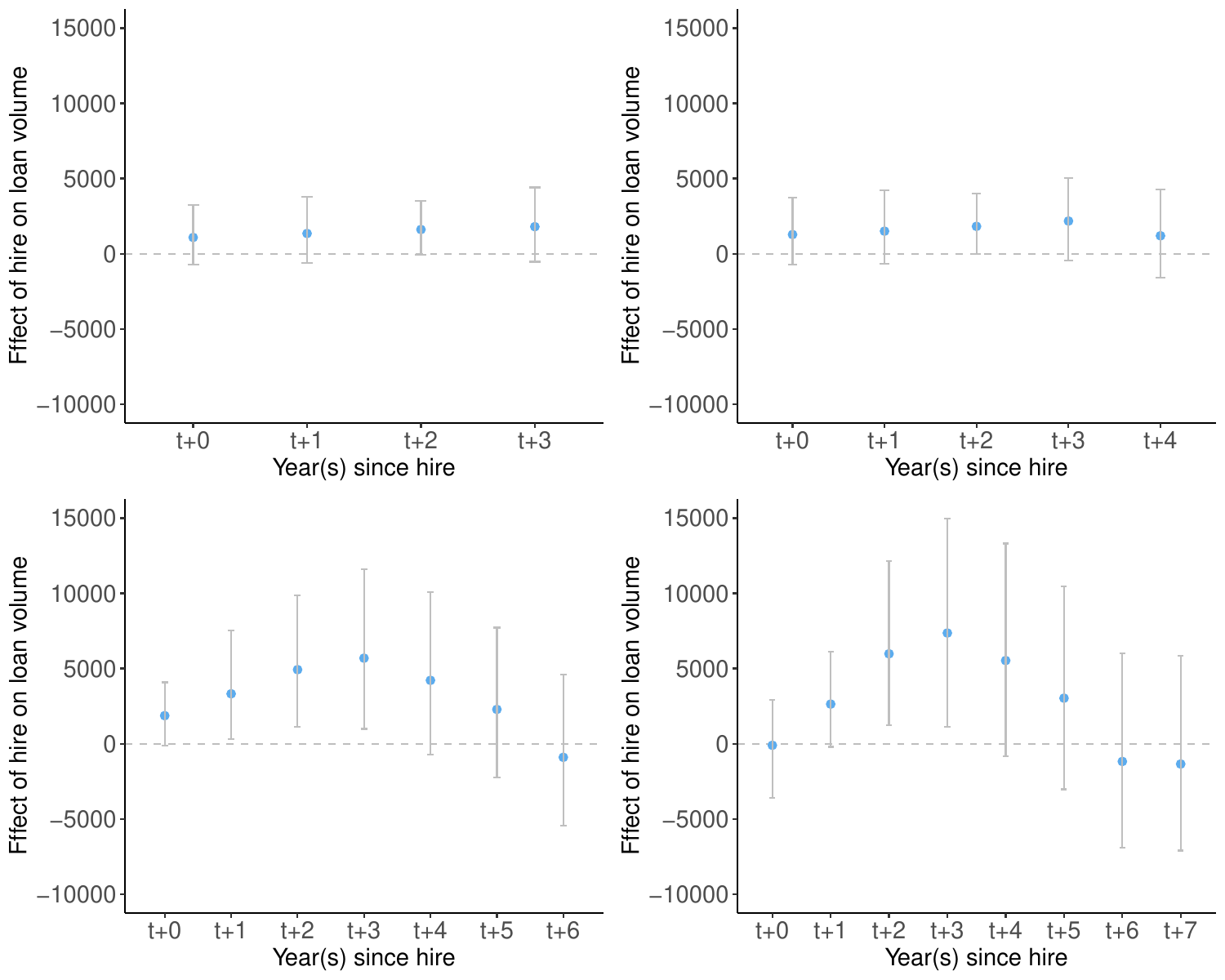}
\vspace{0.2cm}
\caption{Estimated effect of bureaucratic hires on size of government loan received, by year after hire and lead window}
\small
\vspace{-0.3cm}
\label{fig: tscs_loan_leads}
\end{centering}
\end{figure}

\begin{figure}[H]
\begin{centering}
\includegraphics[width = 0.8\textwidth]{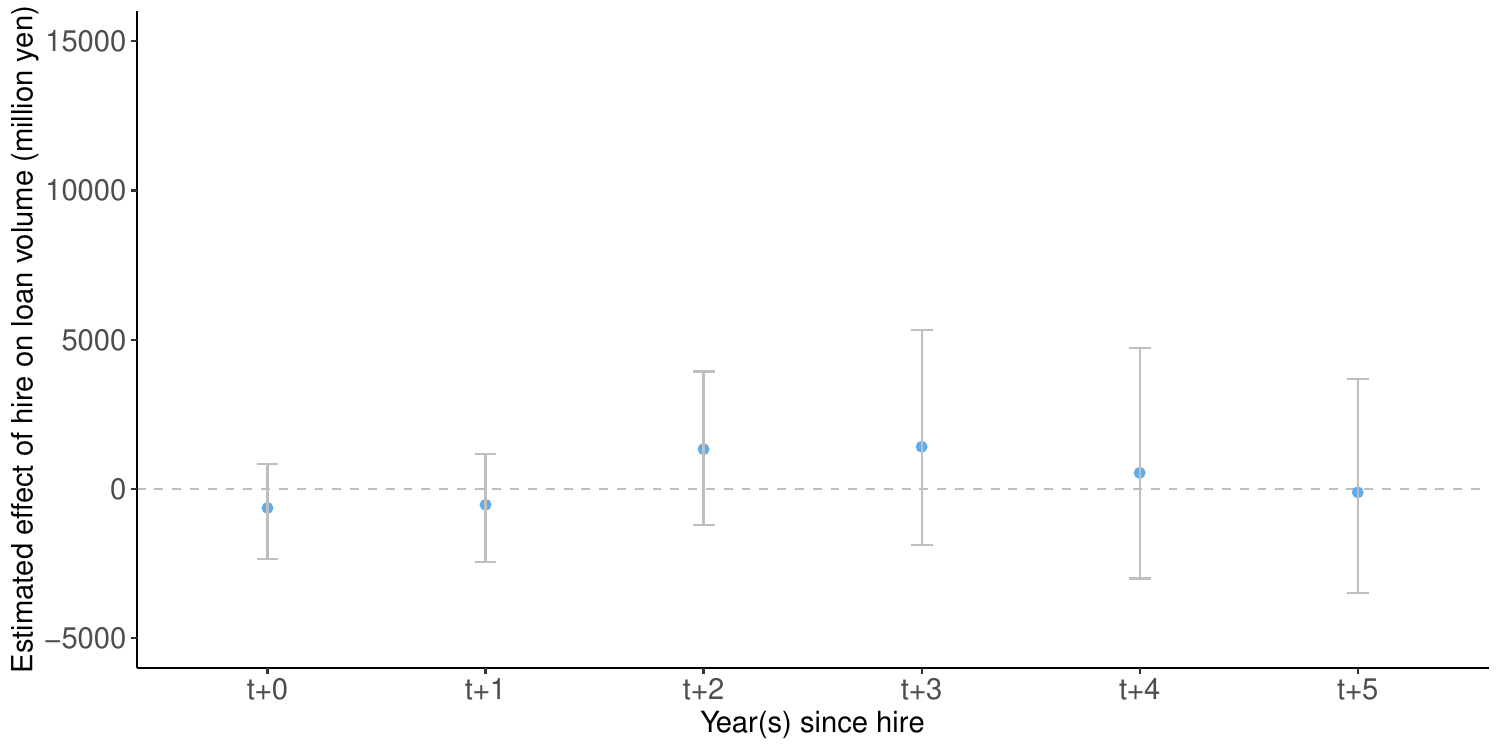}
\vspace{0.2cm}
\caption{Estimated effect of bureaucratic hires on size of government loan received, by year after hire (restricted to matches in two periods prior to treatment)}
\small
\vspace{-0.3cm}
\label{fig: tscs_loan2}
\end{centering}
{\footnotesize
\raggedright Note: Tabular results can be found in \autoref{tab: tscs_loan_lag}. \par}
\end{figure}

\begin{figure}[!htb]
\begin{centering}
\includegraphics[width = \textwidth]{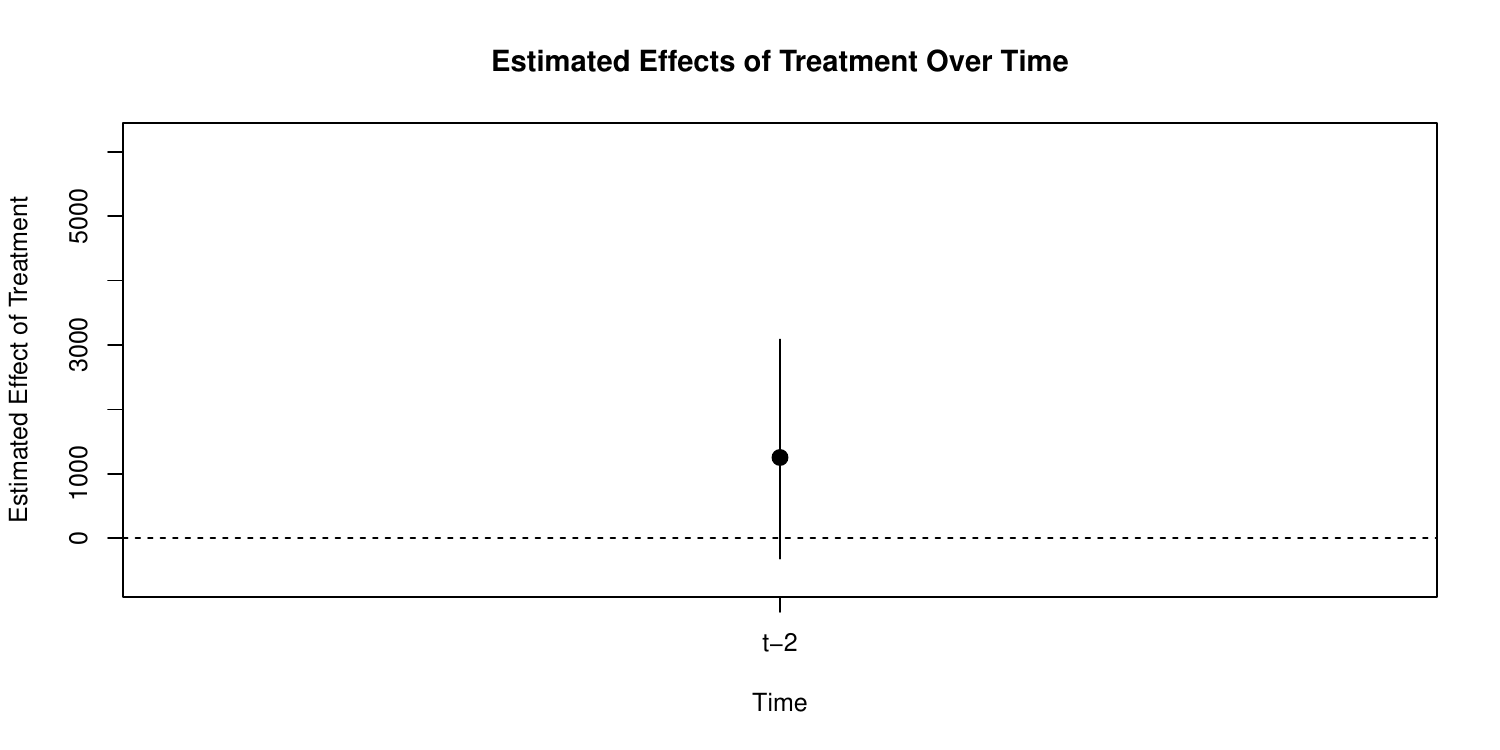}
\vspace{0.2cm}
\caption{Placebo test of effect of bureaucratic hires on size of government loan received, by year before hire (restricted to matches in two periods prior to treatment)}
\small
\vspace{-0.3cm}
\label{fig: all_ministries_2lags_placebo}
\end{centering}
\end{figure}

\begin{figure}[!htb]
\centering

\begin{subfigure}[t]{0.48\textwidth}
    \includegraphics[width=\textwidth]{tscs_loan_METI_MOF.pdf}
    \caption{1-period match}
    \label{fig:tscs_loan_METI_MOF}
\end{subfigure}
\hfill
\begin{subfigure}[t]{0.48\textwidth}
    \includegraphics[width=\textwidth]{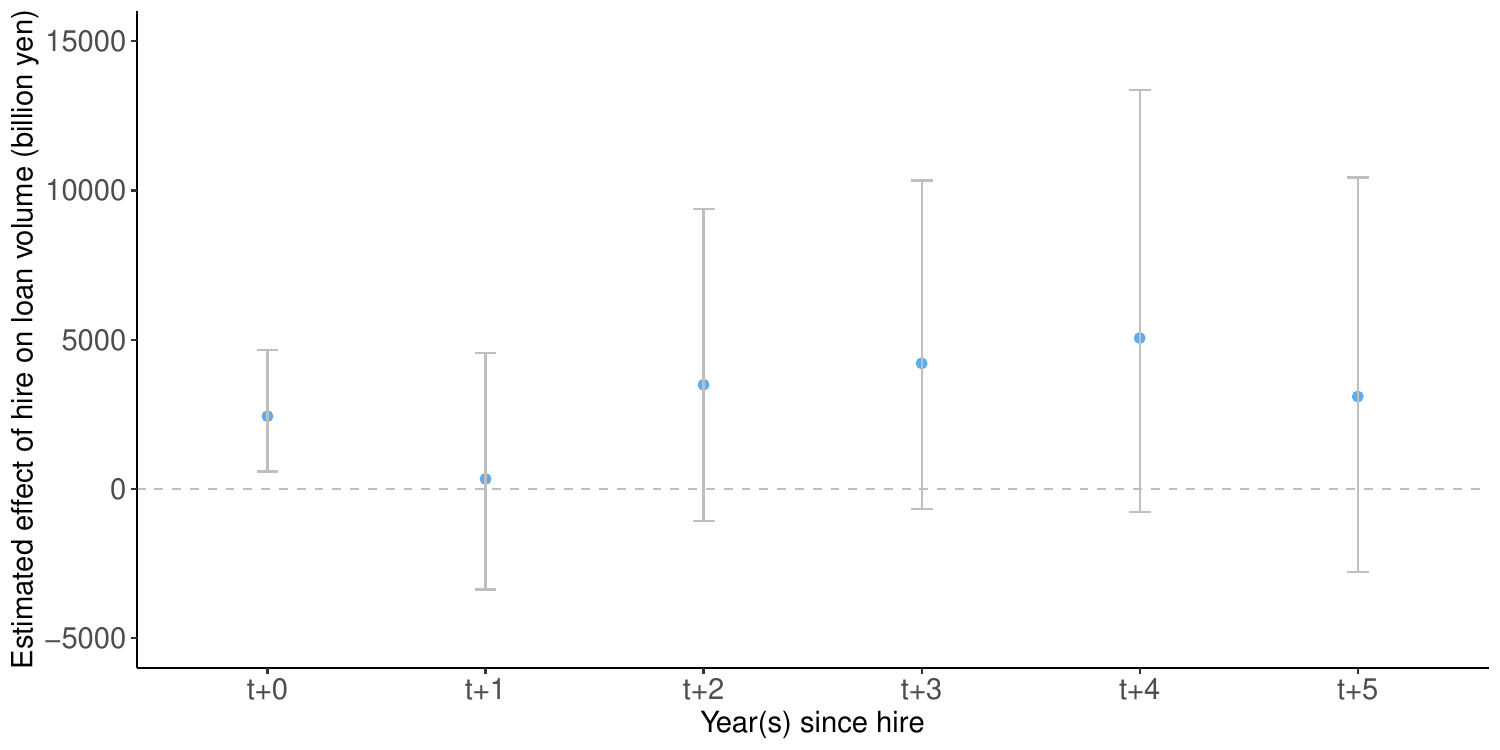}
    \caption{2-period match}
    \label{fig:tscs_loan2_meti_mof}
\end{subfigure}

\vspace{0.5cm}

\begin{subfigure}[t]{0.48\textwidth}
    \includegraphics[width=\textwidth]{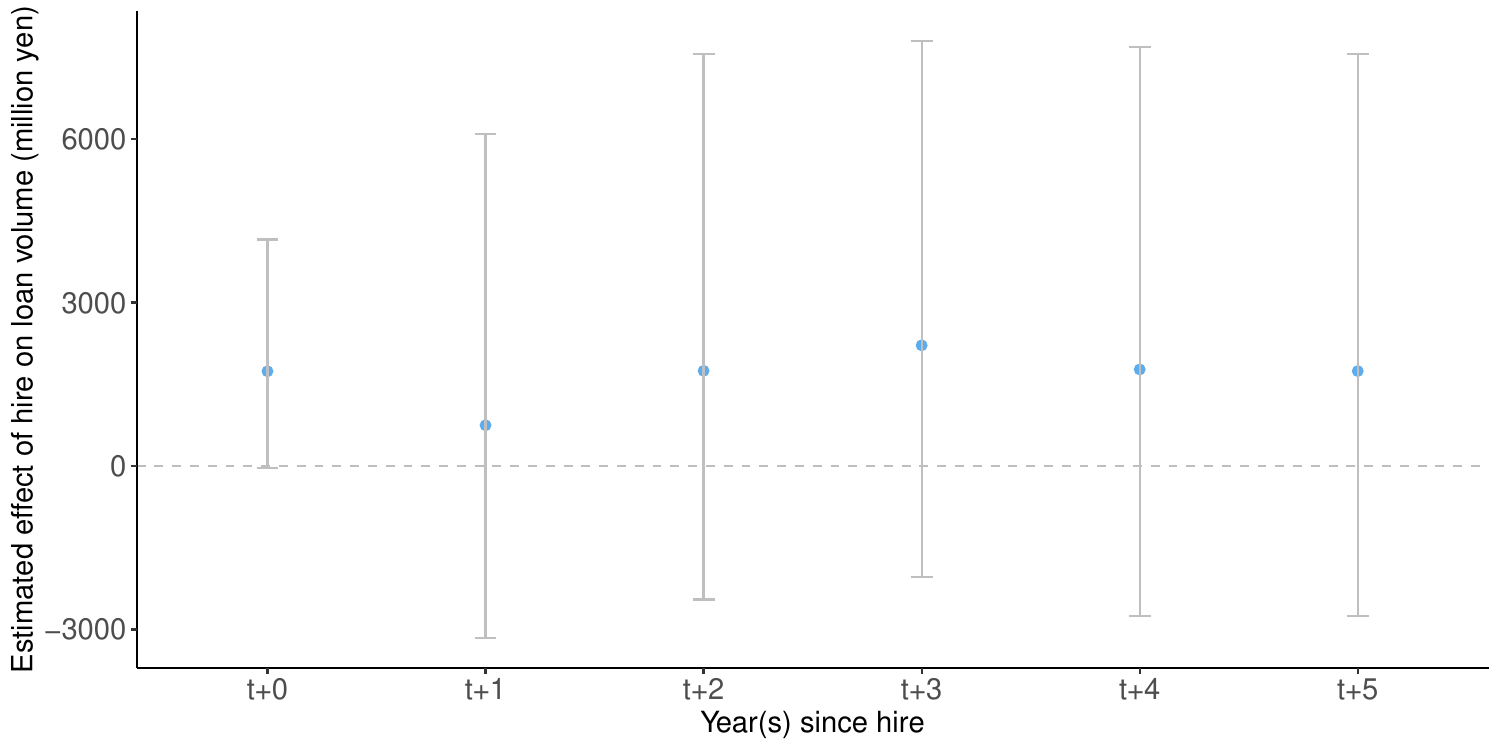}
    \caption{3-period match}
    \label{fig:meti_mof_3lags}
\end{subfigure}
\hfill
\begin{subfigure}[t]{0.48\textwidth}
    \includegraphics[width=\textwidth]{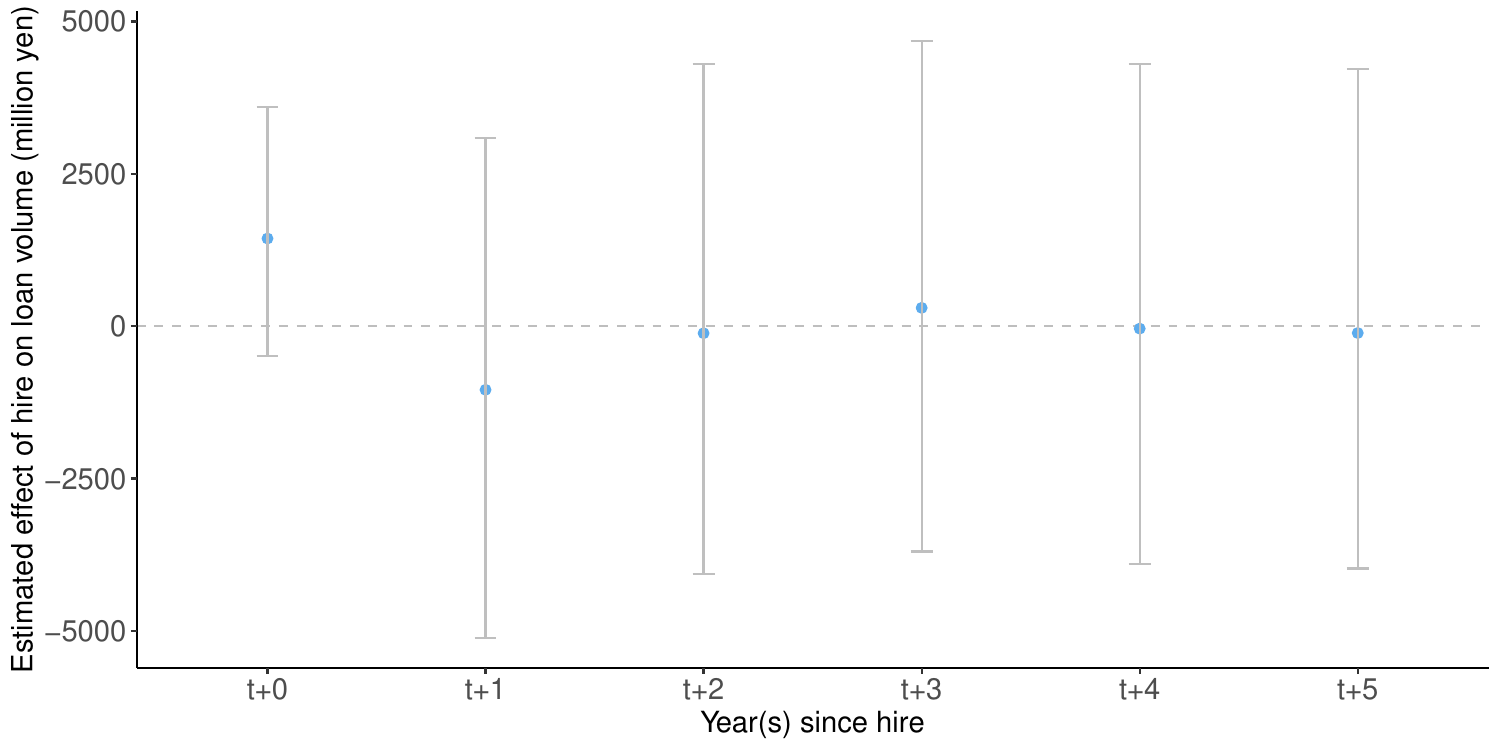}
    \caption{4-period match}
    \label{fig:meti_mof_4lags}
\end{subfigure}

\caption{Estimated effects of METI and MOF hires on size of government loan received, by year after hire, with varying pre-treatment matching windows (1 to 4 periods).}
\label{fig:meti_mof_effects_all}
\end{figure}

\begin{figure}[!h]
\centering

\begin{subfigure}[t]{0.48\textwidth}
    \includegraphics[width=\textwidth]{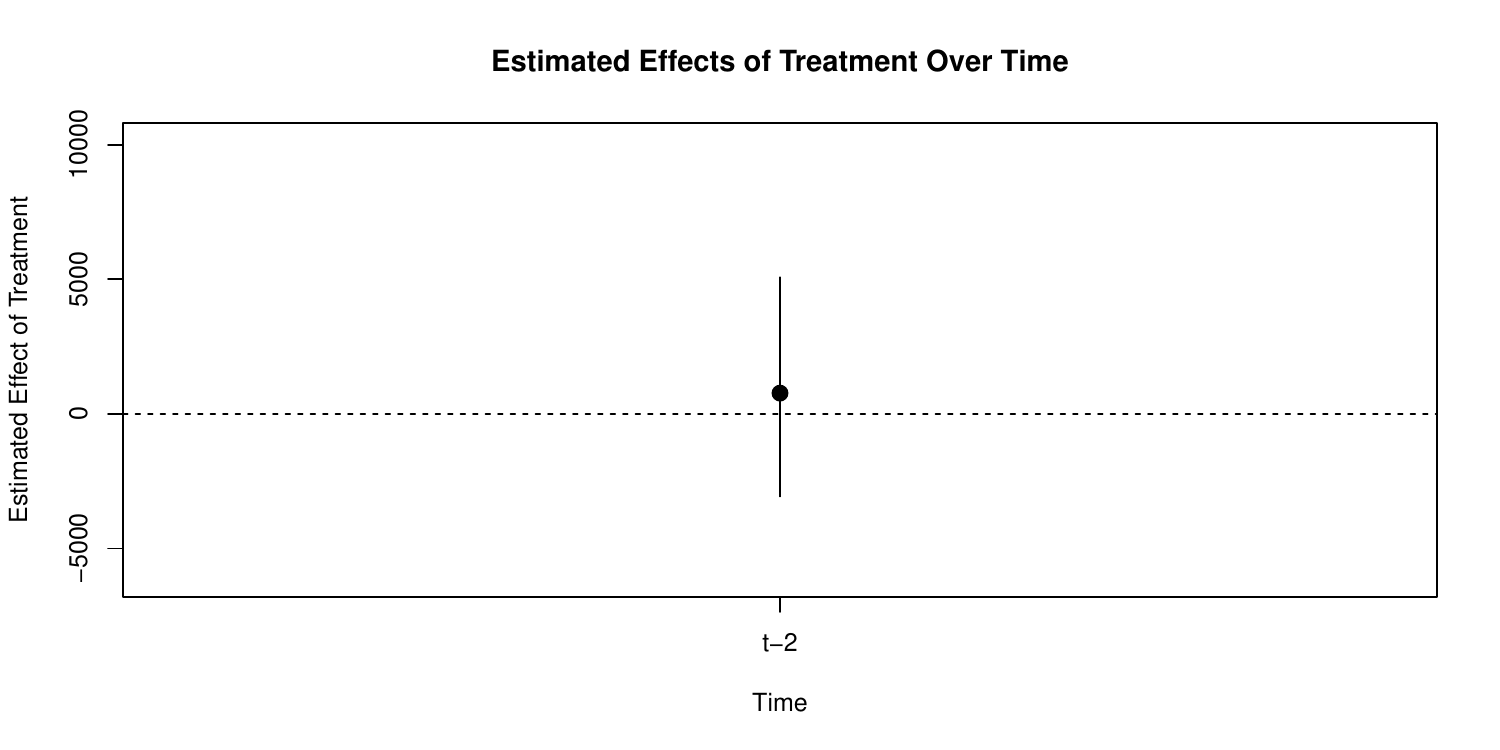}
    \caption{2-period placebo}
    \label{fig:meti_mof_2lags_placebo}
\end{subfigure}
\hfill
\begin{subfigure}[t]{0.48\textwidth}
    \includegraphics[width=\textwidth]{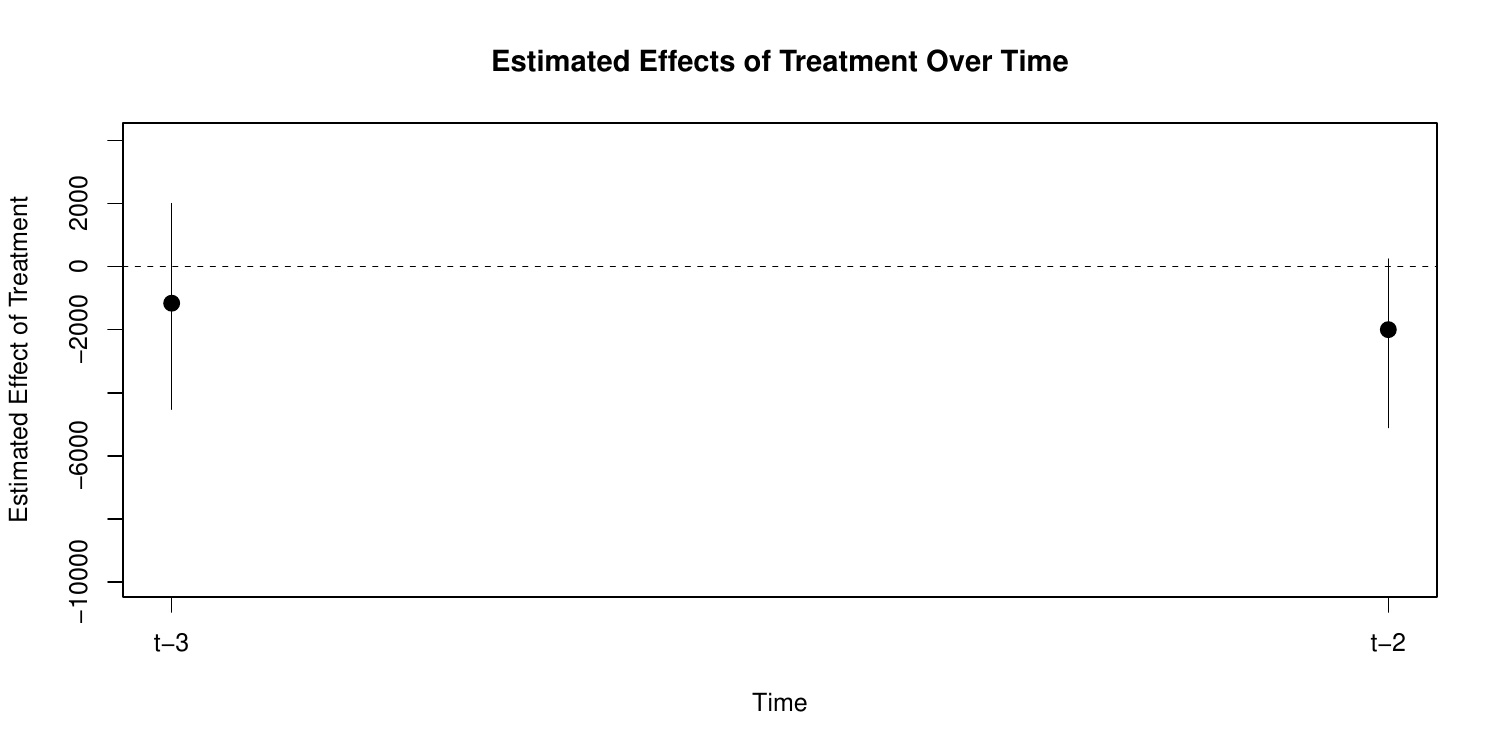}
    \caption{3-period placebo}
    \label{fig:meti_mof_3lags_placebo}
\end{subfigure}

\vspace{0.5cm}

\begin{subfigure}[t]{0.6\textwidth}
    \centering
    \includegraphics[width=\textwidth]{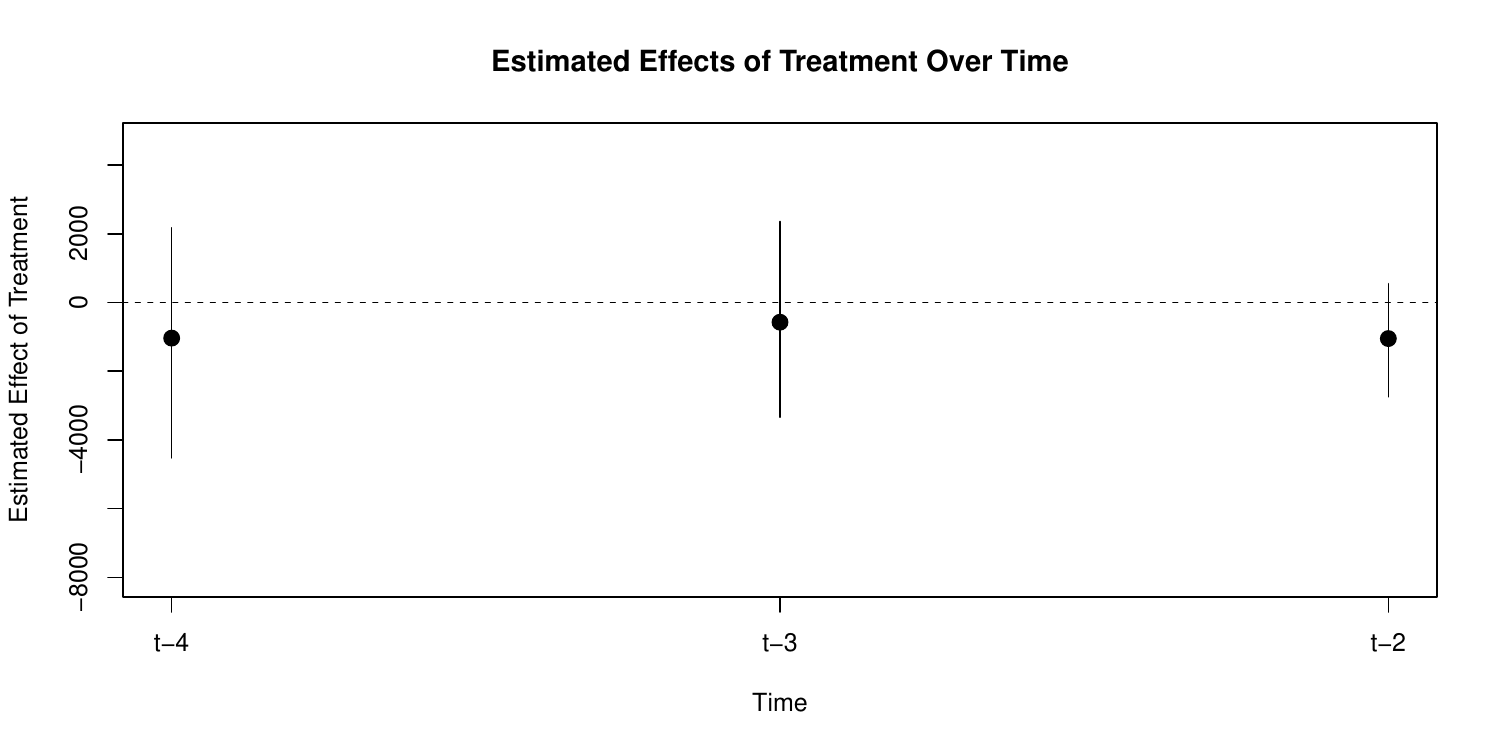}
    \caption{4-period placebo}
    \label{fig:meti_mof_4lags_placebo}
\end{subfigure}

\caption{Placebo tests of effect of METI and MOF hires on size of government loan received, by year before hire (varying number of pre-treatment periods).}
\label{fig:meti_mof_placebo_faceted}
\end{figure}

\begin{figure}[H]
\begin{centering}
\includegraphics[width = 0.8\textwidth]{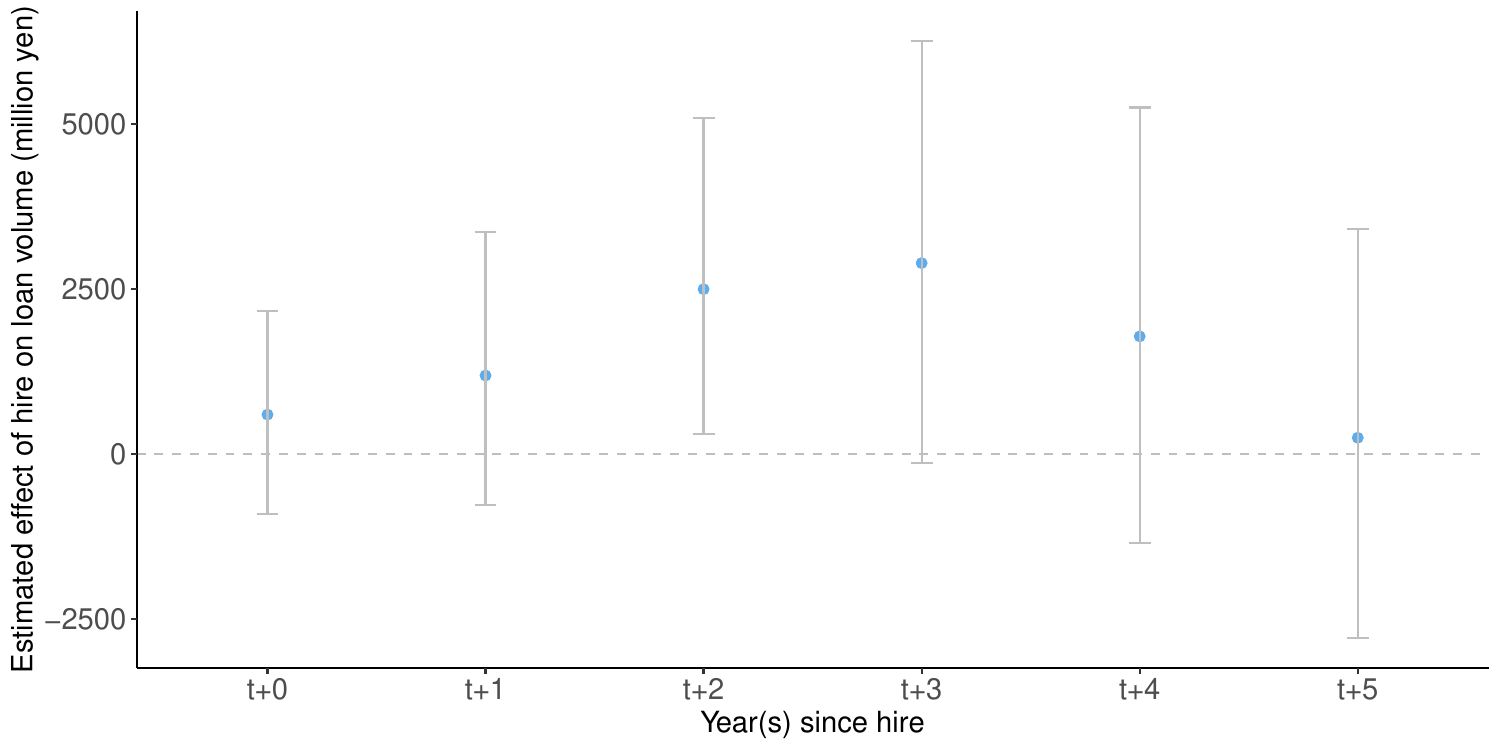}
\vspace{0.2cm}
\caption{Estimated effect of bureaucratic hires on size of government loan received, by year after hire, including additional plausibly post-treatment covariates (leverage, reserve ratio, roe, roi)}
\small
\vspace{-0.3cm}
\label{fig: tscs_loan_posttreatment}
\end{centering}
\end{figure}

\begin{figure}[!htb]
\begin{centering}
\includegraphics[width = 0.8\textwidth]{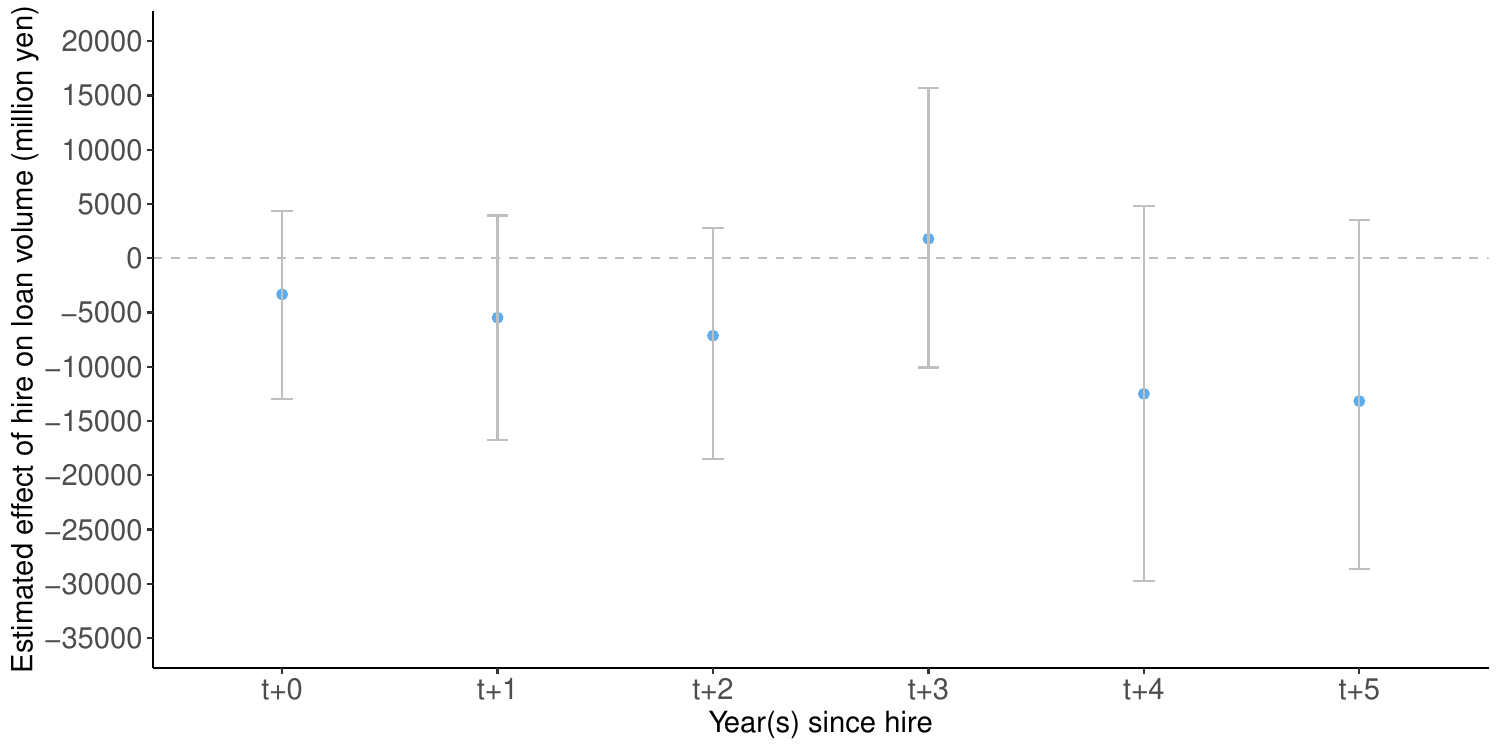}
\vspace{0.2cm}
\caption{Estimated effect of bureaucratic hires on size of private loans received, by year after hire}
\small
\vspace{-0.3cm}
\label{fig: tscs_loan_private}
\end{centering}
{\footnotesize
\raggedright Note: Tabular results can be found in \autoref{tab: tscs_loan_private}. \par}
\end{figure}

\begin{figure}[!htb]
\centering

\begin{subfigure}[t]{0.48\textwidth}
    \includegraphics[width=\textwidth]{tscs_loan.pdf}
    \caption{Mahalanobis matching}
    \label{fig:tscs_loan}
\end{subfigure}
\hfill
\begin{subfigure}[t]{0.48\textwidth}
    \includegraphics[width=\textwidth]{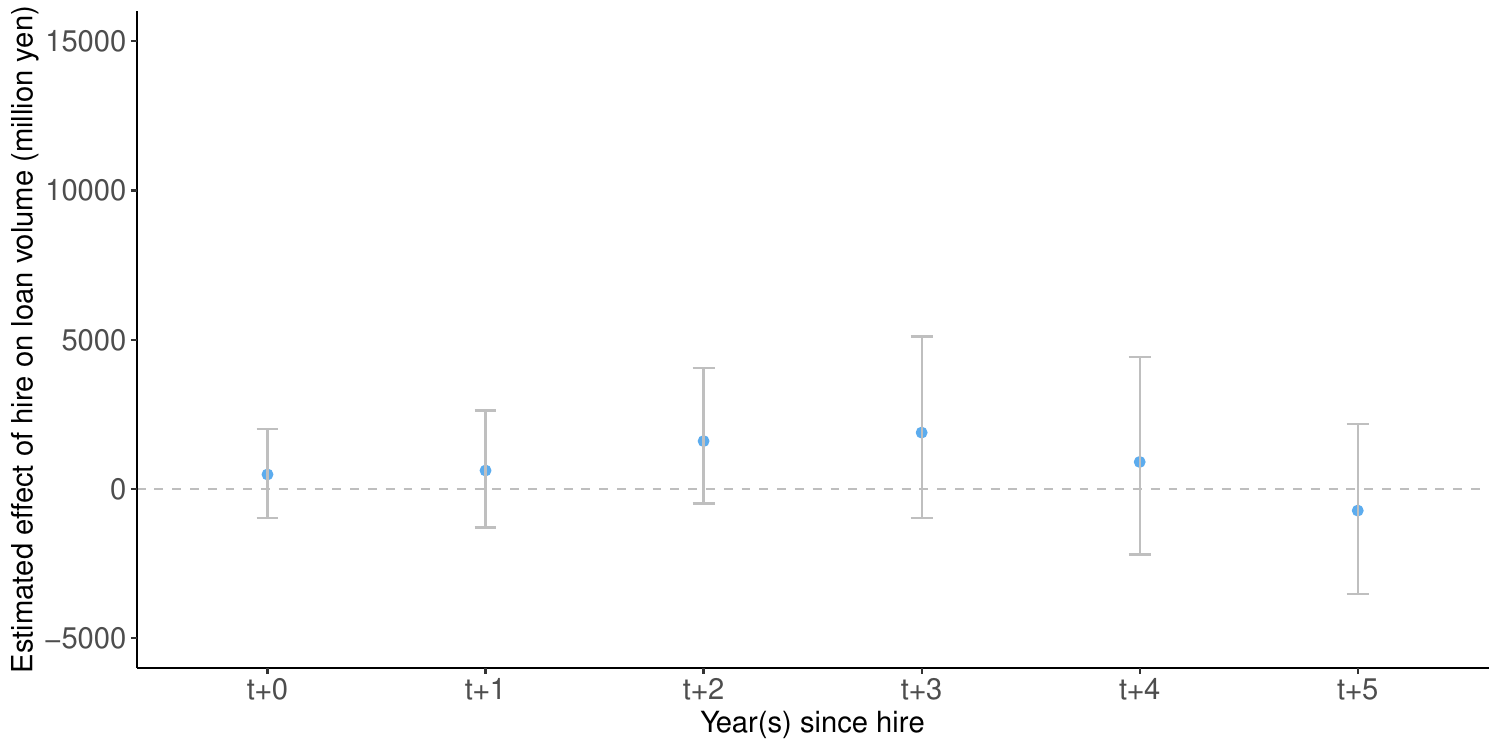}
    \caption{Propensity score matching}
    \label{fig:tscs_loan_ps}
\end{subfigure}

\vspace{0.5cm}

\begin{subfigure}[t]{0.48\textwidth}
    \includegraphics[width=\textwidth]{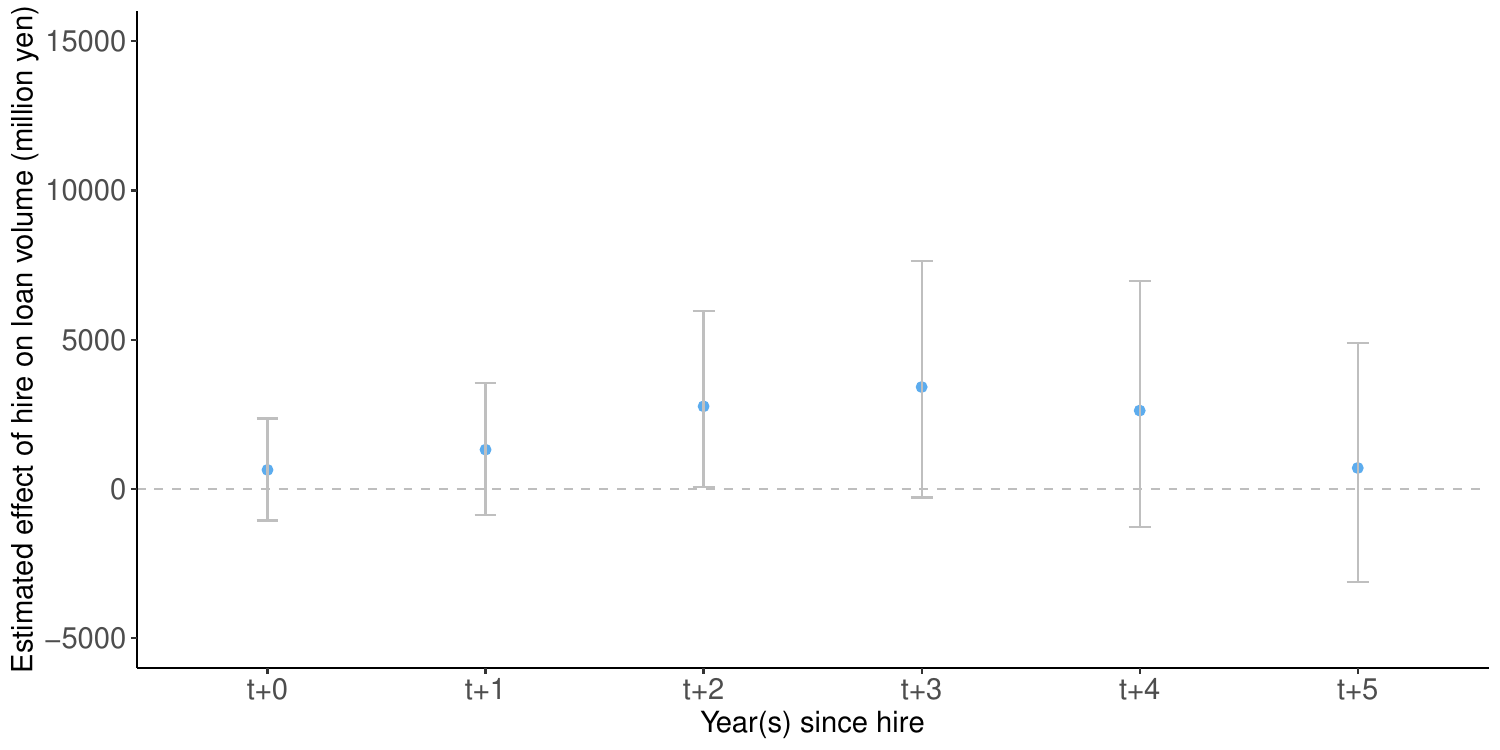}
    \caption{Covariate balanced propensity scores (CBPS)}
    \label{fig:tscs_loan_cbps}
\end{subfigure}
\hfill
\begin{subfigure}[t]{0.48\textwidth}
    \includegraphics[width=\textwidth]{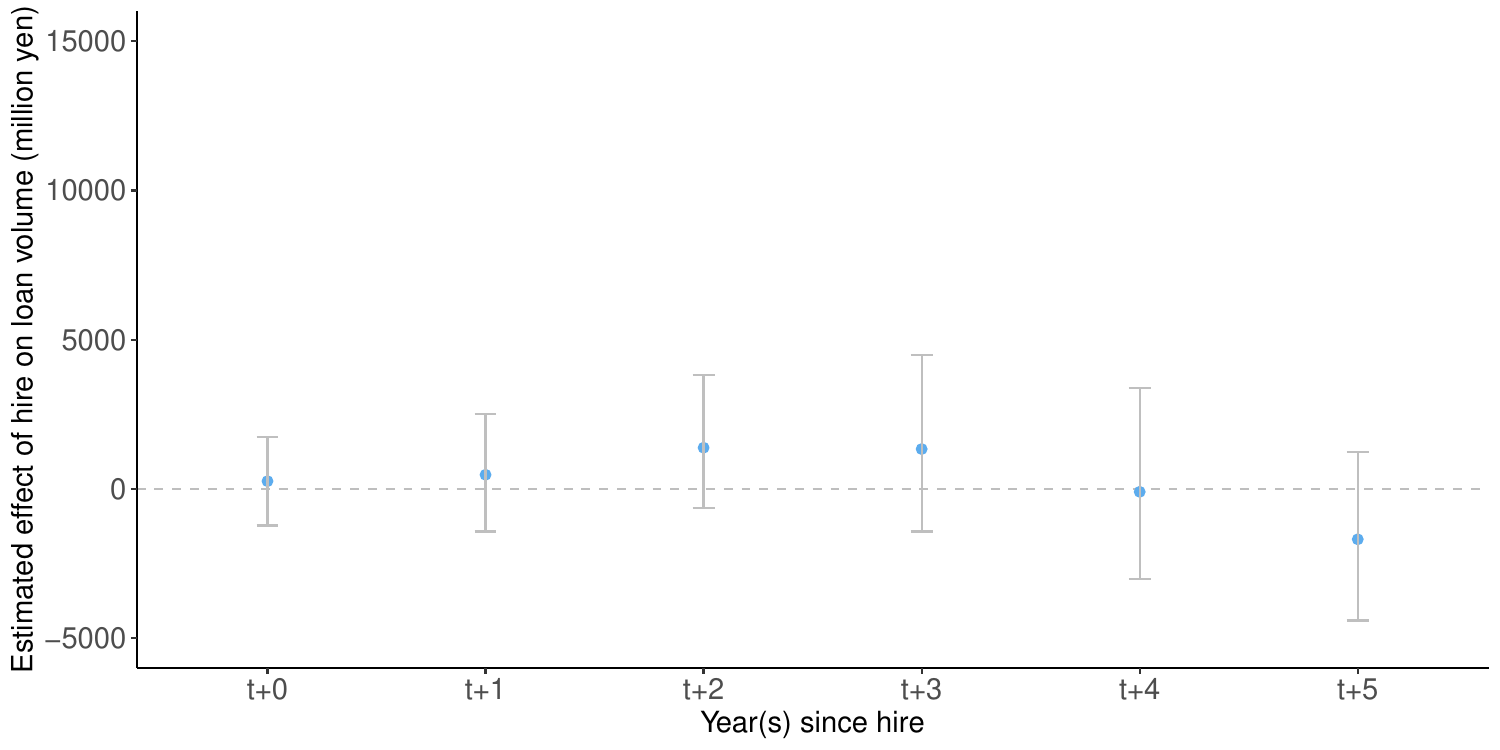}
    \caption{CBPS with marginal structural model (MSM)}
    \label{fig:tscs_loan_cbps_msm}
\end{subfigure}

\caption{Estimated effects of bureaucratic hires on size of government loans received, by year after hire, using alternative matching methods.}
\label{fig:matching_methods_all}
\end{figure}

\begin{figure}[!htb]
\centering

\begin{subfigure}[t]{0.48\textwidth}
    \includegraphics[width=\textwidth]{tscs_loan_METI_MOF.pdf}
    \caption{Mahalanobis matching}
    \label{fig:tscs_loan_METI_MOF}
\end{subfigure}
\hfill
\begin{subfigure}[t]{0.48\textwidth}
    \includegraphics[width=\textwidth]{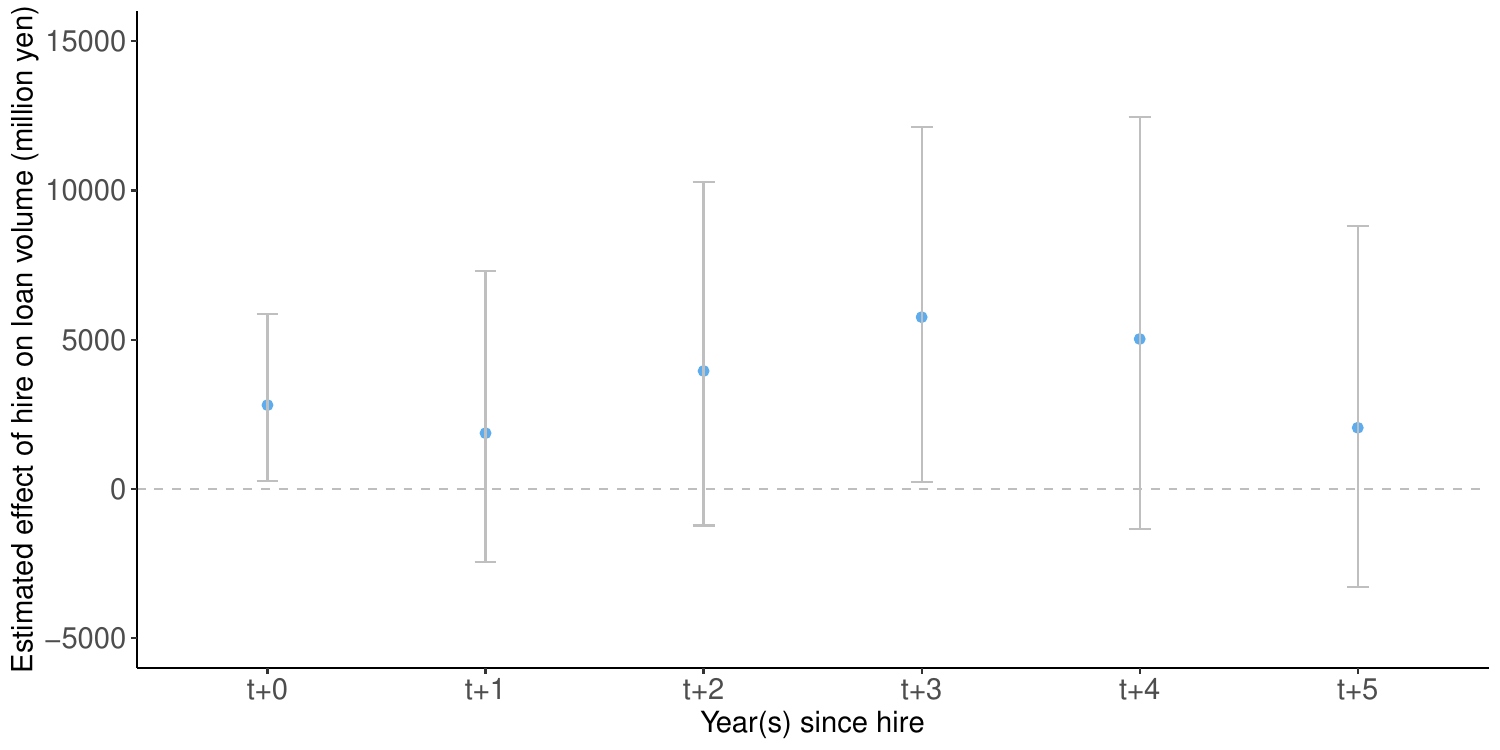}
    \caption{Propensity score matching}
    \label{fig:tscs_loan_ps_meti_mof}
\end{subfigure}

\vspace{0.5cm}

\begin{subfigure}[t]{0.48\textwidth}
    \includegraphics[width=\textwidth]{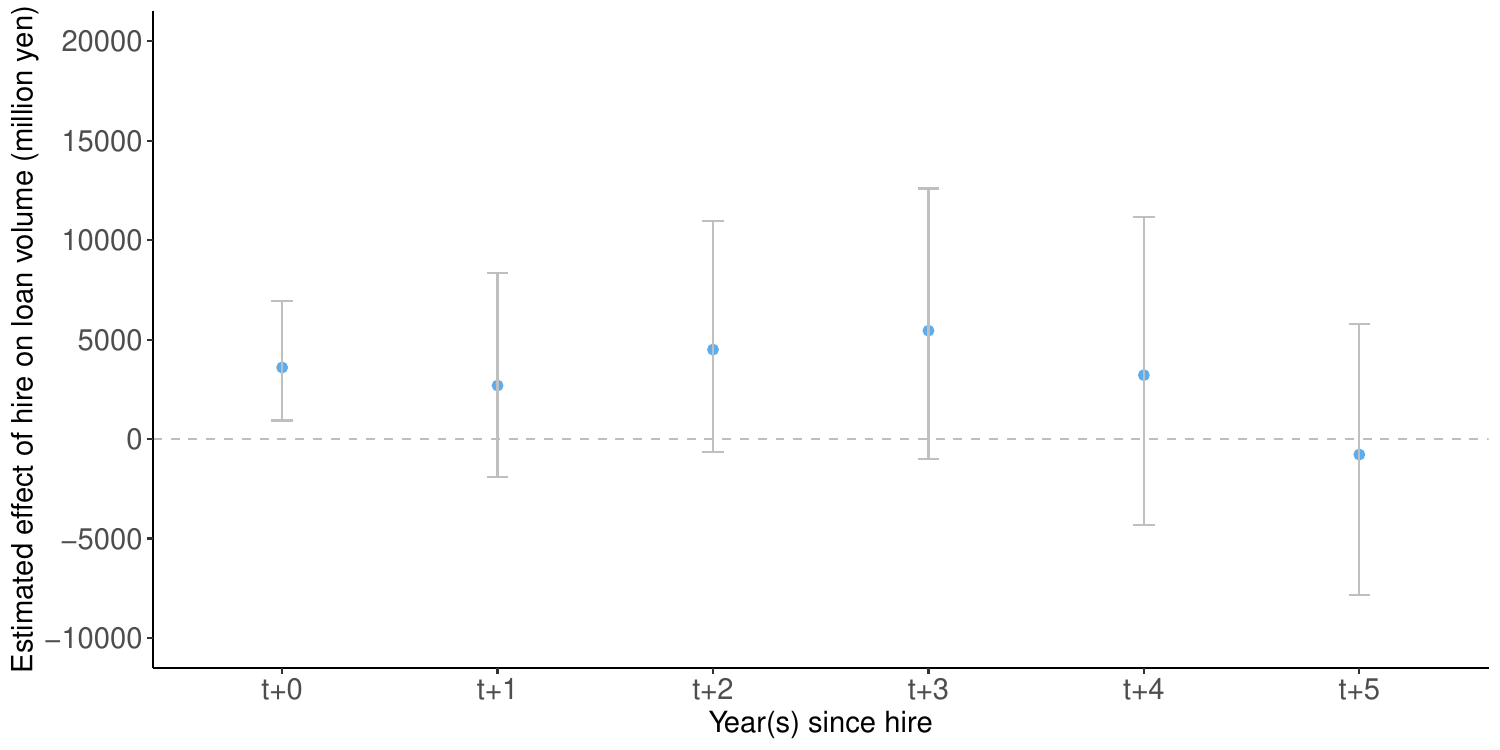}
    \caption{CBPS}
    \label{fig:tscs_loan_cbps_meti_mof}
\end{subfigure}
\hfill
\begin{subfigure}[t]{0.48\textwidth}
    \includegraphics[width=\textwidth]{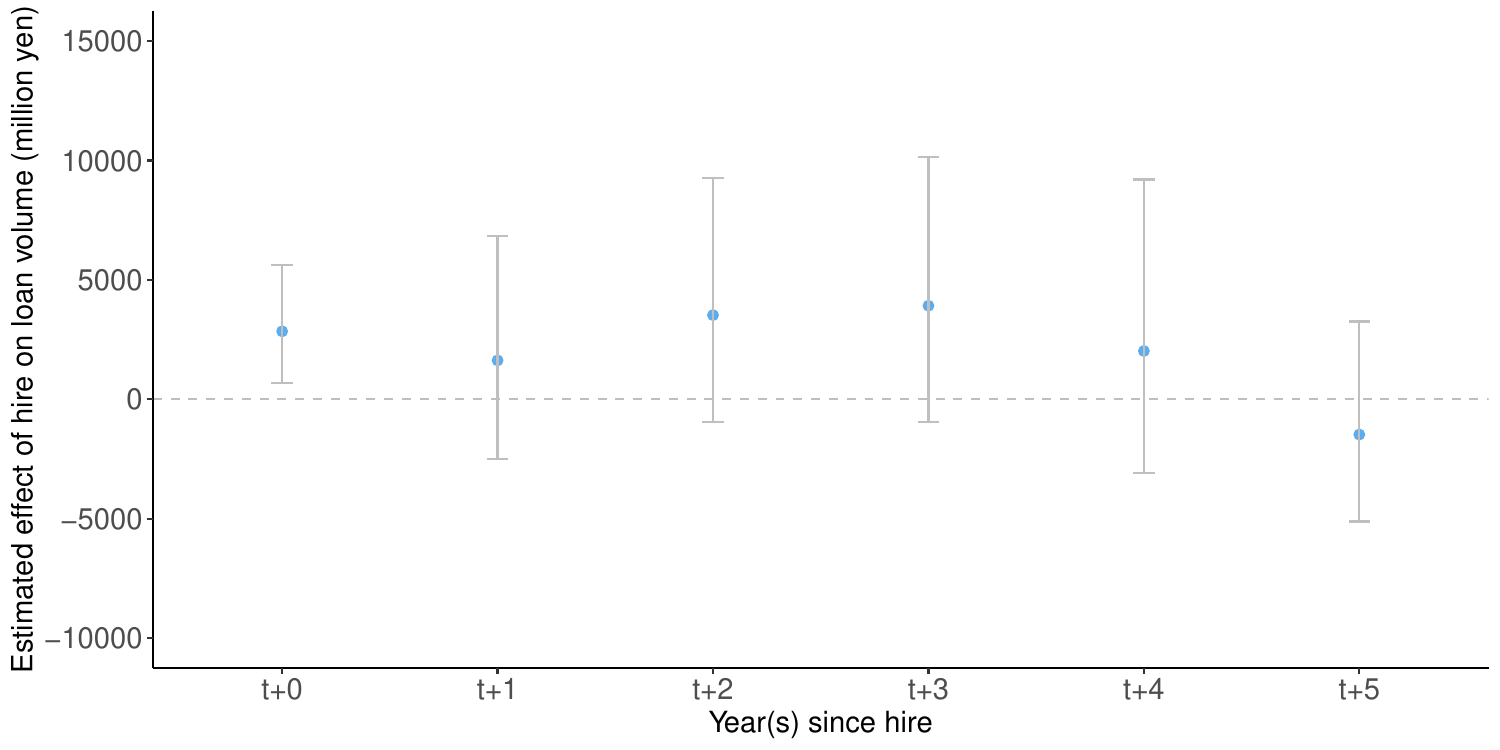}
    \caption{CBPS with MSM}
    \label{fig:tscs_loan_cbps_msm_meti_mof}
\end{subfigure}

\caption{Estimated effects of METI and MOF hires on size of government loan received, by year after hire, using alternative matching methods.}
\label{fig:meti_mof_matching_methods}
\end{figure}

\begin{figure}[!htb]
\centering

\begin{subfigure}[t]{0.48\textwidth}
    \includegraphics[width=\textwidth]{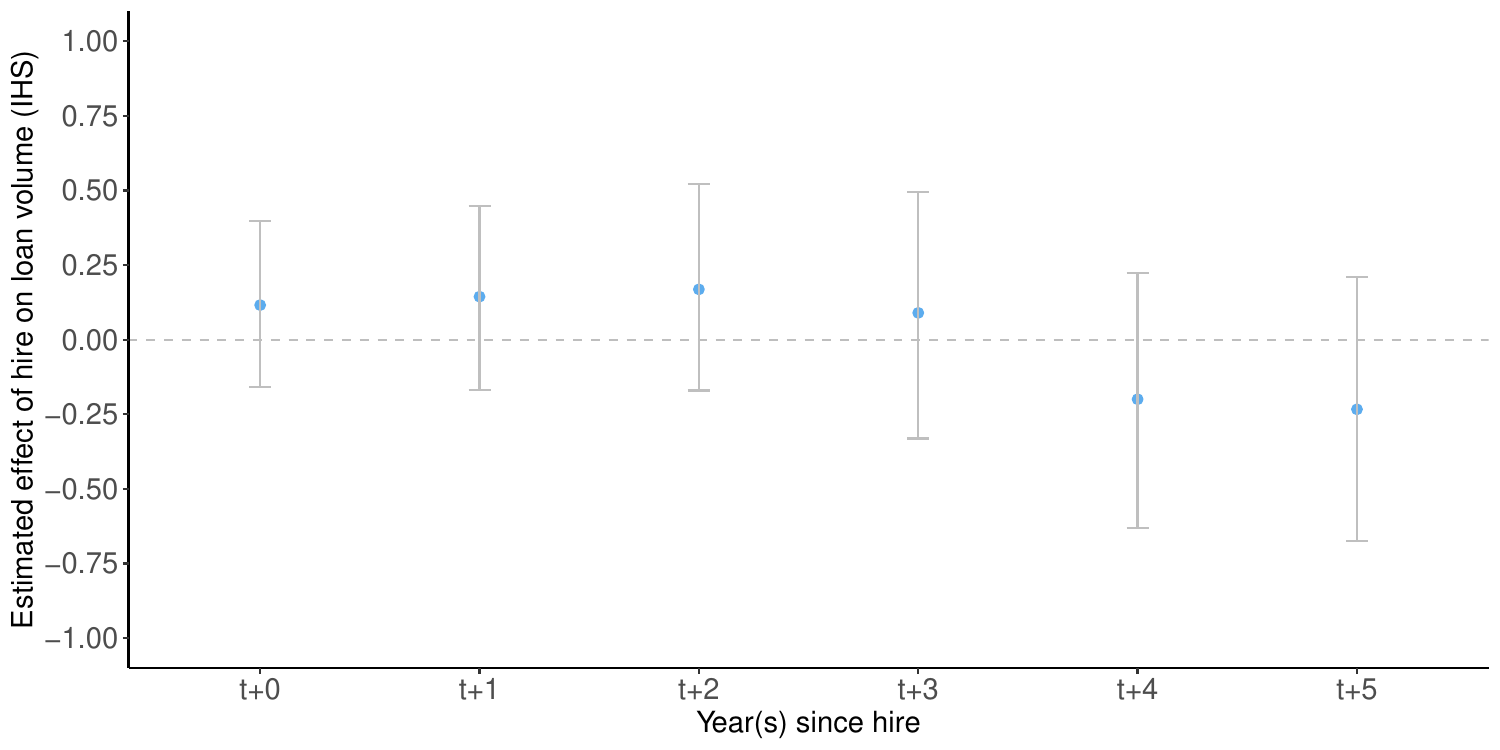}
    \caption{All hires, inverse hyperbolic sine}
    \label{fig:tscs_loan_sine}
\end{subfigure}
\hfill
\begin{subfigure}[t]{0.48\textwidth}
    \includegraphics[width=\textwidth]{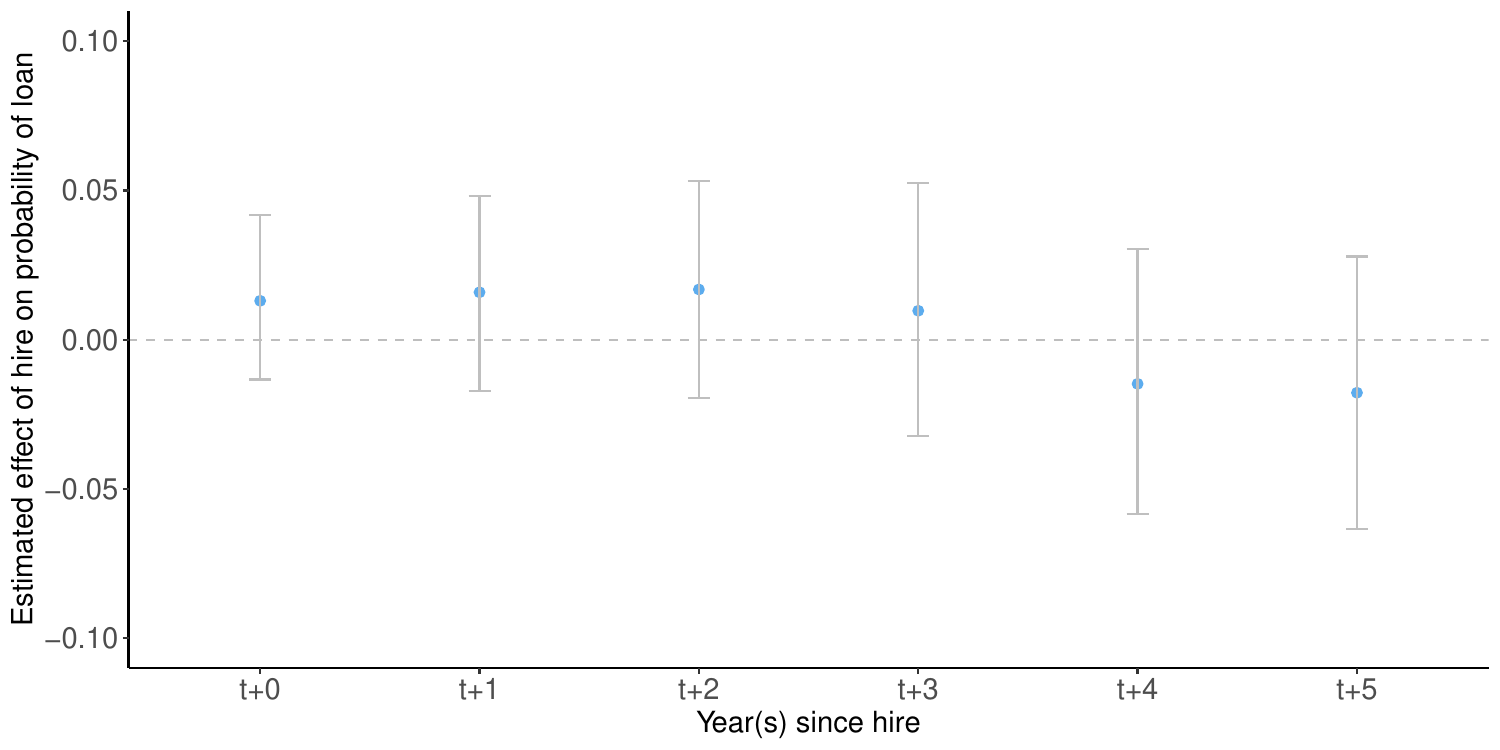}
    \caption{All hires, binary outcome}
    \label{fig:tscs_loan_binary}
\end{subfigure}

\vspace{0.5cm}

\begin{subfigure}[t]{0.48\textwidth}
    \includegraphics[width=\textwidth]{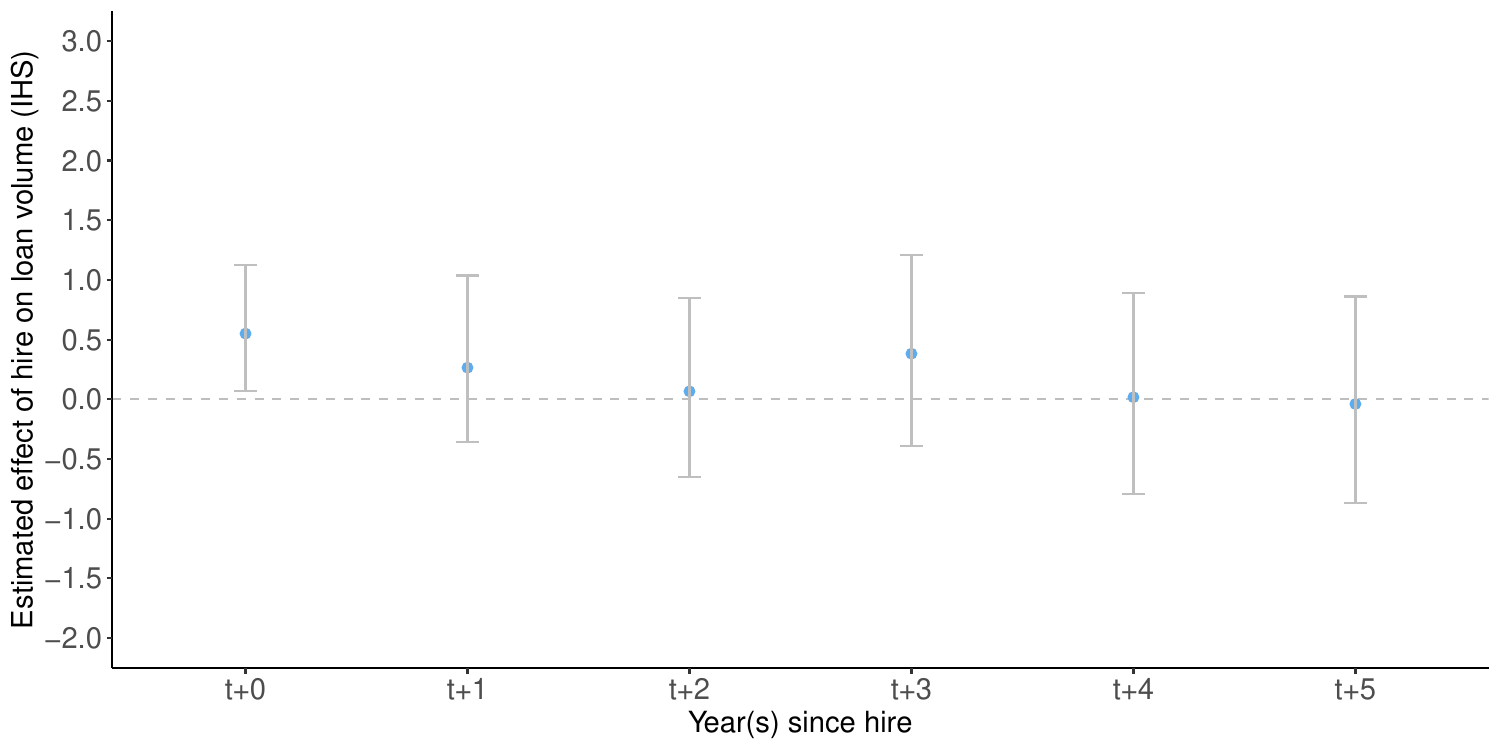}
    \caption{METI/MOF hires, inverse hyperbolic sine}
    \label{fig:tscs_loan_meti_mof_sine}
\end{subfigure}
\hfill
\begin{subfigure}[t]{0.48\textwidth}
    \includegraphics[width=\textwidth]{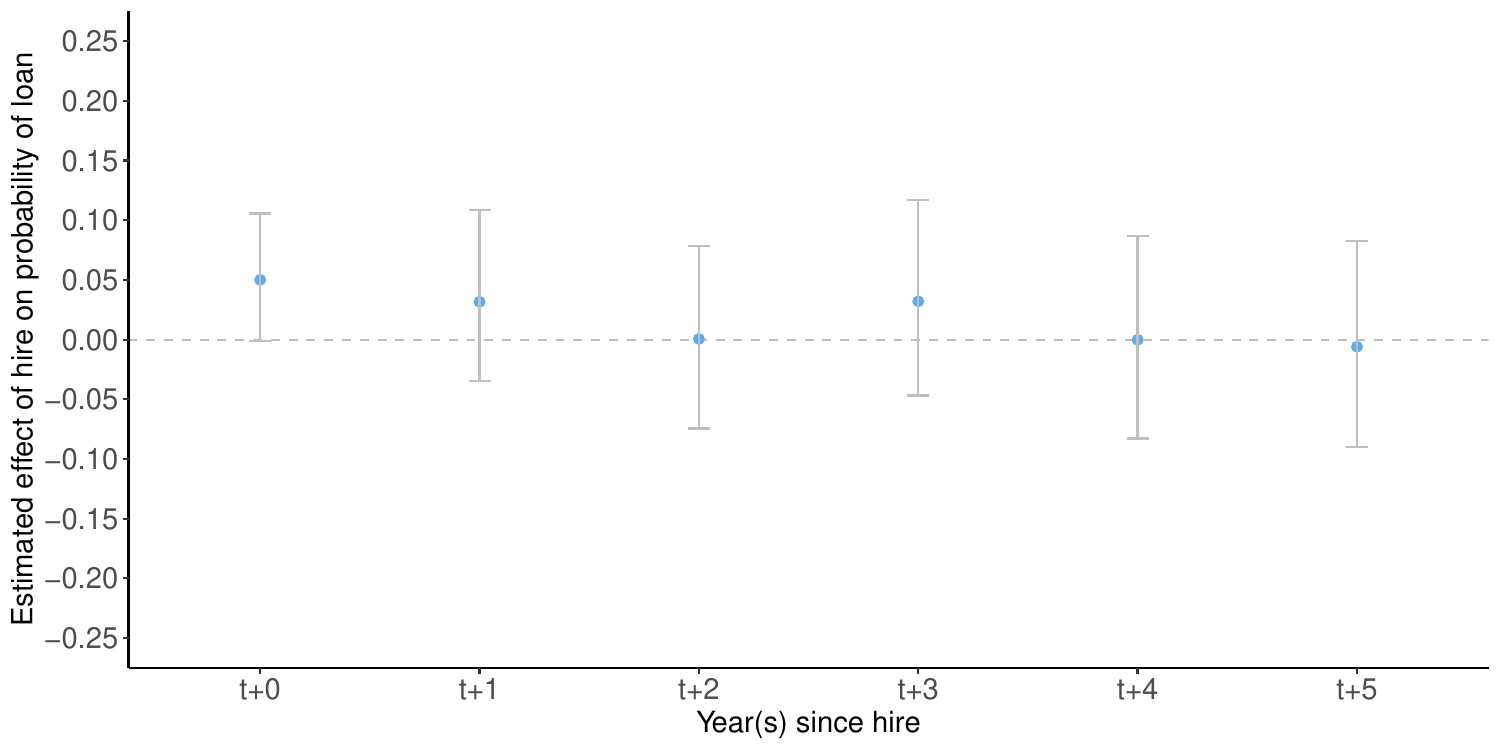}
    \caption{METI/MOF hires, binary outcome}
    \label{fig:tscs_loan_meti_mof_binary}
\end{subfigure}

\caption{Estimated effects of bureaucratic hires on size of government loan received, by year after hire, using alternative outcome variable transformations.}
\label{fig:outcome_transformations}
\end{figure}

\pagebreak
\clearpage
\subsection{Matching balance}  \label{loan_diagnostics}

\begin{figure}[!htb]
\begin{centering}
\includegraphics[width = 0.8\textwidth]{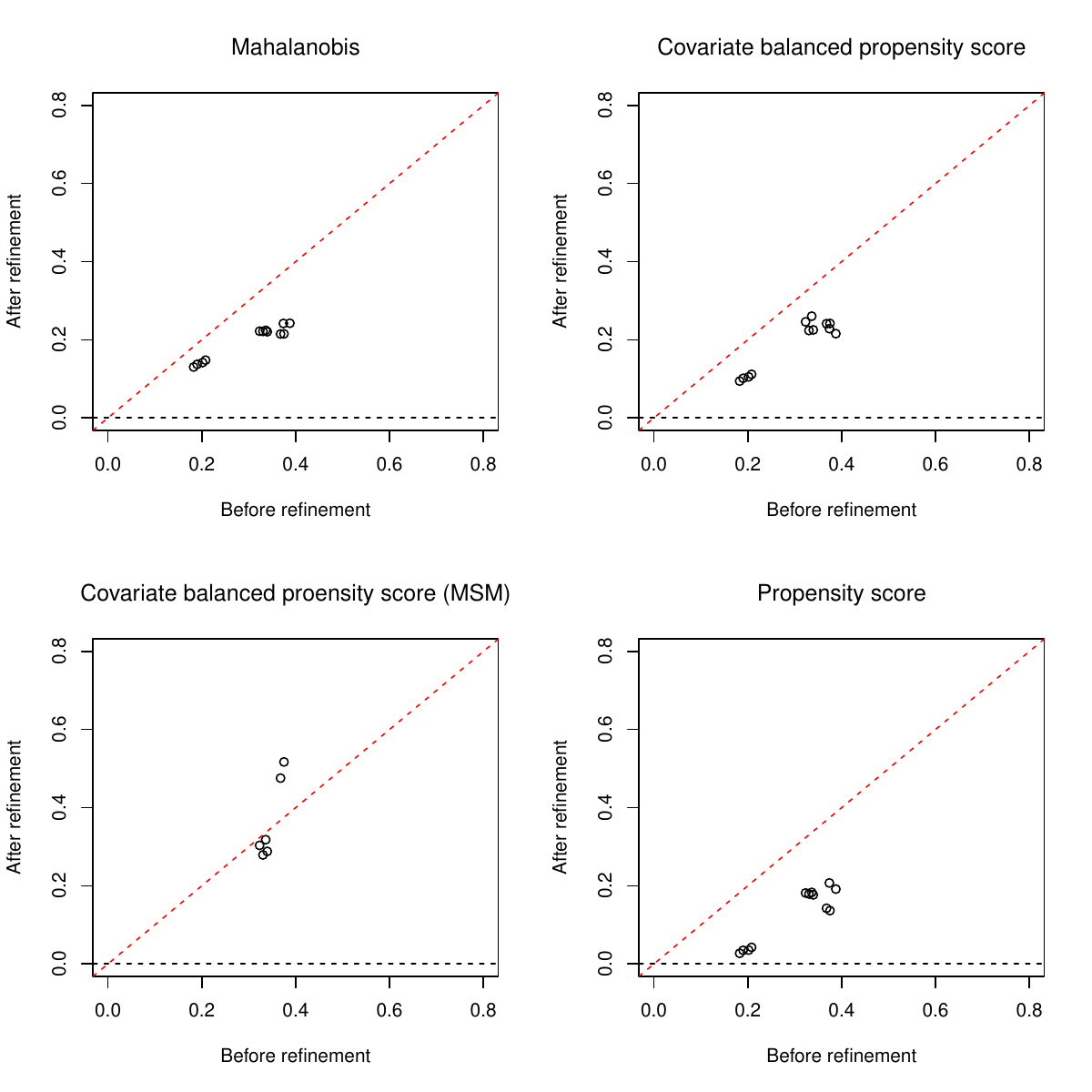}
\vspace{0.2cm}
\caption{Balance of firm financials before and after matching}
\small
\vspace{-0.3cm}
\label{fig: tscs_balance}
\end{centering}
\end{figure}

\begin{figure}[!htb]
\begin{centering}
\includegraphics[width = 0.8\textwidth]{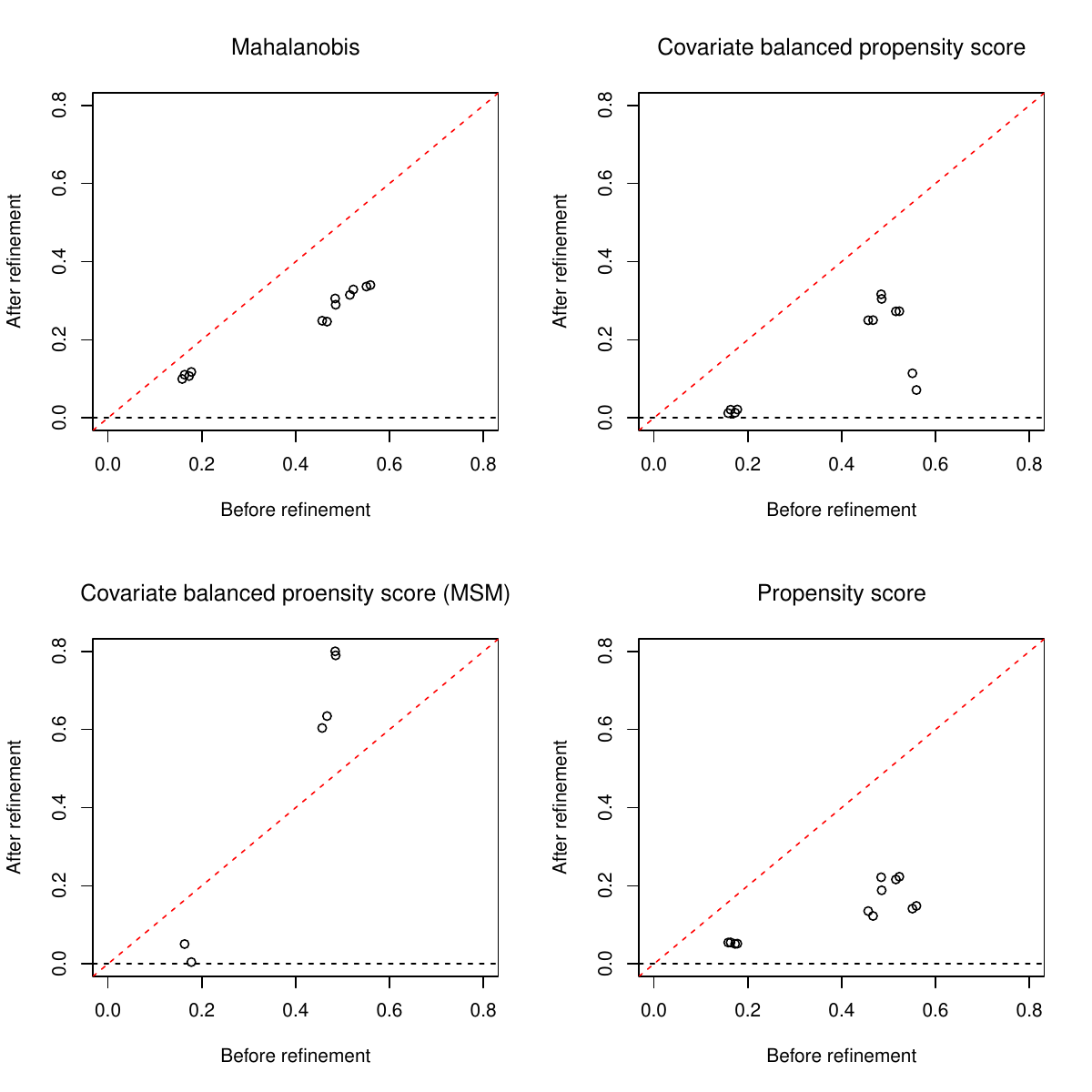}
\vspace{0.2cm}
\caption{Balance of firm financials before and after matching (METI and MOF hires only)}
\small
\vspace{-0.3cm}
\label{fig: tscs_balance_meti_mof}
\end{centering}
\end{figure}

\begin{figure}[!htb]
\begin{centering}
\includegraphics[width = \textwidth]{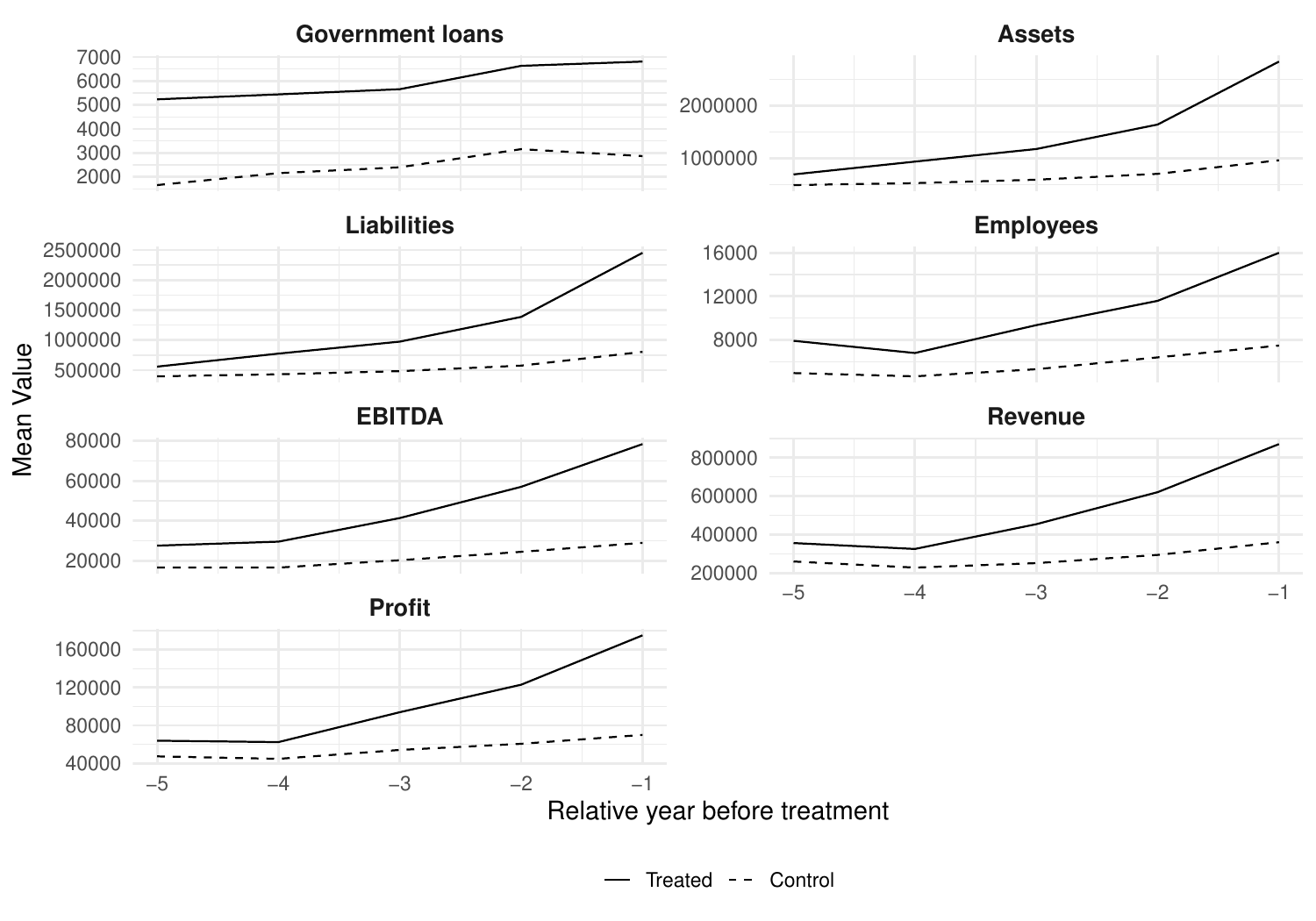}
\vspace{0.2cm}
\caption{Mean value of government loans and covariates in pre-treatment periods, post 1-period lag matching}
\small
\vspace{-0.3cm}
\label{fig: pre_trends}
\end{centering}
\end{figure}

\begin{figure}[!htb]
\begin{centering}
\includegraphics[width = \textwidth]{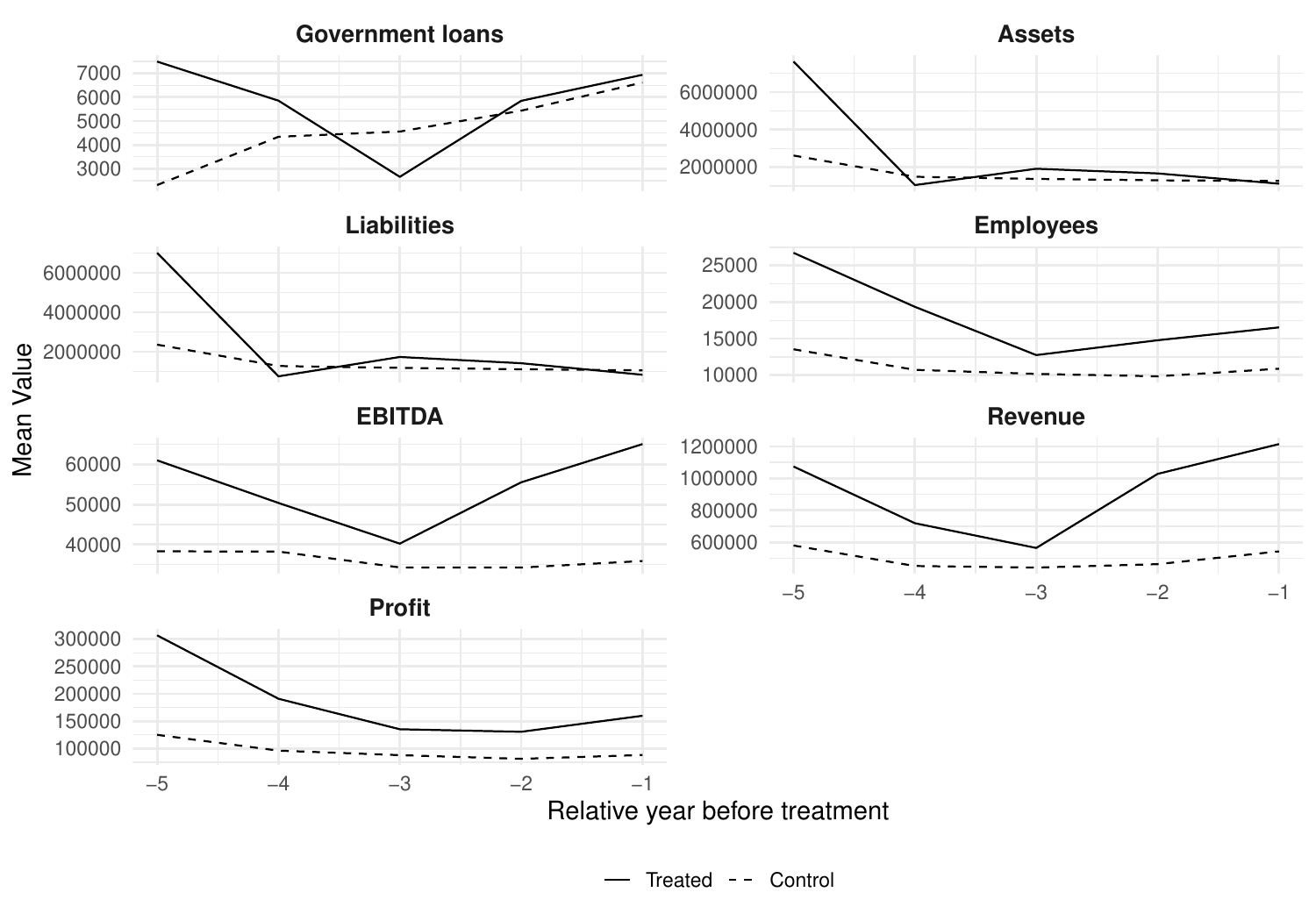}
\vspace{0.2cm}
\caption{Mean value of government loans and covariates in pre-treatment periods, post 1-period lag matching (METI and MOF hires only)}
\small
\vspace{-0.3cm}
\label{fig: pre_trends_meti_mof}
\end{centering}
\end{figure}


\newpage
\clearpage
\section{Event studies} \label{sec: event_study_appendix}

\subsection{Tabular results}

\begin{table}[!htbp] \centering 
  \caption{Cumulative abnormal returns from assistant vice-minister and vice-minister appointments} 
  \label{tab: event_study_all} 
\footnotesize 
\begin{tabular}{@{\extracolsep{5pt}} cccc} 
\\[-1.8ex]\hline 
\hline \\[-1.8ex] 
Event Day & Estimate & 95\% CI lower & 95\% CI upper \\ 
\hline \\[-1.8ex] 
-7 & 0 & 0 & 0 \\ 
-6 & -0.1 & -0.42 & 0.22 \\ 
-5 & 0.17 & -0.29 & 0.65 \\ 
-4 & 0.25 & -0.31 & 0.81 \\ 
-3 & 0.23 & -0.55 & 1.01 \\ 
-2 & 0.02 & -1.01 & 0.96 \\ 
-1 & 0.69 & -0.39 & 1.75 \\ 
0 & 1.24 & -0.07 & 2.57 \\ 
1 & 1.31 & -0.18 & 2.83 \\ 
2 & 1.37 & -0.19 & 3.09 \\ 
3 & 1.4 & -0.23 & 2.99 \\ 
4 & 1.33 & -0.33 & 3.08 \\ 
5 & 0.75 & -0.93 & 2.53 \\ 
6 & 1.01 & -0.77 & 2.87 \\ 
7 & 0.83 & -1.08 & 2.74 \\ 
8 & 0.97 & -1.03 & 2.86 \\ 
\hline \\[-1.8ex] 
\multicolumn{4}{l}{Note: Total events = 47} \\ 
\end{tabular} 
\end{table}

\begin{table}[!htbp] \centering 
  \caption{Cumulative abnormal returns from assistant vice-minister and vice-minister appointments, outside director appointments} 
  \label{tab: event_study_director} 
\footnotesize 
\begin{tabular}{@{\extracolsep{5pt}} cccc} 
\\[-1.8ex]\hline 
\hline \\[-1.8ex] 
Event Day & Estimate & 95\% CI lower & 95\% CI upper \\ 
\hline \\[-1.8ex] 
-7 & 0 & 0 & 0 \\ 
-6 & -0.19 & -0.7 & 0.33 \\ 
-5 & 0.21 & -0.45 & 0.95 \\ 
-4 & 0.55 & -0.37 & 1.47 \\ 
-3 & 0.46 & -1.03 & 1.92 \\ 
-2 & -0.54 & -2.55 & 1.27 \\ 
-1 & 0.19 & -1.86 & 2.37 \\ 
0 & 0.1 & -2.37 & 2.41 \\ 
1 & 0.14 & -2.68 & 2.93 \\ 
2 & 0.05 & -2.68 & 2.81 \\ 
3 & 0.09 & -2.71 & 2.8 \\ 
4 & -0.06 & -3.01 & 2.85 \\ 
5 & -0.88 & -3.92 & 2.15 \\ 
6 & -0.28 & -3.5 & 2.95 \\ 
7 & -0.4 & -4.08 & 2.98 \\ 
8 & -0.12 & -3.92 & 3.44 \\ 
\hline \\[-1.8ex] 
\multicolumn{4}{l}{Note: Total events = 19} \\ 
\end{tabular} 
\end{table}

\begin{table}[!htbp] \centering 
  \caption{Cumulative abnormal returns from assistant vice-minister and vice-minister appointments, internal appointments} 
  \label{tab: event_study_internal} 
\footnotesize 
\begin{tabular}{@{\extracolsep{5pt}} cccc} 
\\[-1.8ex]\hline 
\hline \\[-1.8ex] 
Event Day & Estimate & 95\% CI lower & 95\% CI upper \\ 
\hline \\[-1.8ex] 
-7 & 0 & 0 & 0 \\ 
-6 & -0.17 & -0.57 & 0.25 \\ 
-5 & -0.16 & -0.73 & 0.41 \\ 
-4 & -0.22 & -0.97 & 0.49 \\ 
-3 & -0.12 & -0.99 & 0.72 \\ 
-2 & 0.35 & -0.75 & 1.38 \\ 
-1 & 0.94 & -0.39 & 2.21 \\ 
0 & 2.17 & 0.81 & 3.5 \\ 
1 & 2.33 & 0.72 & 3.9 \\ 
2 & 2.4 & 0.52 & 4.25 \\ 
3 & 2.26 & 0.34 & 4.15 \\ 
4 & 2.08 & 0.03 & 4.08 \\ 
5 & 1.75 & -0.51 & 3.66 \\ 
6 & 1.79 & -0.56 & 3.77 \\ 
7 & 1.49 & -0.98 & 3.55 \\ 
8 & 1.55 & -0.76 & 3.53 \\ 
\hline \\[-1.8ex] 
\multicolumn{4}{l}{Note: Total events = 25} \\ 
\end{tabular} 
\end{table}

\begin{table}[!htbp] \centering 
  \caption{Cumulative abnormal returns after hiring former vice-ministers as consultants} 
  \label{tab: event_study_advisor_vm} 
\footnotesize 
\begin{tabular}{@{\extracolsep{5pt}} cccc} 
\\[-1.8ex]\hline 
\hline \\[-1.8ex] 
Event Day & Estimate & 95\% CI lower & 95\% CI upper \\ 
\hline \\[-1.8ex] 
-7 & 0 & 0 & 0 \\ 
-6 & -0.41 & -1.07 & 0.21 \\ 
-5 & -0.5 & -1.9 & 0.66 \\ 
-4 & -0.37 & -2.01 & 1.06 \\ 
-3 & -0.05 & -1.88 & 1.56 \\ 
-2 & -0.36 & -2.53 & 1.8 \\ 
-1 & 1.12 & -1.55 & 3.77 \\ 
0 & 2.14 & -0.18 & 4.7 \\ 
1 & 3.2 & 0.32 & 5.95 \\ 
2 & 3.47 & -0.16 & 6.72 \\ 
3 & 3.07 & -0.62 & 6.35 \\ 
4 & 2.75 & -1.08 & 5.92 \\ 
5 & 2.5 & -1.39 & 5.53 \\ 
6 & 1.58 & -2.87 & 5.1 \\ 
7 & 0.72 & -3.87 & 3.75 \\ 
8 & 0.7 & -3.73 & 3.74 \\ 
\hline \\[-1.8ex] 
\multicolumn{4}{l}{Note: Total events = 9} \\ 
\end{tabular} 
\end{table}

\begin{table}[!htbp] \centering 
  \caption{Cumulative abnormal returns from assistant vice-minister and vice-minister appointments, METI appointments} 
  \label{tab: event_study_meti} 
\footnotesize 
\begin{tabular}{@{\extracolsep{5pt}} cccc} 
\\[-1.8ex]\hline 
\hline \\[-1.8ex] 
Event Day & Estimate & 95\% CI lower & 95\% CI upper \\ 
\hline \\[-1.8ex] 
-7 & 0 & 0 & 0 \\ 
-6 & 0.05 & -0.34 & 0.47 \\ 
-5 & 0.26 & -0.21 & 0.76 \\ 
-4 & 0.35 & -0.29 & 1.04 \\ 
-3 & 0.7 & -0.22 & 1.69 \\ 
-2 & 0.72 & -0.29 & 1.82 \\ 
-1 & 1.25 & -0.15 & 2.71 \\ 
0 & 2.35 & 0.72 & 4.02 \\ 
1 & 2.42 & 0.49 & 4.51 \\ 
2 & 2.23 & 0.31 & 4.42 \\ 
3 & 2.09 & 0.22 & 4.2 \\ 
4 & 1.85 & -0.1 & 4.08 \\ 
5 & 1.23 & -0.73 & 3.39 \\ 
6 & 1.92 & -0.09 & 4.25 \\ 
7 & 2.14 & 0.09 & 4.62 \\ 
8 & 2.47 & 0.42 & 4.76 \\ 
\hline \\[-1.8ex] 
\multicolumn{4}{l}{Note: Total events = 27} \\ 
\end{tabular} 
\end{table}

\begin{table}[!htbp] \centering 
  \caption{Cumulative abnormal returns from assistant vice-minister and vice-minister appointments, appointments from ministries other than METI} 
  \label{tab: event_study_other} 
\footnotesize 
\begin{tabular}{@{\extracolsep{5pt}} cccc} 
\\[-1.8ex]\hline 
\hline \\[-1.8ex] 
Event Day & Estimate & 95\% CI lower & 95\% CI upper \\ 
\hline \\[-1.8ex] 
-7 & 0 & 0 & 0 \\ 
-6 & -0.3 & -0.81 & 0.22 \\ 
-5 & 0.04 & -0.78 & 0.91 \\ 
-4 & 0.1 & -0.79 & 1.04 \\ 
-3 & -0.4 & -1.64 & 0.67 \\ 
-2 & -0.92 & -2.66 & 0.6 \\ 
-1 & -0.07 & -1.72 & 1.39 \\ 
0 & -0.25 & -2.18 & 1.42 \\ 
1 & -0.2 & -2.42 & 1.74 \\ 
2 & 0.2 & -2.17 & 2.25 \\ 
3 & 0.46 & -2.13 & 2.86 \\ 
4 & 0.63 & -2.01 & 3.25 \\ 
5 & 0.11 & -2.83 & 2.68 \\ 
6 & -0.21 & -3.23 & 2.41 \\ 
7 & -0.93 & -4.25 & 1.89 \\ 
8 & -1.05 & -4.51 & 1.92 \\ 
\hline \\[-1.8ex] 
\multicolumn{4}{l}{Note: Total events = 20} \\ 
\end{tabular} 
\end{table}

\newpage
\clearpage
\subsection{Subgroup effects} 

\begin{figure}[!htb]
\includegraphics{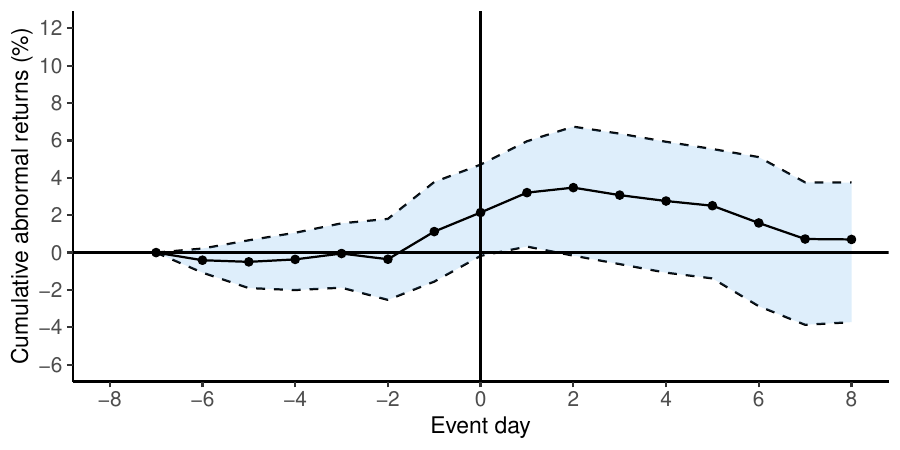}
\vspace{0.2cm}
\caption{Cumulative abnormal returns after hiring former vice-ministers as consultants}
\small
\vspace{-0.3cm}
\label{fig: event_study_advisor_vm}
{\footnotesize
\raggedright Note: Tabular results can be found in \autoref{tab: event_study_advisor_vm}. \par}
\end{figure}

\begin{figure}[!htb]
\includegraphics{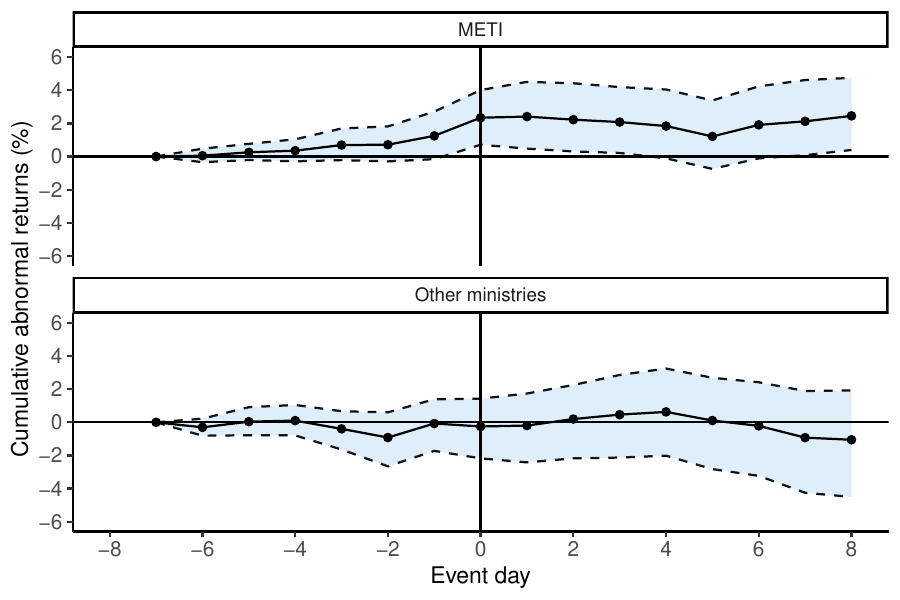}
\vspace{0.2cm}
\caption{Cumulative abnormal returns from assistant vice-minister and vice-minister appointments from METI vs. other ministries}
\small
\vspace{-0.3cm}
\label{fig: event_study_meti}
{\footnotesize
\raggedright Note: Tabular results can be found in \autoref{tab: event_study_meti} and \autoref{tab: event_study_other}. \par}
\end{figure}

\clearpage
\pagebreak
\subsection{Stock robustness} 

\begin{figure}[!htb]
\includegraphics{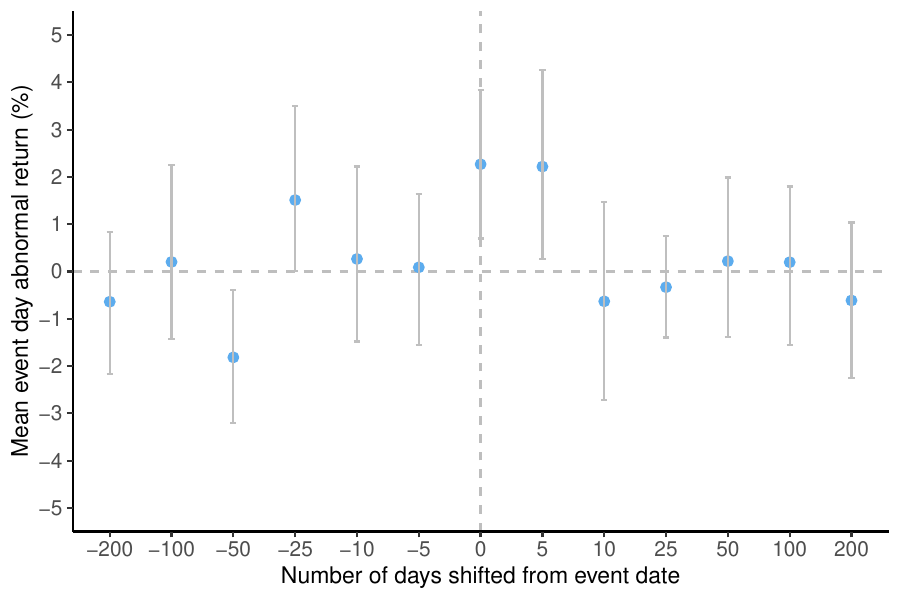}
\vspace{0.2cm}
\caption{Time-shifted placebo sensitivity analysis of mean event day abnormal return for internal hires}
\small
\vspace{-0.3cm}
\label{fig: event_study_placebo}
\end{figure}

\begin{figure}[!htb]
\includegraphics{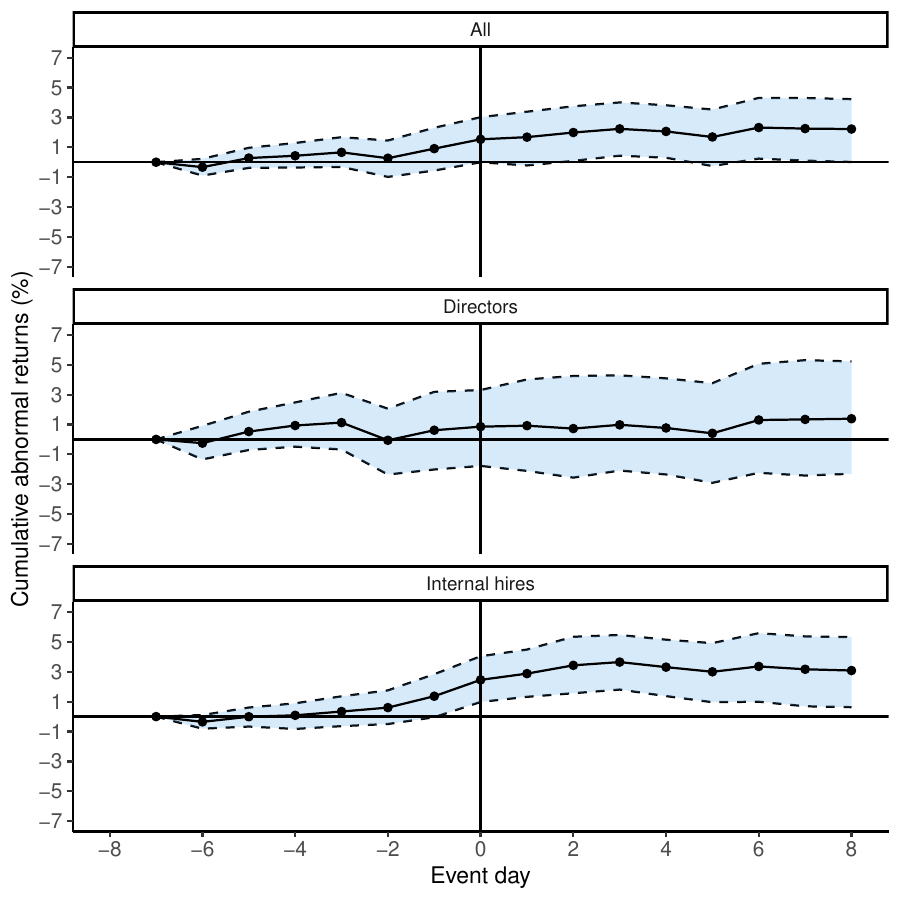}
\vspace{0.2cm}
\caption{Cumulative abnormal returns from assistant vice-minister and vice-minister appointments (constant mean return model)}
\small
\vspace{-0.3cm}
\label{fig: event_study_cmr}
\end{figure}

\clearpage

\begin{figure}[!htb]
\includegraphics{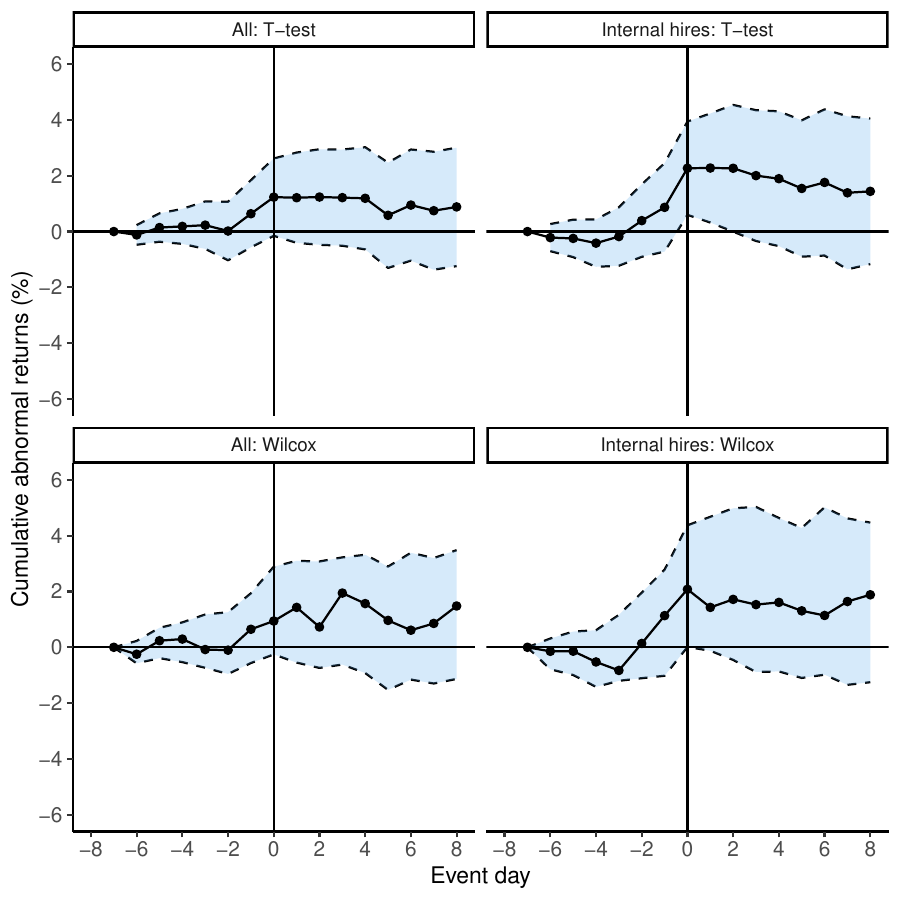}
\vspace{0.2cm}
\caption{Cumulative abnormal returns from assistant vice-minister and vice-minister appointments (95\% CIs from t-test and Wilcoxon rank test)}
\small
\vspace{-0.3cm}
\label{fig: event_study_classic_wilcox}
\end{figure}
\vspace{-0.5cm}
\noindent
\small
Note: Wilcoxon rank test charts plot median CARs rather than mean. 
\normalsize

\begin{figure}[!htb]
\includegraphics{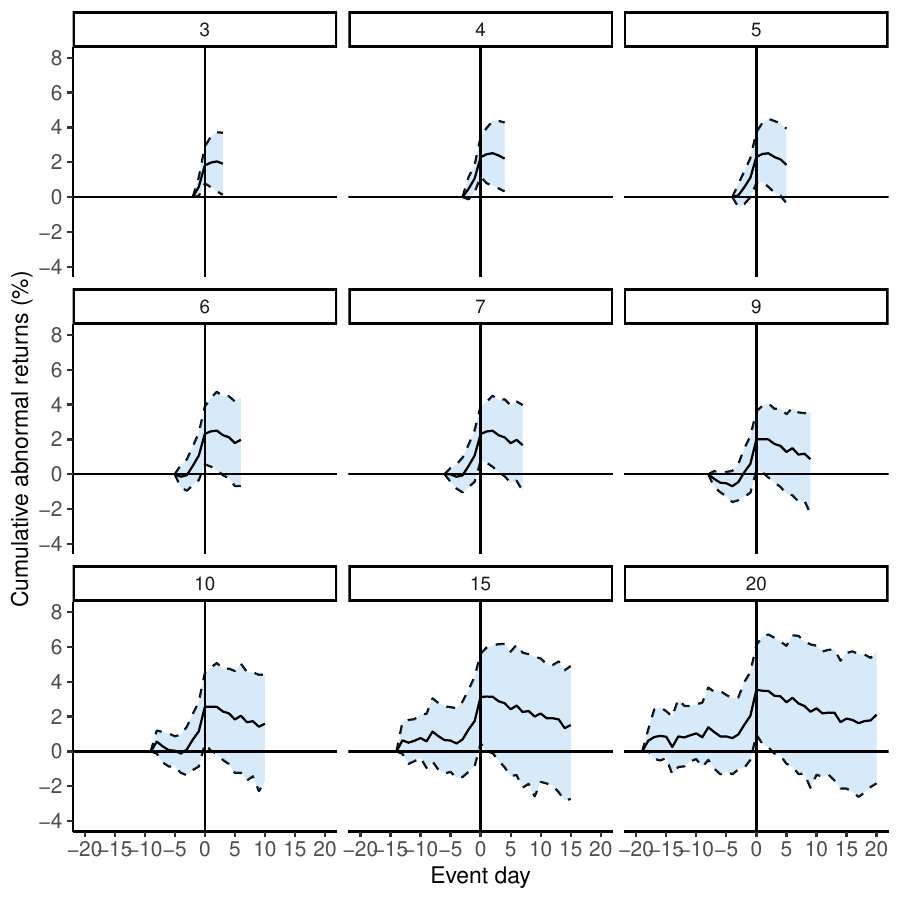}
\vspace{0.2cm}
\caption{Cumulative abnormal returns from internal assistant vice-minister and vice-minister appointments, by event window}
\small
\vspace{-0.3cm}
\label{fig: event_study_event_windows}
\end{figure}

\clearpage
\newpage
\section{Nonprofit contract value} \label{sec: nonprofit_appendix}

\subsection{Tabular results}

\begin{table}[!htbp] \centering 
  \caption{Effect of amakudari appointments on NPO negotiated contract value (binary outcome)} 
  \label{tab: npo_binary} 
\footnotesize 
\begin{tabular}{@{\extracolsep{5pt}} cccc} 
\\[-1.8ex]\hline 
\hline \\[-1.8ex] 
time & Effect & SE & N \\ 
\hline \\[-1.8ex] 
3 & -0.28 & 0.28 & 728 \\ 
2 & 0.3 & 0.19 & 1150 \\ 
1 & 0.17 & 0.15 & 1723 \\ 
0 & 0.68 & 0.11 & 2865 \\ 
\hline \\[-1.8ex] 
\end{tabular} 
\end{table}

\begin{table}[!htbp] \centering 
  \caption{Effect of amakudari appointments on NPO negotiated contract value (continuous outcome)} 
  \label{tab: npo_cont} 
\footnotesize 
\begin{tabular}{@{\extracolsep{5pt}} cccc} 
\\[-1.8ex]\hline 
\hline \\[-1.8ex] 
time & Effect & SE & N \\ 
\hline \\[-1.8ex] 
3 & -0.21 & 0.19 & 590 \\ 
2 & 0.07 & 0.11 & 988 \\ 
1 & 0.03 & 0.09 & 1508 \\ 
0 & 0.26 & 0.05 & 2711 \\ 
\hline \\[-1.8ex] 
\end{tabular} 
\end{table}

\clearpage
\newpage
\subsection{NPO robustness} 

\begin{figure}[!htb]
\begin{centering}
\includegraphics[width = \textwidth]{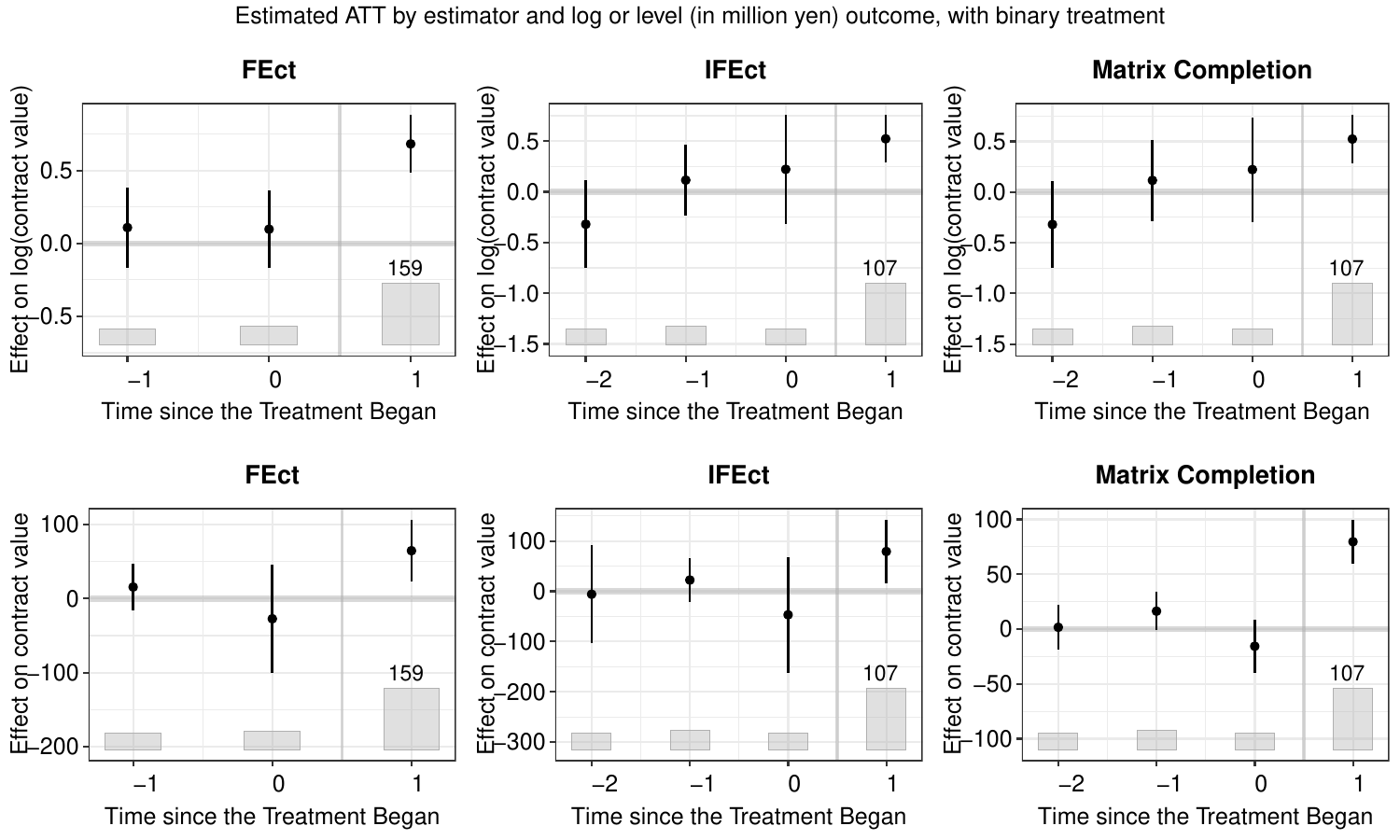}
\vspace{0.2cm}
\caption{FEct, IFEct, and MC estimators: effect of \textit{amakudari} appointments on NPO negotiated contract value, monthly aggregated data}
\small
\vspace{-0.3cm}
\label{fig: npo_robust_binary}
\end{centering}
\end{figure}

\begin{table}[!htbp] \centering 
\begin{tabular}{@{\extracolsep{5pt}}lcccc} 
\\[-1.8ex]\hline 
\hline \\[-1.8ex] 
 & \multicolumn{4}{c}{\textit{Dependent variable:}} \\ 
\cline{2-5} 
\\[-1.8ex] & \multicolumn{2}{c}{Contract value (log)} & \multicolumn{2}{c}{Contract value (million yen)} \\ 
\\[-1.8ex] & (1) & (2) & (3) & (4)\\ 
\hline \\[-1.8ex] 
 Hires (binary) & 0.416$^{***}$ &  & 31.755$^{*}$ &  \\ 
  & (0.071) &  & (18.378) &  \\ 
  & & & & \\ 
Hires (continuous) &  & 0.092$^{***}$ &  & $-$0.550 \\ 
  &  & (0.019) &  & (4.789) \\ 
  & & & & \\
\hline \\[-1.8ex] 
Observations & 6,575 & 6,575 & 6,575 & 6,575 \\ 
\hline 
\hline \\[-1.8ex] 
\textit{Note:}  & \multicolumn{4}{r}{$^{*}$p$<$0.1; $^{**}$p$<$0.05; $^{***}$p$<$0.01} \\ 
\end{tabular} 
\caption{Two-way fixed effects estimates of \textit{amakudari} appointments on NPO negotiated contract value, monthly aggregation}
  \label{tab: twfe_month} 
\end{table} 

\begin{figure}[!htb]
\begin{centering}
\includegraphics{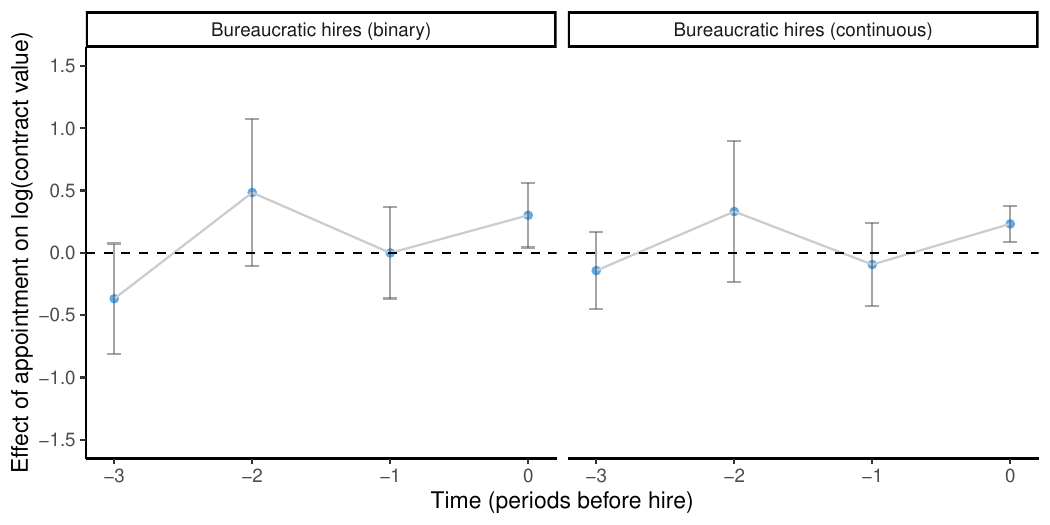}
\vspace{0.2cm}
\caption{$DID_M$ estimator effect of \textit{amakudari} appointments on log(NPO negotiated contract value), yearly aggregated data}
\small
\vspace{-0.3cm}
\label{fig: npo_amakudari_yearly}
\end{centering}
\end{figure}

\begin{figure}[!htb]
\begin{centering}
\includegraphics[width = \textwidth]{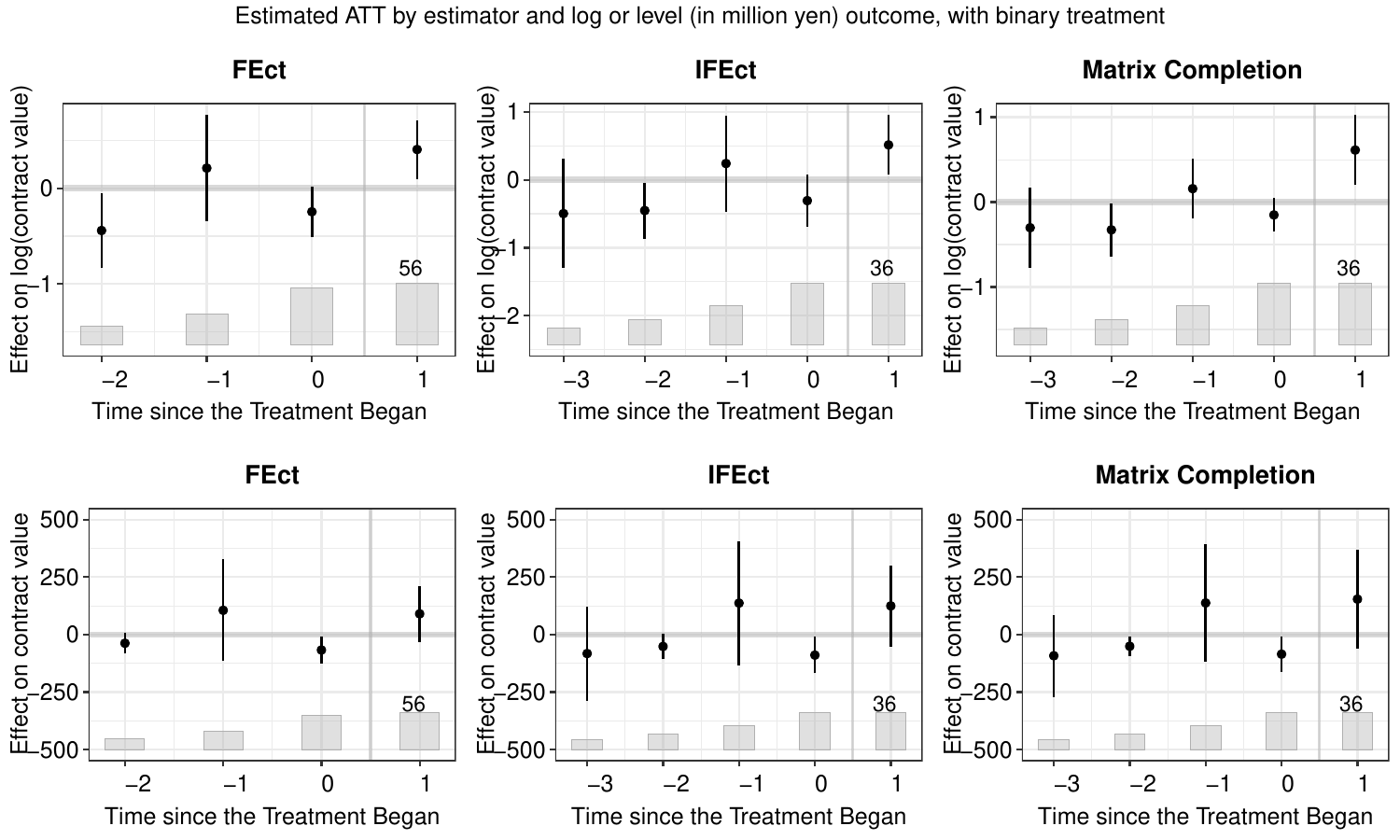}
\vspace{0.2cm}
\caption{FEct, IFEct, and MC estimators: effect of \textit{amakudari} appointments on NPO negotiated contract value, yearly aggregated data}
\small
\vspace{-0.3cm}
\label{fig: npo_twfe}
\end{centering}
\end{figure}

\pagebreak

\begin{table}[!htbp] \centering 
\begin{tabular}{@{\extracolsep{5pt}}lcccc} 
\\[-1.8ex]\hline 
\hline \\[-1.8ex] 
 & \multicolumn{4}{c}{\textit{Dependent variable:}} \\ 
\cline{2-5} 
\\[-1.8ex] & \multicolumn{2}{c}{Contract value (log)} & \multicolumn{2}{c}{Contract value (million yen)} \\ 
\\[-1.8ex] & (1) & (2) & (3) & (4)\\ 
\hline \\[-1.8ex] 
 Hires (binary) & 0.274$^{***}$ &  & 6.180 &  \\ 
  & (0.085) &  & (41.165) &  \\ 
  & & & & \\ 
 Hires (continuous) &  & 0.082$^{***}$ &  & 4.352 \\ 
  &  & (0.019) &  & (9.318) \\ 
  & & & & \\ 
\hline \\[-1.8ex] 
Observations & 3,480 & 3,480 & 3,480 & 3,480 \\ 
\hline 
\hline \\[-1.8ex] 
\textit{Note:}  & \multicolumn{4}{r}{$^{*}$p$<$0.1; $^{**}$p$<$0.05; $^{***}$p$<$0.01} \\ 
\end{tabular} 
\caption{Two-way fixed effects estimates of \textit{amakudari} appointments on NPO negotiated contract value, yearly aggregation}
  \label{tab: twfe_year} 
\end{table}


\clearpage
\pagebreak
\subsection{Benford's Law}

\begin{figure}[!htb]
\centering
\begin{subfigure}{\textwidth}
\centering
\includegraphics[width = 0.33\textwidth]{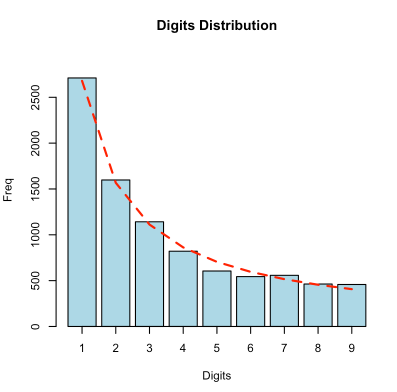}
\caption{Competitive bid contracts} 
\label{fig: benford_bid}
\end{subfigure}
\begin{subfigure}{\textwidth}
\centering
\includegraphics[width = 0.33\textwidth]{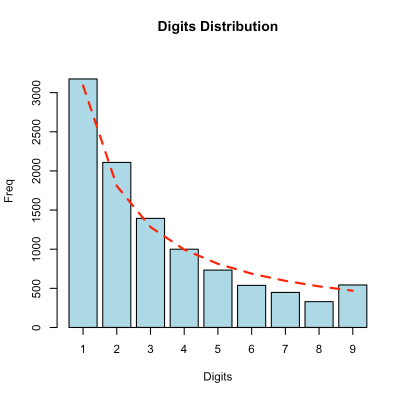}
\caption{Negotiated contracts}
\label{fig: benford_neg}
\end{subfigure}
\begin{subfigure}{\textwidth}
\centering
\includegraphics[width = 0.33\textwidth]{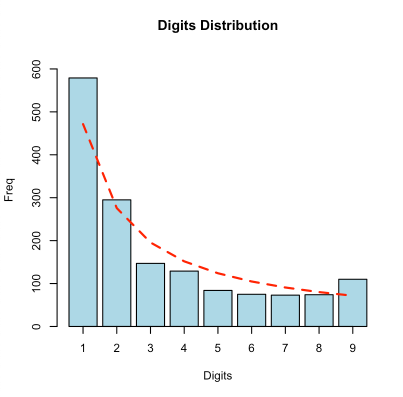}
\caption{Negotiated contracts when former bureaucrat in director position}
\end{subfigure}
\caption{Distribution of first digits: actual distribution in blue and predicted distribution according to Benford's Law in red}
\label{fig: benford}
\end{figure}

\end{document}